\documentclass[iop,apj,twocolappendix,letterpaper,twocolumn]{emulateapj}

\usepackage{graphicx}
\usepackage{txfonts}
\usepackage{amssymb}
\usepackage[breaklinks, colorlinks, citecolor=blue]{hyperref}
\usepackage{natbib}
\usepackage{booktabs}
\newcommand \CO {$^{12}$CO}
\newcommand{\kms} {\, \rm km\,s^{-1}}
\newcommand \pc {{\,\rm pc}}
\newcommand \yr {{\,\rm yr}}
\newcommand \s {{\,\rm s}}
\newcommand \erg {{\, \rm erg}}
\newcommand{\be} {\begin{equation}}
\newcommand{\ee} {\end{equation}}
\renewcommand{\apjl} {ApJL}

\shorttitle{Physical properties of molecular clouds for the entire Milky Way disk} 
\shortauthors{Miville-Desch\^enes, Murray, \& Lee}

\begin{document}

\title{Physical properties of molecular clouds for the entire Milky Way disk}

\author{Marc-Antoine Miville-Desch\^enes\altaffilmark{1}, Norman Murray\altaffilmark{2,3}, Eve J. Lee\altaffilmark{4}}

\altaffiltext{1}{Institut d'Astrophysique Spatiale, CNRS, Univ. Paris-Sud, Universit\'e Paris-Saclay, B\^at. 121, F-91405 Orsay, France; mamd@ias.u-psud.fr}
\altaffiltext{2}{Canadian Institute for Theoretical Astrophysics, 60 St. George Street, University of Toronto, Toronto, ON M5S 3H8, Canada; murray@cita.utoronto.ca}
\altaffiltext{3}{Canada Research Chair in Astrophysics}
\altaffiltext{4}{Astronomy Department, University of California, Berkeley, CA 94720, USA; evelee@berkeley.edu}

\begin{abstract}
This study presents a catalog of 8107 molecular clouds that covers the entire Galactic plane and includes 98\% of the $^{12}$CO emission observed within $b\pm5^\circ$. The catalog was produced using a hierarchical cluster identification method applied to the result of a Gaussian decomposition of the Dame et al. data. The total H$_2$ mass in the catalog is $1.2\times10^9$\,M$_\odot$, in agreement with previous estimates. We find that 30\% of the sight lines intersect only a single cloud, with another 25\% intersecting only two clouds. The most probable cloud size is $R\sim30$\,pc. We find that $M\propto\,R^{2.2\pm0.2}$, with no correlation between the cloud surface density, $\Sigma$, and $R$. In contrast with the general idea, we find a rather large range of values of $\Sigma$, from 2 to $300\,M_\odot$\,pc$^{-2}$, and a systematic decrease with increasing Galactic radius, $R_{\rm gal}$. The cloud velocity dispersion and the normalization $\sigma_0=\sigma_v/R^{1/2}$ both decrease systematically with $R_{\rm gal}$. When studied over the whole Galactic disk, there is a large dispersion in the line width-size relation, and a significantly better correlation between $\sigma_v$ and $\Sigma\,R$. The normalization of this correlation is constant to better than a factor of two for $R_{\rm gal}<20$\,kpc. This relation is used to disentangle the ambiguity between near and far kinematic distances. We report a strong variation of the turbulent energy injection rate. In the outer Galaxy it may be maintained by accretion through the disk and/or onto the clouds, but neither source can drive the 100~times higher cloud-averaged injection rate in the inner Galaxy. 
\end{abstract}

\keywords{ISM: clouds---Galaxy: local interstellar matter---Methods: data analysis}

\section{Introduction}

Molecular gas accounts for a small fraction of the baryonic matter in nearby galaxies, typically 10\% by mass, and an even smaller fraction in galaxy halos, at any redshift. Despite this fact, astronomers have devoted enormous amounts of telescope time to observations of molecular material. The primary reason for this effort is because of the crucial role of molecular gas in star formation and hence in galaxy evolution. For example, stars form in molecular gas \citep{myers1986,scoville1989}. Further, the interactions of newly formed stars with their (initially molecular) surroundings, collectively know as ``feedback,'' are believed to moderate both the rate of star formation and, via expulsion of molecular, atomic, and ionized gas, the cumulative stellar mass in the host galaxy. In other words, the amount, distribution, and kinematics of the molecular gas in a galaxy provide important constraints on the physics of star formation and galaxy evolution. 

Since its discovery by \citet{wilson1970}, 
CO line emission has been the primary tracer of molecular gas in galaxies.
In the Milky Way, the large-scale properties of molecular emission and the associated H$_2$ gas were deduced from two main sets of observations. First, data obtained with the CfA 1.2m telescopes \citep[one in New York and one in Chile;][]{cohen1980,dame1985,dame1987,bronfman1989a} were combined by \citet{dame2001} to produce a data set covering most of the sky with significant CO emission with an angular resolution of about 8.5\,arcmin. Various subsets of these data were used by \citet{grabelsky1987,grabelsky1988}, \citet{bronfman1988}, \citet{rosolowsky2006}, \citet{nakanishi2006} and \citet{rice2016} to infer global properties of the molecular medium. 

The second large observational effort was conducted by the Boston University group using the 14\,m FCRAO telescope. Their data sets \citep[available in $^{12}$CO and in $^{13}$CO;][]{clemens1986,sanders1986,heyer1998,jackson2006} cover a smaller area on the sky but at higher angular resolution (46''). 
These data are the basis of the analysis of Galactic molecular clouds by \citet{solomon1987}, \citet{heyer2001}, \citet{brunt2003b}, \citet{rathborne2009} and \citet{heyer2009}. 

The analysis of molecular emission, especially CO, has been done in two main ways: (1) statistical properties of the entire emission, or (2) segmentation of the data into what are thought to be relevant structures. The statistical analysis of the emission of targeted regions provides highly valuable information about the nature of the processes involved in the structure and dynamics of star-forming regions \citep{falgarone1992a,miesch1994,heithausen1998,ossenkopf2002}. This approach relies on the understanding of projection effects to relate the observed statistical properties (intensity, centroid velocity) to physical three-dimensional properties. Often, this type of analysis makes use of a comparison with numerical simulations \citep[e.g.,][]{ostriker2001,ossenkopf2002,brunt2003a}.

The second approach relies on algorithms to identify coherent structures in the data.
Over most of the Galactic plane, the \CO\ emission can often be described as a sum of a few isolated velocity components. In a position--position--velocity (PPV) cube, the molecular emission appears clumpy. This is particularly true toward the outer Galaxy and in the solar neighborhood.
As highlighted by \citet{grabelsky1987}, this is very different from what is seen in 21\,cm PPV cubes, where the emission is much more spread out in velocity space; the large velocity spread in 21\,cm emission is believed to be due principally  to the high temperature and large volume filling factor of the warm neutral medium (WNM). The CO molecule is heavier than either H or H$_2$, and the observed excitation temperature is $<100\,$K, so that its thermal line width is much smaller than that of H or H$_2$. In general, CO emission generally appears in cold and dense structures of the interstellar medium (ISM), so the CO emission line has a relatively low velocity dispersion, which facilitates the identification of isolated structures. 

Given the fragmented structure of the molecular emission, several attempts have been made to identify coherent structures from the CO data.
One technique is to set a threshold in CO brightness temperature, $T_{\rm B}^{\rm min}$, then identify all the connected emission structures in PPV space that have $T_{\rm B} > T_{\rm B}^{\rm min}$. The choice of $T_{\rm B}^{\rm min}$ has a significant impact on the result. It can be a factor $n$ times the noise level or, in areas with strong emission, a value that allows for the identification of structures above a confusion background. One way to improve on this is to look for hierarchical coherence within structures identified with different threshold levels, e.g., {\em clumpfind} \citep{williams1994} or the dendrogram technique \citep{rosolowsky2008}.
Another method is to use algorithms designed to identify structures with predefined shape like {\em gaussclump} \citep{stutzki1990}. In {\em gaussclump} the shape of a structure is assumed to be an ellipsoid in PPV space. 

No method gives a perfectly satisfying solution, in part because unrelated structures might appear connected in a PPV cube if they have similar (projected) velocities. This is particularly true in the inner Galaxy, where velocity crowding mixes emission from different clouds along the line of sight, making a unique identification almost impossible. In addition, the boundaries of clouds become more blurred as the volume filling factor of molecular matter increases, especially in the ``molecular ring,'' seen between 3 and 6\,kpc from the Galactic center. 

The structure of the molecular ISM itself makes the identification difficult. Molecular clouds have been described in varying ways, ranging from ``spherical gravitationally bounded structures'' to ``a clumpy, turbulent and multi-phase medium.'' In the latter case, the boundaries of a cloud are ill-defined, and the identification of clouds can be somewhat arbitrary. The latter picture is favored by what higher angular resolution CO data have revealed. On the other hand, large coherent molecular structures are readily identified in low-resolution data \citep[e.g.,][]{solomon1987}. These are the so-called Giant Molecular Clouds or GMCs. When observed at higher angular resolution, these large puffy clouds appear fragmented with abundant small-scale structures \citep{jackson2006,goldsmith2008}. The identification of ``clouds'' from  high-resolution data results in a much larger number of smaller structures. 

Despite all those difficulties, it is important to try to decompose molecular emission in isolated structures, even if they are somewhat arbitrary. For example, \citet{scoville1989} and \citet{myers1986} both found that most of the star formation in the first  galactic quadrant of the Milky Way takes place in readily identified GMCs. This strongly suggests that GMCs are physically relevant objects, even if we have a hard time identifying them in a unique manner.

Applying different detection schemes to the same data set leads to different cloud catalogs. Yet, these different sets of clouds follow surprisingly concordant statistical properties, e.g., similar distributions of cloud mass and similar size-line width relations.

In this context, why produce another catalog of clouds?

Most early studies of the Milky Way's molecular medium were confined to a limited range in Galactic longitude, so they were clearly not complete catalogs.  More recently, \citet{nakanishi2006}, \citet{pohl2008} and \citet{rice2016} exploited the full coverage of the Galactic plane of the \citet{dame2001} \CO\ atlas.
\citet{nakanishi2006} established some large-scale properties of the molecular gas (scale height, surface density, average density) as a function of galactocentric radius using a global fit of the whole \citet{dame2001} data set. They assumed that the spatial distribution of the molecular gas in the Galactic disk is the sum of two axisymetric Gaussian distributions as a function of Galactic height $z$. \citet{rice2016} identified clouds using a dendrogram technique. 
However, previous studies that identified clouds from CO data managed to include at most 40\% of the CO emission in their clouds \citep{solomon1987,solomon1989,rice2016}.

In this paper we propose an alternative analysis of the \citet{dame2001} data to estimate the large-scale physical properties of molecular clouds of the Milky Way. Our analysis is based on a combination of the classical Gaussian decomposition of the CO emission spectra with a cluster analysis method used to identify coherent structures in the data.
The method presented here is able to recover more than 90\% of the CO emission. Based on this new catalog of clouds, the most complete done to date, we are able to study the dynamical properties of GMCs and their spatial variations in the Milky Way disk.

The paper is organized as follows.
The data segmentation and clustering methods are described in Sect.~\ref{sec:clouds}. 
The physical properties given in the molecular cloud catalog are detailed in Sect.~\ref{sec:properties}.
The main results of the paper are presented in Sect.~\ref{sec:results} and discussed in Sect.~\ref{sec:discussion}.

\section{From CO Emission to Clouds}

\label{sec:clouds}

\subsection{Data}

The data set used here is that of \citet{dame2001}. Those authors combined observations obtained over a period of 20\,yr with two telescopes, one in the north (first located in New York City and then in Cambridge, Massachusetts) and one in the south (Cerro Tololo, Chile). These 1.2\,m telescopes have an angular resolution of $\sim 8.5'$ at 115\,GHz, the frequency of the $^{12}$CO\,1--0 line. 
For the current study we used the data set covering the whole Galactic plane with $\pm 5^\circ$ in Galactic latitude.\footnote{Available at https://www.cfa.harvard.edu/rtdc/CO} We used the {\em raw} data set where no interpolation was done, neither in $v$ nor in Galactic coordinates $l-b$. The channel width of the whole data set is 1.3\,km\,s$^{-1}$. 
Due to the combination of data from different eras and different instruments, the noise level of the map is not uniform over the data set. Figure~\ref{fig:PDF_noise} in Appendix~\ref{sec:gaussfit} shows the histogram of the noise level. There are three peaks in the distribution at 0.06, 0.10, and 0.19\,K.

\subsection{The Cloud Identification Method}

The cloud identification method presented here uses a clustering hierarchical algorithm to identify coherent structures in PPV space. In that sense, it is in the spirit of {\em clumpfind} \citep{williams1994} or of the dendrogram technique \citep{rosolowsky2008}. On the other hand, the analysis proposed here differs from previous ones as the clustering algorithm is not applied on the observed brightness, $T_{\rm B}$. Instead, coherent structures are looked for in a more sparse description of the data resulting from a Gaussian decomposition of the CO spectra. 
This has the advantage of significantly reducing the confusion due to cloud overlap in PPV space and therefore facilitating the identification of isolated structures. The projection of the data onto a set of Gaussian components also limits the effect of noise on the structure identification process, and it permits us to include the column density of clouds down to the sensitivity limit of the survey. It also allows us to identify faint clouds that would be missed by methods based on $T_{\rm B}$ thresholds. 

The approach we present here results in a catalog of molecular clouds that includes 98\% of the CO emission of the \citet{dame2001} survey, more than a factor of two higher than any other cloud identification method has done to date.

The overall cloud identification procedure is divided into two main steps. First, the entire data cube is decomposed into a set of Gaussian functions. Second, coherent structures are identified using a hierarchical cluster analysis scheme. 

\subsection{Gaussian Decomposition}

For each sky position of the cube, the CO spectrum, $T_{\rm B}(v)$, is described as a sum of Gaussian components:
\begin{equation}
  \label{eq:gaussian}
T_{\rm B}^\prime(v) = \sum_{i=1}^N A_i \, \exp( (v-v_i)^2/2\sigma_i^2)
\end{equation}
where $A_i$, $v_i$ and $\sigma_i$ are the amplitude, centroid, and width of each Gaussian, respectively.
Decomposing $T_{\rm B}(v)$ into a sum of Gaussian components is a way to compress the information. 
A spectrum of typically a few hundred samples with complex \CO\ emission can be described by only a few tens of Gaussian coefficients. This decomposition is valid even if the lines are not Gaussian; it is simply another way of describing the data. In addition, this compression of information has the potential to reduce the noise. Most of the noise is at high (spatial) frequency. An isolated Gaussian (i.e., emission not seen in neighboring positions at the same central velocity) with a width of one or two channels and an amplitude lower than 3 times the noise level is probably noise. 

\begin{figure}
\centering
\includegraphics[draft=false, angle=0]{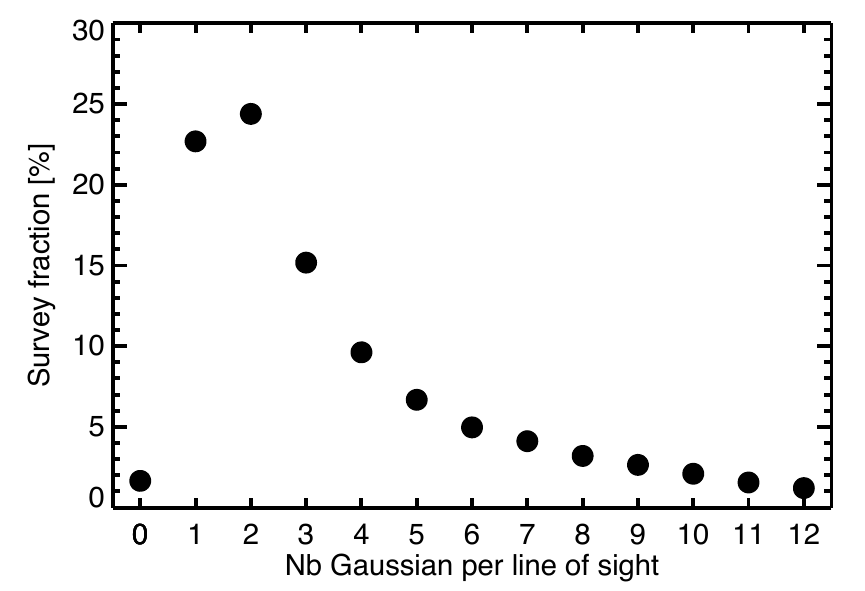}
\caption{\label{fig:PDF_nbGauss} Histogram of the number of Gaussian components used to fit a CO profile. The $y$ axis is given in fraction of the total number of pixels observed in the \citet{dame2001} survey for $-5^\circ < b < 5^\circ$.}
\end{figure}

The choice of a Gaussian decomposition is motivated by the fact that, on most of the sky, the \CO\ spectra are relatively simple and show only single or double components (Figure~\ref{fig:example_spectra} top panel). On lines of sight in the vicinity of the Galactic center, the \CO\ spectra are much more complex, but they can also be described by the sum of a limited number of Gaussian functions (Figure~\ref{fig:example_spectra} bottom panel). For these more complex spectra the decomposition on a Gaussian basis is clearly not unique. To guide the search, we look for solutions that have a spatial coherence; an iterative process is being used to favor solutions that are close to what is found in neighboring positions. The algorithm used is described in Appendix~\ref{sec:gaussfit}. 

Each spectrum was fitted with a maximum of $N=12$ Gaussian components. Given the sensitivity of the \citet{dame2001} data, this appears to be enough to describe the morphological complexity of the $^{12}$CO spectra, even in the inner part of the Galactic plane. Examples of Gaussian decompositions are shown in Figure~\ref{fig:example_spectra}.

This data segmentation resulted in the identification of about $5.4 \times 10^5$ Gaussian components.  The occurrence of the number of Gaussian components per spectrum is given in Figure~\ref{fig:PDF_nbGauss}. About two-thirds of the lines of sight could be decomposed using only one, two, or three Gaussian components. A mere 1.7\% of the sky positions have zero Gaussian components, and only 1.2\% of the spectra needed 12 components to describe the emission down to the noise level. 

\begin{figure*}
\centering
\includegraphics[draft=false, angle=0, width=15cm]{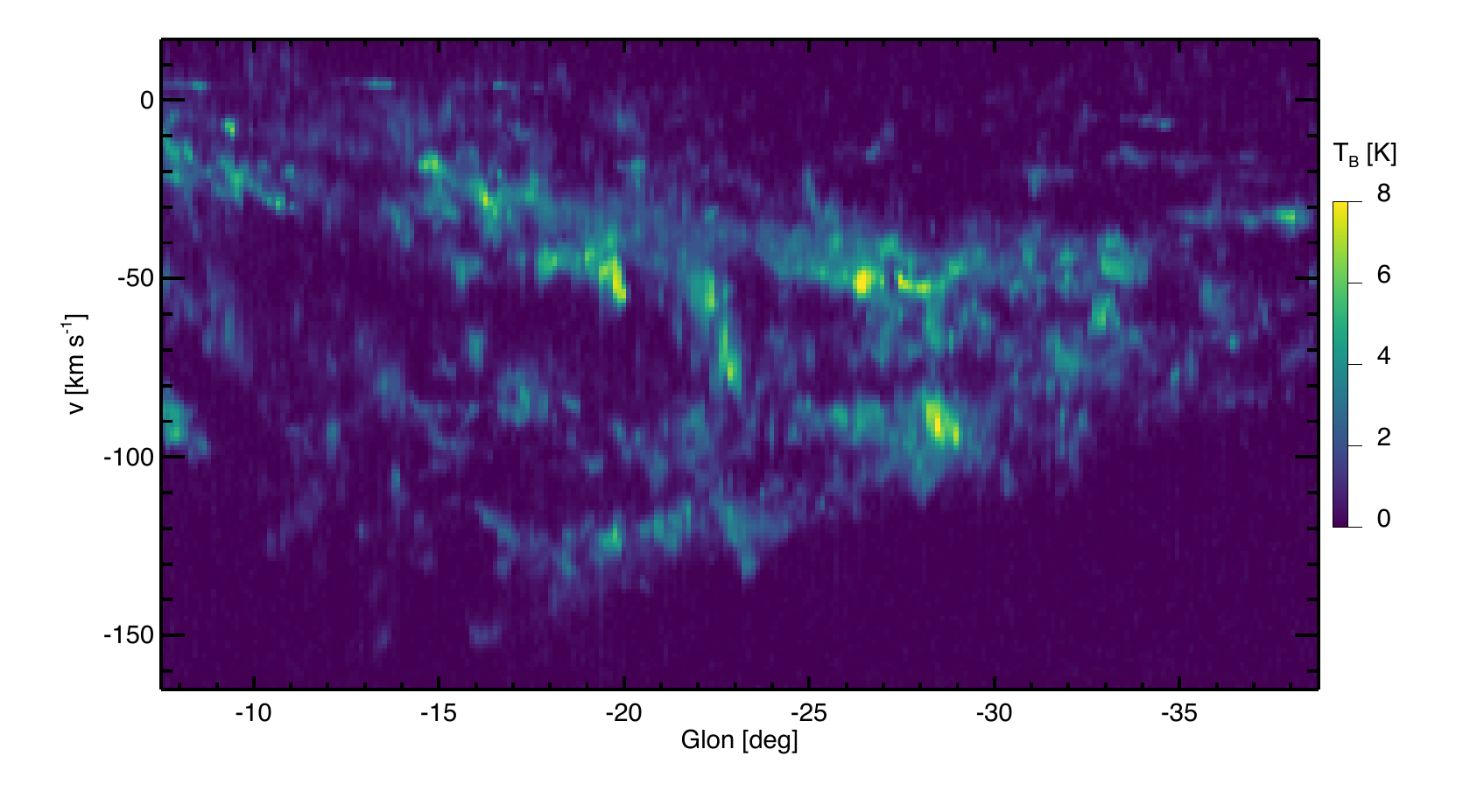}
\includegraphics[draft=false, angle=0, width=15cm]{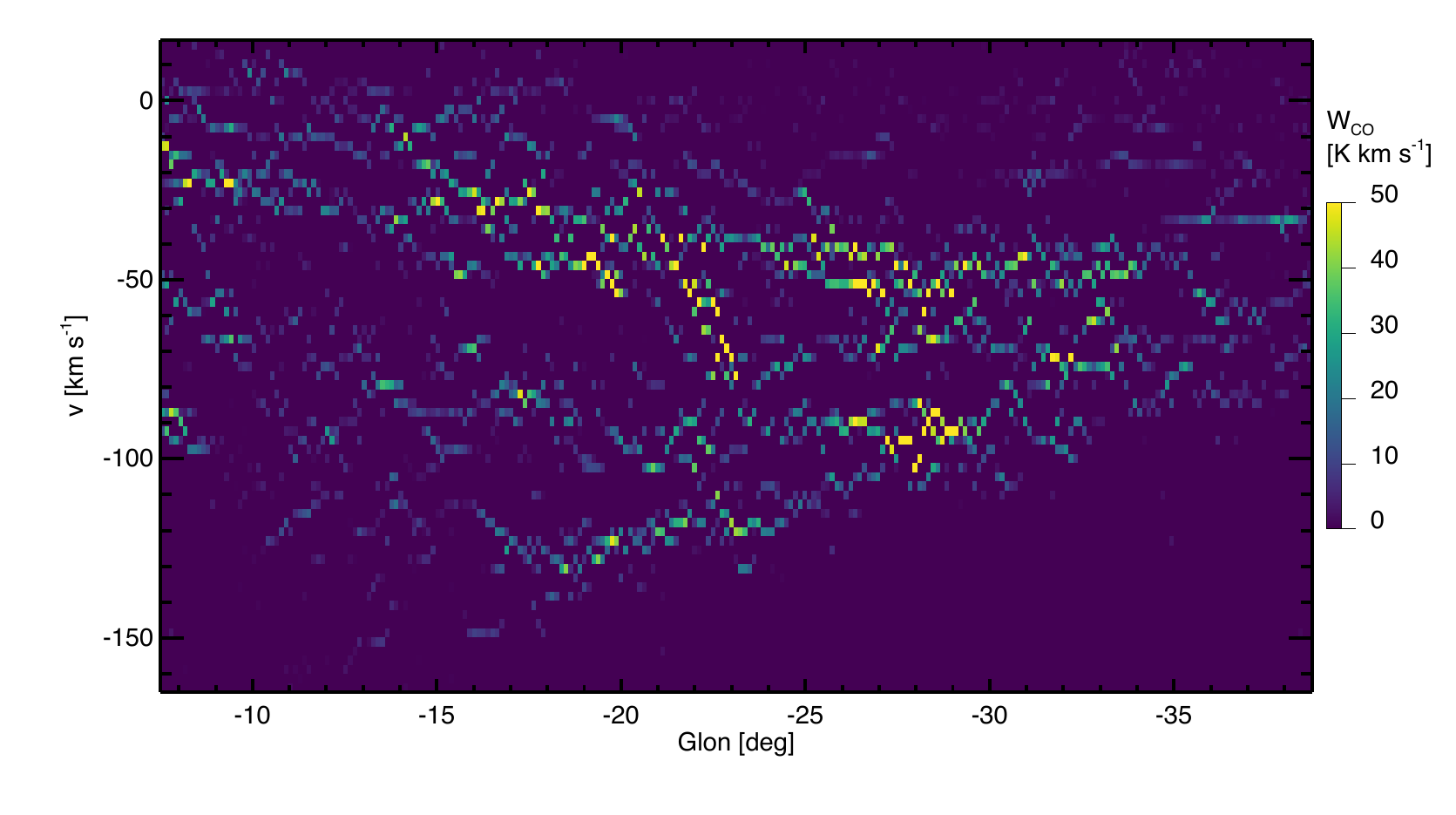}
\caption{\label{fig:lv_c_c2w} Longitude--velocity diagrams at GLAT$\,=0^\circ$. {Top}: log of brightness temperature, $T_{\rm B}$, in units of K. {Bottom}: integrated emission of the Gaussian components concentrated at their central velocities, $W_{\rm CO}$ -- see Eq.~\ref{eq:wco_cube}, in units of K\,km\,s$^{-1}$. }
\end{figure*}

\subsection{Cluster Analysis}

\begin{figure}
\centering
\includegraphics[draft=false, angle=0]{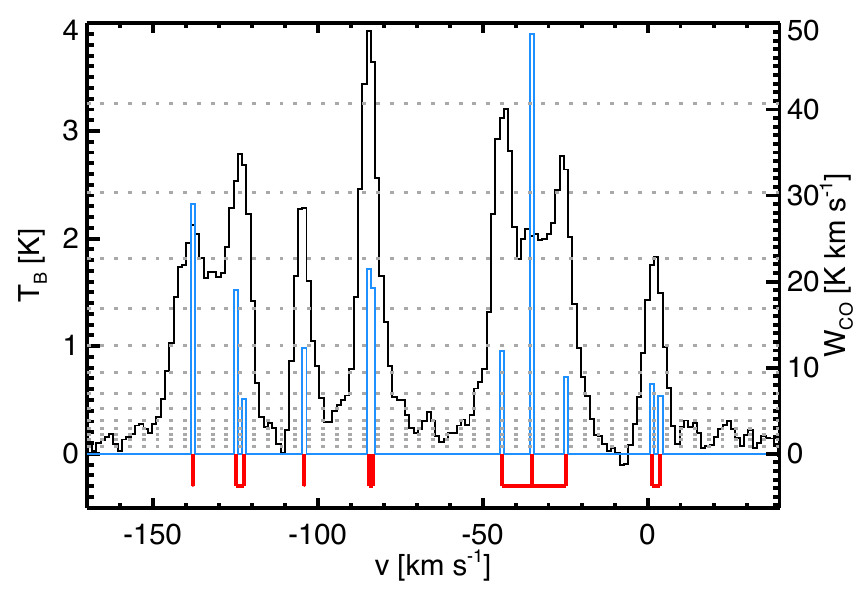}
\caption{\label{fig:example_spectra_cluster} Example of cluster identification for a complex CO spectrum located at $(l,b)=(-18.5^\circ, 0^\circ)$. The solid black line is the observed spectrum, in brightness temperature in kelvin (left vertical axis). The vertical blue line segments indicate the integrated emission, in $W_{\rm CO}$, units of K\,km\,s$^{-1}$ (right vertical axis), of each Gaussian component. The red lines indicate how the Gaussian components were grouped into clouds. This spectrum was decomposed into 11 Gaussian components that were grouped into six different clouds. The horizontal dotted lines indicate the $W_{\rm CO}$ levels (logarithmically spaced) used in the threshold descent.}
\end{figure}

\subsubsection{The Integrated Emission Cube}

The usefulness of the Gaussian decomposition in the identification of molecular clouds is best seen in Figures~\ref{fig:lv_c_c2w} and \ref{fig:example_spectra_cluster}. The top panel of Figure~\ref{fig:lv_c_c2w} presents a 2D cut through the CO data cube for a constant value of Galactic latitude (a so-called $l-v$ diagram). Here we show a particularly complex region of the sky in the fourth quadrant. This representation of $T_{\rm B}(l,b,v)$ allows us to identify coherent structures that blend in an integrated intensity map. Even though individual clouds can be identified by eye, there is significant confusion and overlap between structures. Looking at this representation, one can appreciate why thresholding methods have been used in the past to isolate structures. But this isolation comes at the expense of leaving out the faint emission, below the lowest threshold level.

The Gaussian decomposition is of great help in clearing some of the confusion. The bottom panel of Figure~\ref{fig:lv_c_c2w} shows the same $l-v$ diagram but for a cube representing the integrated emission of the Gaussian components:
\begin{equation}
\label{eq:wco_cube}
W_{\rm CO}(l,b,v) = \sqrt{2\pi} \,A(v) \,\sigma(v).
\end{equation}
    The cube $W_{\rm CO}(l,b,v)$ contains the integrated emission of every Gaussian component put at the central velocity position ($v$). In some cases, more than one Gaussian component will have the same central velocity $v$ at a given ($l$, $b$) position. In such cases, $W_{\rm CO}(l,b,v)$ contains the sum of the column densities of all the components.

    This cube contains all the emission of the original data cube described by the Gaussian decomposition, but without the spread of the velocity dispersion of each component. 
Figure~\ref{fig:example_spectra_cluster} provides an example of a complex spectrum of the cube, where the black solid line is the observed spectrum, and the vertical blue solid line segments represent the integrated emission ($W_{\rm CO}$ - right axis) of each Gaussian component at its respective velocity. This spectral representation and the $l-v$ cut of $W_{\rm CO}$ (Figure~\ref{fig:lv_c_c2w}) illustrate how the Gaussian decomposition helps in separating blended emission and in identifying coherent structures.

\subsubsection{Hierarchical Cluster Identification}

The term `clustering analysis' refers to the task of identifying coherent groups in a data set.
The appropriate clustering algorithm and parameter settings depend on the individual data set and on the intended use of the results.
Here the cluster identification was done on the $W_{\rm CO}$ cube using a classical hierarchical algorithm with a threshold descent. The identification of coherent structures or clusters is done in position ($l$, $b$) and velocity ($v$) space at the same time. Because of this, and because it uses a threshold descent, the algorithm presented here has some similarity with {\em clumpfind} \citep{williams1994}.

In detail, a first set of clusters was identified by grouping neighboring pixels with $W_{\rm CO}$ higher than a threshold value. This selection on $W_{\rm CO}$ creates islands in PPV space. The threshold value is then lowered and a new set of pixels in $W_{\rm CO}$ are selected. Lowering the threshold reveals the outskirts of already-defined clusters, as well as finding new islands with fainter peak emission. Each Gaussian component is either attached to a preexisting cluster, if one is nearby, or it is identified as the first component of a new cluster. This descent is continued for a number $N_{\rm th}$ of thresholds down to a minimum value.

In the descent we used $N_{\rm th}=15$ threshold values, spaced logarithmically; see Figure~\ref{fig:example_spectra_cluster}, where the horizontal gray lines depict the thresholds. The highest threshold value was set to $W_{\rm CO} = 100$~K\,km\,s$^{-1}$, while the lowest threshold value was $W_{\rm CO}=0.8$~K\,km\,s$^{-1}$.
The lowest threshold value corresponds to a Gaussian amplitude close to the noise level. For a narrow component with $\sigma_i=1.3$\,km\,s$^{-1}$, the lowest threshold corresponds to $A_i=0.25$\,K; for 93\% of the spectra the noise level is lower than $0.15$\,K (see Figure~\ref{fig:PDF_noise}). 

At each step of the process, we look for clusters that should be merged. We applied a strict criterion for cluster merging in order to avoid any runaway process that would lead to immense structures. Such overlinking is a common flaw of friends-of-friends algorithms. It is also known as the `chaining phenomenon' in single linkage clustering methods, where two clusters are linked together even if only a single element in each cluster is close to each other.

A cluster $i$ is merged to a cluster $j$ if its central coordinate is enclosed within the volume of $j$:
\begin{eqnarray}
|l_i-l_j| & < & 2 \, \sigma_{l,j} \\ 
|b_i-b_j| & < & 2 \, \sigma_{b,j} \\ 
|\langle v_i \rangle - \langle v_j \rangle| & < & 2 \, \sigma_{v,j},
\end{eqnarray}
where $\sigma_{l,j}$ is the $W_{\rm CO}$ weighted standard deviation of $l$ for cluster $j$, and similarly for $b$, as defined in equation (\ref{eqn: standard deviation}) below.
We also only allow the merging of clusters with less than 50 Gaussian components.
The properties of the bigger clusters of the sample depend slightly on the merging procedure. If the merging criterion is too loose, big clusters merge into unrealistically large structures. The criteria selected here allow us to avoid this problem.

 We explored several different values of $N_{\rm th}$, and different minimum and maximum threshold values. It appears that these details do not significantly affect the identification. The number of clusters identified depends mostly on the lowest threshold level. As we will describe later, the data revealed a large number of small, isolated, and faint clusters. 

\subsubsection{Basic Results}

\begin{figure}
\centering
\includegraphics[width=\linewidth, draft=false, angle=0]{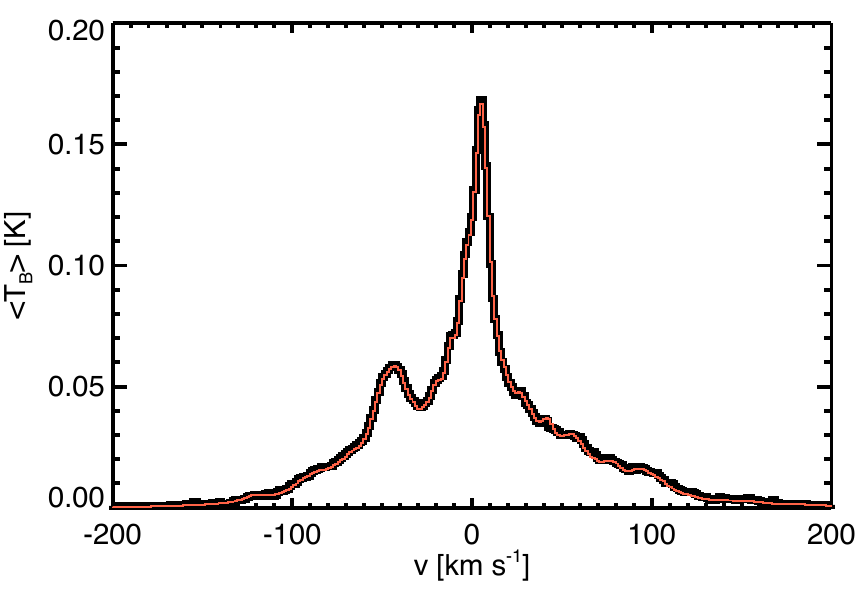}
\caption{\label{fig:WCO_in_clouds} $T_{\rm B}$ spectrum averaged over the full range in longitude and latitude considered here ($-180^\circ < l < 180^\circ$, $-5^\circ < b < 5^\circ$). The black line is the data. The thinner red line represents the emission included in clouds. }
\end{figure}

This method identified 8107 clusters that have at least 5 pixels on the sky. In the following we refer to these clusters as clouds.
These clouds are composed of 89\% of the Gaussian components ($5.8\times10^4$ Gaussian components are not associated with any cloud), which corresponds to 98\% of the CO emission of the \citet{dame2001} survey. The fact that the cluster identification includes almost the totality of the observed emission is illustrated in Figure~\ref{fig:WCO_in_clouds}, which shows the average $T_{\rm B}$ spectrum of the whole data set (black) and of the emission recovered in clouds (red). 

\begin{figure}
\centering
\includegraphics[draft=false, angle=0]{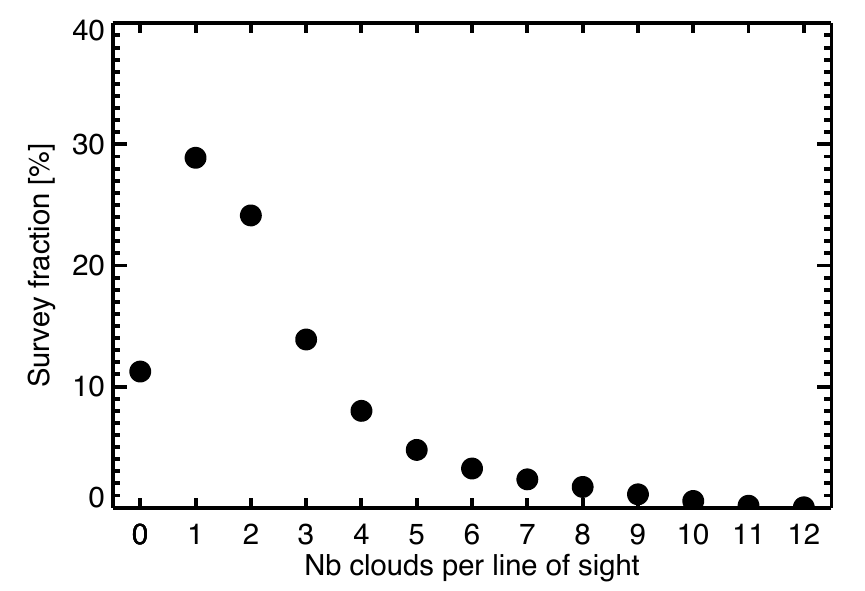}
\caption{\label{fig:PDF_nbCloud} Histogram of the number of clouds found on each line of sight. The $y$-axis is given in fraction of the total number of pixels observed in the \citet{dame2001} survey for $-5^\circ < b < 5^\circ$.}
\end{figure}

The histogram of the number of clouds per line of sight is shown in Figure~\ref{fig:PDF_nbCloud}.
On 88\% of the sky positions, there are four or fewer clouds on the line of sight; on 30\% of the sky there is only one cloud on the line of sight. Roughly 11.5\% of the surveyed sky was cloud-free using our criteria.

\section{Physical Properties of Molecular Clouds}

\label{sec:properties}

Each cloud we identify is represented by a set of Gaussian components that provides a description of the cloud in $(l,b,v)$ space. We remind the reader that in a given spectrum, it is possible that multiple Gaussian components may belong to a single cloud. In other words, we do not assume that the $^{12}$CO spectrum of a given cloud on a given sky position is Gaussian; the CO spectrum can be of any shape, and in particular, it need not be Gaussian. This is illustrated in Figure~\ref{fig:example_spectra_cluster} where the red tick marks indicate the result of the cluster identification. In this specific example, the 11 Gaussian components needed to decompose the spectrum were associated with six different clouds. 

In the following subsections we define a set of quantities for each cloud. Taken together, the quantities listed in Table~\ref{tab:catalog} compose the cloud catalog produced in this study.

\subsection{Basic Properties}

Each Gaussian component is defined by six parameters $[l_i, b_i, A_i, v_i, \sigma_i, C_i]$
where $[l_i, b_i]$ are the Galactic longitude and latitude, $[A_i, v_i, \sigma_i]$ are the Gaussian spectral parameters (see Eq.~\ref{eq:gaussian}) and $C_i$ is the index number of the cloud with which the Gaussian is associated.
A cloud with a given number $C$ defines a subset of $N_{\rm comp}$ Gaussian components. The cloud also consists of a number of pixels on the sky, $N_{\rm pix}$ such that $N_{\rm pix} \leq N_{\rm comp}$. The angular area of a cloud is simply $A = N_{\rm pix} \, d\Omega$ where $d\Omega$ is the solid angle of a single pixel ($56.25$\,arcmin$^2$ or $4.76 \times 10^{-6}$\,sr). 

\subsubsection{Position}
\label{sec:position}
For each cloud we define its central position as the intensity-weighted mean coordinate:
the Galactic longitude, $l$, and latitude, $b$, are
\begin{equation} 
l = \frac{\sum_i W_{\rm CO}^i\,l_i}{W_{\rm CO}^{\rm tot}}
\end{equation}
\begin{equation}
b = \frac{\sum_i W_{\rm CO}^i\,b_i}{W_{\rm CO}^{\rm tot}}
\end{equation}
where the sum is done over the $N_{\rm comp}$ Gaussian components, $W_{\rm CO}^i=\sqrt{2\pi} \, A_i \, \sigma_i \, dv$ is the integrated emission of a single Gaussian component, and $W_{\rm CO}^{\rm tot} = \sum_i W_{\rm CO}^i$ the total integrated emission of all components.

Similarly, the standard deviations along $l$ and $b$ are
\begin{eqnarray}\label{eqn: standard deviation}
\sigma_l & = & \sqrt{\frac{\sum_i W_{\rm CO}^i\, (l_i - l)^2}{W_{\rm CO}^{\rm tot}}}\\
\sigma_b & = & \sqrt{\frac{\sum_i W_{\rm CO}^i\, (b_i - b)^2}{W_{\rm CO}^{\rm tot}}}.
  \end{eqnarray}
These quantities provide an estimate of the angular extent of each cloud along the Galactic longitude and latitude directions.

For each cloud we also compute the angle $\theta$ with respect to the Galactic plane ($b=0^\circ$). We first compute the slope $p$ of the linear regression between $b_i$ and $l_i$, weighted by $W_{\rm CO}^i$. Then the angle is simply $\theta = \tan^{-1}(p) $.

\subsubsection{Velocity}

For a given cloud, one can define the total CO spectrum by adding up all the Gaussian components: 
\begin{equation}
  \label{eq:total_spectrum}
T_{\rm B}^{\rm tot}(v) = \sum_{i=1}^{N_{\rm comp}} A_i \, \exp( (v-v_i)^2/2\sigma_i^2).
\end{equation}

Using this total cloud CO spectrum, we compute the emission-weighted mean velocity $\langle v \rangle$ and the velocity dispersion $\sigma_v$ of each cloud:
\begin{eqnarray}
  \label{eq:avgv}
\langle v \rangle & = & \frac{1}{W_{\rm CO}^{\rm tot}} \sum_v v \, T_{\rm B}^{\rm tot}(v) \, dv\\
\sigma_v^2  & = & \frac{1}{W_{\rm CO}^{\rm tot}} \sum_v v^2 \, T_{\rm B}^{\rm tot}(v) \, dv - \langle v \rangle^2
\end{eqnarray}
where $dv$ is the channel width.

\subsubsection{Surface Density}

The average H$_2$ column density, $N_{\rm H2}$, and surface density, $\Sigma$, are defined as
\begin{eqnarray}
  N_{\rm H2} & = & \frac{W_{\rm CO}^{\rm tot}\, X_{\rm CO}}{N_{\rm pix}}\\
  \Sigma & = & \frac{N_{\rm H2} \, 2\, \mu \, m_{\rm H} \, {\rm pc}^2}{M_{\odot}}
\end{eqnarray}
where $m_{\rm H}$ is the mass of the hydrogen atom, $\mu=1.36$ is to take into account the contribution of helium and metals, and $M_\odot$ is the mass of the Sun. Throughout this paper we use $X_{\rm CO} = 2 \times 10^{20}$\,cm$^{-2}$\,(K\,km\,s$^{-1}$)$^{-1}$ \citep{bolatto2013}.

\subsection{Cloud Angular Radius}

\label{sec:radius}

The angular radius of a cloud is often defined as the minimum effective size of a cloud $R_{\rm ang} = \sqrt{A/\pi}$, where in this expression $A$ is the angular area, i.e., the sum of the angular area of all the pixels in the cloud. This amounts to making the approximation that clouds are spherical. At the other end of the spectrum, \citet{larson1981} defined the typical scale $L$ as the maximum extent of the cloud on the sky, i.e., the maximum distance between pixels.

An alternative definition of the typical scale of a cloud is the brightness-weighted radius, defined as
\begin{equation}
R_{\rm ang} = \frac{\sum_i W_{\rm CO}^i\, \sqrt{ (l_i- l )^2 + (b_i- b)^2}}{W_{\rm CO}^{\rm tot}}
\end{equation}
where the sum is over all the Gaussian components.
This method is similar to what \citet{solomon1987} and \citet{bolatto2008} used. 

We used a slightly different implementation of the brightness-weighted radius based on the eigenvalues of the inertia matrix of the on-sky emission:
\begin{equation}
\label{eq:inertia}
\psi =  \left[ \begin{array}{cc}
      \sigma_l^2 & \sigma^2_{lb}\\
      \sigma^2_{lb} & \sigma_b^2
      \end{array} \right]
\end{equation}
where
\begin{equation}
\sigma_{lb} =\sqrt{\frac{\sum_i W_{\rm CO}^i\, (l_i - l) (b_i-b)}{W_{\rm CO}^{\rm tot}}}.
\end{equation}
The maximum and minimum eigenvalues of $\psi$ provide the largest and smallest half-axis of the projected structure,  $R_{\rm max}$ and $R_{\rm min}$, respectively.
We find structures that are somewhat elongated, with an axis ratio $R_{\rm max} / R_{\rm min}$ of about 1.5 on average.

To allow for the likelihood that clouds are not spherical, we adopted the following definition of the angular radius:
   \begin{equation}
     \label{eq:rang}
     R_{\rm ang} = \left( R_{\rm max}\, R_{\rm min}\, R_{\rm min} \right)^{1/3}.
\end{equation}
This is based on the assumption that it is statistically more likely that the depth along the line of sight is closer to the smallest axis seen in projection on the sky.

\begin{table}
  \begin{center}
\caption{\label{tab:catalog} Entries of the Molecular Cloud Catalog.}
\tabskip=0pt
    \begin{tabular}{lll}\specialrule{\lightrulewidth}{0pt}{0pt} \specialrule{\lightrulewidth}{1.5pt}{\belowrulesep}
    Entry & Units & Description\\ \hline
    $C$ & ... & Cloud number\\
    $N_{\rm comp}$ & ... & Number of Gaussian components\\
    $N_{\rm pix}$ & ... & Number of pixels on the sky\\
    $A$ & deg$^{2}$ & Angular area\\
    $l$ & deg & Baricentric Galactic longitude\\
    $\sigma_l$ & deg & Galactic longitude standard deviation\\
    $b$ & deg & Baricentric Galactic latitude\\
    $\sigma_b$ & deg & Galactic latitude standard deviation\\
    $\theta$ & deg & Angle with respect to $b=0^\circ$\\
    $W_{\rm CO}$ & K\,km\,s$^{-1}$ & Integrated CO emission\\
    $N_{\rm H2}$ & cm$^{-2}$ & Average column density\\
    $\Sigma$ & $M_\odot$\,pc$^{-2}$ & Surface density\\ 
    $v_{\rm cent}$ & km\,s$^{-1}$ & Centroid velocity\\
    $\sigma_v$  & km\,s$^{-1}$ & Velocity standard deviation\\
    $R_{\rm max}$ & deg & Largest eigenvalue of the inertia matrix\\
    $R_{\rm min}$ & deg & Smallest eigenvalue of the inertia matrix\\
    $R_{\rm ang}$ & deg & Angular size\\
    $R_{\rm gal}$ & kpc & Galactocentric radius\\
    $I_{\rm NF}$ & ... & Near or far distance flag\\
    $D$ & kpc & Kinematic distance\\
    $z$ & kpc & Distance to Galactic midplane\\
    $S$ & pc$^{2}$ & Physical area \\
    $R$ & pc & Physical size\\
    $M$ & $M_\odot$ & Mass \\ \bottomrule[\lightrulewidth]
      \end{tabular}
\end{center}
{{\bf Notes.} For clouds located in the inner Galaxy, two values are given for $z$, $S$, $R$ and $M$ corresponding to the near and far kinematic distances. The index $I_{\rm NF}$ gives an estimate of which distance is more likely based on the $\sigma_v - \Sigma R$ relation (Eq.~\ref{eq:sigv_SigmaR}).}
  \end{table}

\subsection{Distance}

None of the quantities described in \S\S \ref{sec:position}-\ref{sec:radius}  depend on the distance to the cloud. However, in order to quantify the physical size, mass, and density, it is necessary to have an estimate of the distance. Here we rely on the kinematic distance, which assumes that the cloud average velocity, $\langle v \rangle$, is dominated by the motion set by Galactic rotation. 

Following \citet{roman-duval2009}, the galactocentric radius of a cloud is
\begin{equation}
R_{\rm gal} = R_0 \sin(l) \frac{V(r)}{v_p + V_0 \sin(l)}
\end{equation}
where $v_p = \langle v \rangle / \cos(b)$ is the velocity component of the cloud in the plane of the Galaxy, and $R_0=8.5$\,kpc and $V_0=220$\,km\,s$^{-1}$ are, respectively, the galactocentric radius and the orbital velocity of the Sun.
In this study we used the rotation curve $V(r)$ defined in \citet{brand1993}.

From the value of $R_{\rm gal}$, the distance from Earth to the cloud is given by
\begin{equation}
D = R_0 \cos(l) \pm \sqrt{R_{\rm gal}^2 - R_0^2\sin^2(l)}.
\end{equation}
In the inner Galaxy, where $R_{\rm gal} < R_0$, there are two solutions for $D$ for a given ($l$, $\langle v \rangle$).
This is referred to as the kinematic distance ambiguity. 

Deciding how to disentangle between the near and far distance has been a major difficulty in estimating the properties of Galactic molecular gas. Different methods have been used to circumvent this difficulty:
(1) association with objects at known distance, such as H\,{\sc ii} regions \citep{kolpak2003}; (2) 21\,cm absorption \citep{roman-duval2009}; (3) the cloud mass--size relationship \citep{roman-duval2010}; 
(4) Galactic axial symmetry \citep{scoville1975,gordon1976,sanders1984,bronfman1988}; and (5) two Gaussian latitude profiles \citep{clemens1988,nakanishi2006}. 

In order to estimate distances for our clouds, we would like a method that does not rely on external observations. Many studies \citep{dame1986,solomon1987,grabelsky1988,garca2014} have used the $\sigma_v-R$ relation to select between near and far kinematic distance in the inner Galaxy. For a given cloud, $\sigma_v$ is observed directly, while the inferred cloud radius $R$ depends on distance. The idea is to select the near or far distance that would give a value of $R$ that is closer to the $\sigma_v-R$ relation, set by \citet{larson1981} for instance. 

This approach relies on the hypothesis that the $\sigma_v-R$ relation is valid everywhere in the Galaxy; this could be the case if the line width--size scaling is due to a turbulent cascade, for example. However, even if the $\sigma_v-R$ scaling arises as part of a turbulent cascade, it may still be different in different Galactic locations, due, for example, to the presence of different driving mechanisms in different locations. In addition, this method makes the assumption that $\sigma_v$ does not depend significantly on any other physical parameter, e.g., the gas surface density $\Sigma$. 
Anticipating our results, we find a large dispersion in the $\sigma_v-R$ relation (Figure~\ref{fig:m_vs_r}, top right panel).

\begin{table*}
  \begin{center}
\caption{\label{tab:avgvalues} Typical Values of Cloud Parameters.}
\tabskip=0pt
\begin{tabular}{llcccccccc}\specialrule{\lightrulewidth}{0pt}{0pt} \specialrule{\lightrulewidth}{1.5pt}{\belowrulesep}
Parameter & Units & \multicolumn{2}{c}{All Clouds} & \multicolumn{2}{c}{Inner Galaxy} & \multicolumn{2}{c}{Outer Galaxy} & \multicolumn{2}{c}{$\alpha_{\rm vir} \leq 3$} \\ 
 & & Avg & Median & Avg & Median & Avg & Median & Avg & Median  \\ \hline
$\Sigma$ & $M_\odot$\,pc$^{-2}$ & 28.6 & 16.5 & 41.9 & 31.6 & 10.4 & 7.0 & 46.1 & 37.1 \\ 
$M$ & $10^4$\,$M_\odot$ & 15.1 & 3.8 & 22.6 & 7.8 & 5.4 & 1.6 & 40.7 & 12.0 \\ 
$R$ & pc  & 31.5 & 25.1 & 30.8 & 25.2 & 32.9 & 24.9 & 37.8 & 29.5\\ 
$n_{\rm H2}$ & cm$^{-3}$ & 24.1 & 9.6 & 33.7 & 16.9 & 11.0 & 3.3 & 30.5 & 19.2 \\ 
$\sigma_v$ & km\,s$^{-1}$ & 4.0 & 3.6 & 4.9 & 4.6 & 2.8 & 2.5 & 3.1 & 2.7 \\ 
$\sigma_0$ & km\,s$^{-1}$ & 0.8 & 0.8 & 1.0 & 0.9 & 0.6 & 0.5 & 0.5 & 0.5 \\ 
$\alpha_{\rm vir}$ & ... & 22.4 & 8.5 & 20.7 & 7.3 & 25.0 & 11.3 & 2.0 & 2.1  \\ 
$\tau_{\rm dyn}$ & $10^6$\,yr & 8.5 & 6.4 & 6.2 & 5.3 & 12.3 & 9.9 & 14.6 & 10.3 \\ 
$\tau_{\rm ff}$ & $10^6$\,yr & 13.1 & 10.1 & 9.4 & 7.6 & 19.0 & 17.2 & 9.5 & 7.1\\ 
$P_{\rm int}$ & $10^4$\,K\,cm$^{-3}$ & 10.9 & 3.0 & 18.7 & 8.5 & 1.7 & 0.6 & 9.8 & 4.0 \\ 
$\dot{E}_{\rm dis}$ & L$_{\rm Sun}$ & 146.9 & 9.9 & 285.1 & 51.5 & 10.5 & 1.7 & 205.8 & 15.3 \\ 
$2\epsilon$ & $10^{-27}$\,erg\,cm$^{-3}$\,s$^{-1}$ & 503.5 & 59.4 & 878.6 & 187.4 & 68.5 & 7.1 & 191.7 & 49.4 \\ 
\bottomrule[\lightrulewidth]
\end{tabular}
  \end{center}
  {\bf Notes.} Average and median values of physical quantities for different subsets of the catalog: all clouds, clouds located at $R_{\rm gal}\leq 8.5$\,kpc, clouds located at  $R_{\rm gal} > 8.5$\,kpc, and clouds with $\alpha_{\rm vir} \leq 3$.
  \end{table*}

In fact, \citet{heyer2009} argued that the observed $\sigma_v-R$ relation also depends on $\Sigma$. If clouds are assumed to be in virial equilibrium, $\sigma_v = (\pi\, G / 5)^{1/2} \, R^{1/2} \, \Sigma^{1/2}$, and the use of $\sigma_v-R$ to distinguish between near and far distance should take into account the surface density of the cloud. Like \citet{heyer2009}, we also find that $\sigma_v$ depends on both $R$ and $\Sigma$, but not exactly to the power of 1/2. Our analysis is compatible with $\sigma_v \propto (R\,\Sigma)^{0.43}$ (see Figure~\ref{fig:m_vs_r}). This is the basis of our method to select between the near and far distance for each cloud in the inner Galaxy. 
In addition, in a few cases, where the far distance would put a cloud to a Galactic height $|z| > 200$\,pc, the near distance was selected. This is justified by the fact that the FWHM (in $z$) of molecular gas is of the order of 100\,pc \citep{ferriere2001}.

\subsection{Mass, Radius, and Density}

Given the distance, $D$, the mass, physical radius, and H$_2$ number density are defined as
\begin{eqnarray}
M & = & \Sigma \, N_{\rm pix} \, D^2 \,d\Omega\\
R & = & D \, \tan(R_{\rm ang}) \\
n_{\rm H2} & = & \frac{3\,M}{4 \,\pi\, \mu\, R^3}
\end{eqnarray}

\begin{figure}
\centering
\includegraphics[width=\linewidth, draft=false, angle=0]{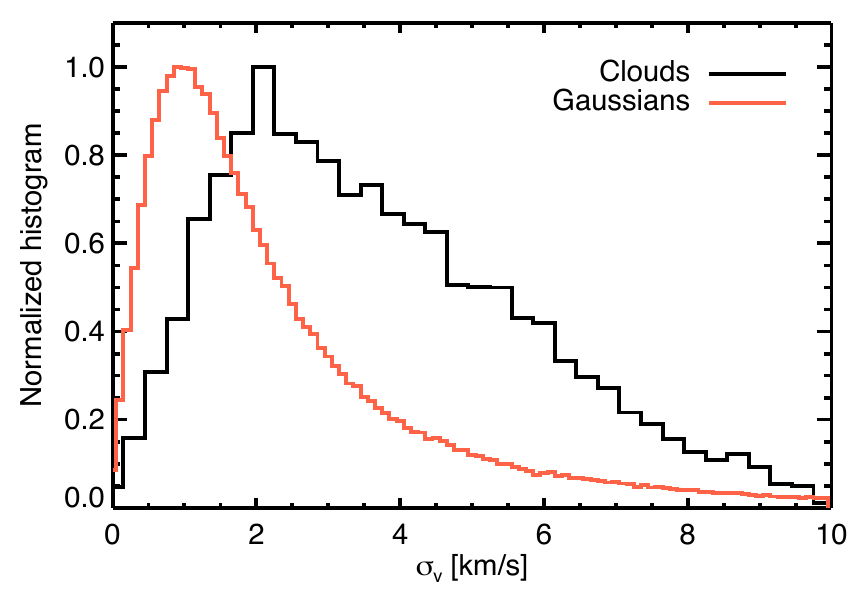}
\caption{\label{fig:PDF_sigmaV} Histogram of the velocity dispersion ($\sigma_v$) for all clouds in the catalog (black) and 
  for all Gaussian components ($\sigma_i$) (red). The FWHM is $4.5\,{\rm km\,s^{-1}}$ for clouds and $2.0\,{\rm km\,s^{-1}}$ for the Gaussian components. In both cases the contribution of the instrumental broadening was removed quadratically.}
\end{figure}

\begin{figure*}
\centering
\includegraphics[draft=false]{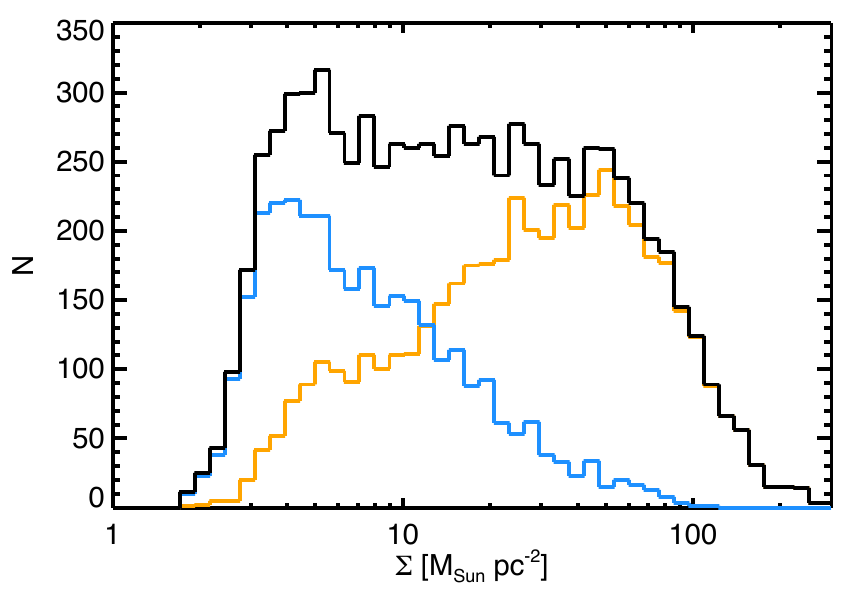}
\includegraphics[draft=false, angle=0]{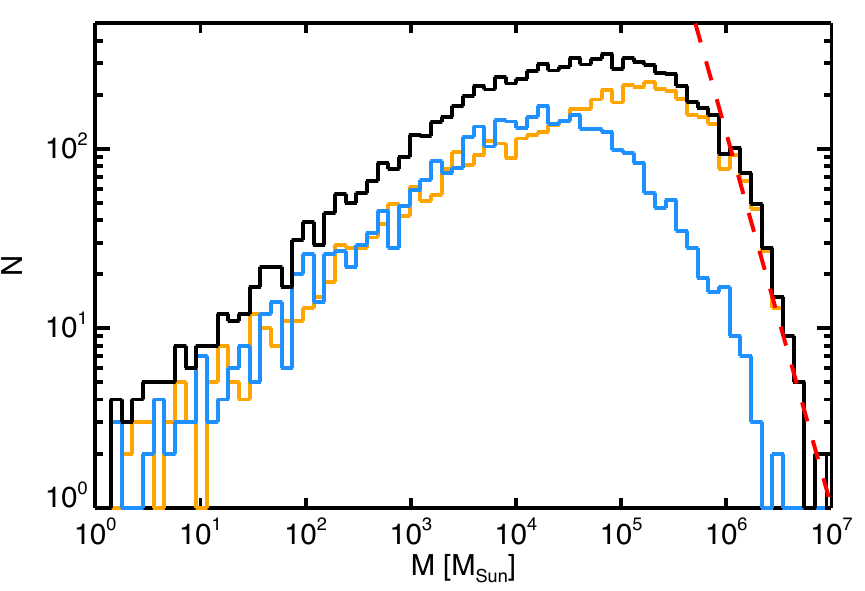}
\includegraphics[draft=false, angle=0]{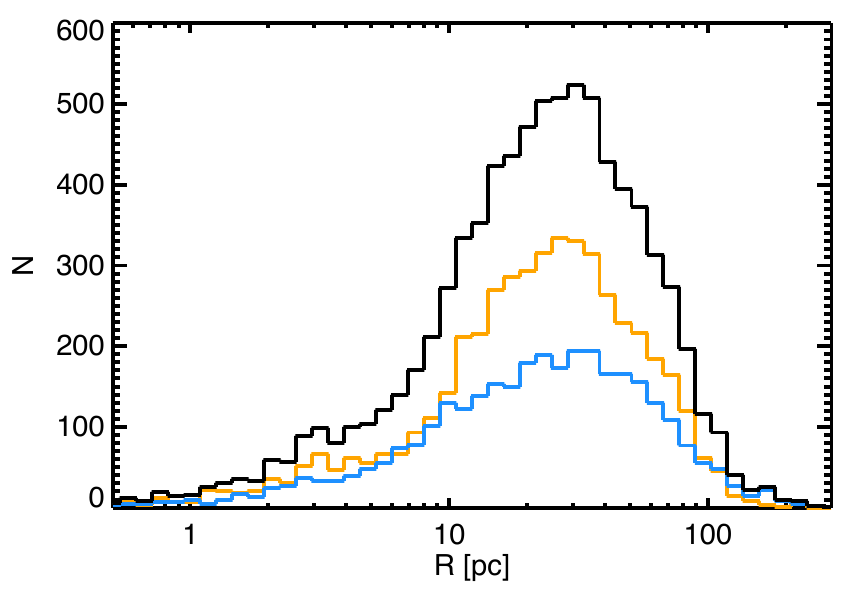}
\includegraphics[draft=false, angle=0]{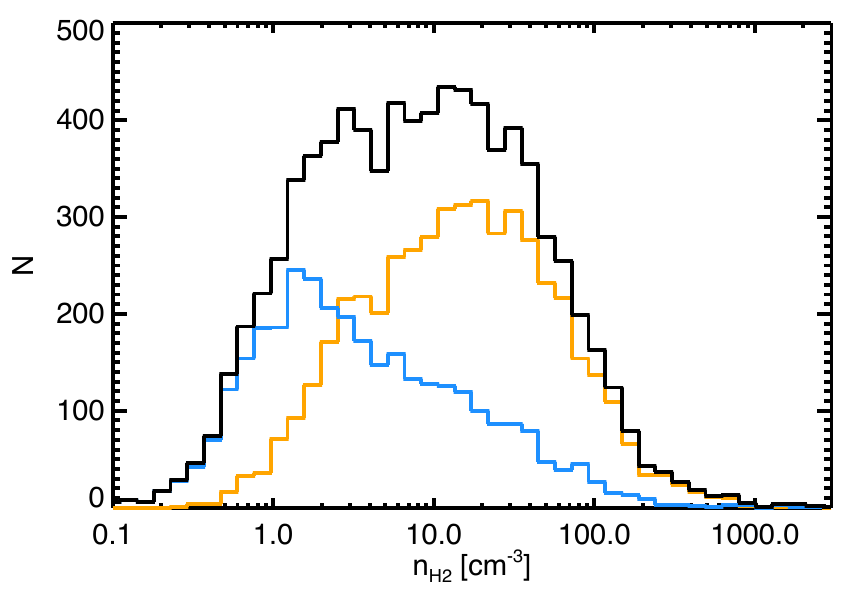}
\caption{\label{fig:PDF_mass} Probability distribution functions (histograms) of $\Sigma$, $M$, $R$, and $n_{\rm H2}$ for all the clouds in the catalog (black histograms), for inner-Galaxy clouds (orange histograms), and outer-Galaxy clouds (blue histograms). Note the dramatically different surface densities of inner- and outer-Galaxy clouds (top left panel); the mode of the outer-Galaxy cloud distribution is at a surface density only 1/10 that of the inner-Galaxy cloud distribution. The dashed red line in the upper right (mass $M$) panel shows a power-law fit to the extreme high-mass end of the distribution, $dN/d\ln M\sim M^{2.0\pm 0.1}$, i.e., the largest clouds hold most of the molecular gas---see Figure \ref{fig:cumulative_mass}. The typical values for these quantities are summarized in Table~\ref{tab:avgvalues}.}
\end{figure*}

\section{Results}

\label{sec:results}

The cloud catalog produced here includes 98\% of the total \CO\ emission of the \citet{dame2001} survey within $b\pm 5^\circ$. 
In the following we describe the statistics of the cloud physical parameters ($M$, $\Sigma$, $n_{\rm H2}$, $R$, and $\sigma_v$), their spatial distribution in the Galactic plane, the mass--size and velocity--size relations, and more advanced physical parameters related to the respective role of gravity and turbulence in their dynamics.

\subsection{Distribution Functions of Physical Parameters}

\subsubsection{Velocity Dispersion}

The identification of clouds and their description as the sum of Gaussian components provide some statistics on the global velocity dispersion of each cloud ($\sigma_v$), as well as on the width of each of the Gaussian components. The latter provides information on the line-of-sight velocity dispersion. 

The histogram of the velocity dispersion $\sigma_i$ of the individual Gaussian components  (Figure~\ref{fig:PDF_sigmaV}, red line) shows a median value of 1.65\,km\,s$^{-1}$ but a most probable value of only 0.85\,km\,s$^{-1}$. The histogram is positively skewed, with values as large as 10\,km\,s$^{-1}$. These large values are likely to be due to the velocity crowding effect in some areas of the inner Galaxy. The histogram is relatively narrow, with ${\rm FWHM}=2.0$\,km\,s$^{-1}$. 

Figure~\ref{fig:PDF_sigmaV} also shows the histogram of $\sigma_v$ for all the clouds in the catalog (black line). On the scale of clouds, the median velocity dispersion is 3.6\,km\,s$^{-1}$ with a most probable value of 1.95\,km\,s$^{-1}$. The histogram is broader (${\rm FWHM}=4.5$\,km\,s$^{-1}$) than that of the Gaussian components. For both $\sigma_i$ and $\sigma_v$, the contribution of the instrumental broadening was removed quadratically.

The histogram of $\sigma_v$ is broader and peaks at a higher value than the histogram of the individual Gaussian components. This is due to the fact that $\sigma_v$ includes both the average line-of-sight velocity dispersion and the variation of the centroid velocity over the cloud. In contrast, the statistics of the Gaussian width $\sigma_i$ sample only part of the line-of-sight component, because the spectrum of a cloud, at a given position on the sky, might be composed of more than one Gaussian component. The ratio of the most probable values is about 2.2, as is the ratio of the median values.

The velocity dispersion of a given cloud is the quadratic sum of the turbulent and thermal broadening. 
Gaussian decomposition is able to partly separate these two contributions to the line broadening. In that context, the ratio of the most probable values of these two histograms gives a lower limit on the Mach number of molecular clouds ($Mach > 2.2$).

\subsubsection{Cloud Mass Surface Density}

The probability distributions of $\Sigma$, $M$, $R$, and $n_{\rm H2}$ are shown in Figure~\ref{fig:PDF_mass}. In each panel, the solid black line shows the full sample, while the orange and blue lines show, respectively, the clouds located in the inner and outer Galaxy. 

The distribution of the mass surface density (Figure~\ref{fig:PDF_mass}, top left) shows a range of $2\lesssim\Sigma \lesssim 300\,M_\odot$\,pc$^{-2}$.
The mean value of $\Sigma$ is $28.6\,M_\odot$\,pc$^{-2}$, with a large standard deviation of $29.3\,M_\odot$\,pc$^{-2}$ and a skewness of $1.5$. The mass surface density of clouds is higher in the inner Galaxy, with a mean value of $41.9\,M_\odot$\,pc$^{-2}$, compared to $10.4\,M_\odot$\,pc$^{-2}$ in the outer Galaxy. All those values are tabulated in Table~\ref{tab:avgvalues}.

The difference in $\Sigma$ between the inner and outer Galaxy could be partly attributed to a variation of the $X_{\rm CO}$ factor that we assumed constant. Because the metallicity decreases with $R_{\rm gal}$ \citep{balser2011}, $X_{\rm CO}$ is expected to increase toward the outer Galaxy. Therefore, assuming a constant $X_{\rm CO}$ systematically underestimates the surface density and mass of clouds in the outer Galaxy. \citet{heyer2015} estimated that this effect could be as high as $\sim 50$\,\%. This effect is potentially large, but it is still smaller than the variation of a factor of four observed here.

In general, the values of $\Sigma$ obtained here are similar to those found by \citet{heyer2009}. Both our mean values and those of \citet{heyer2009}  are much smaller than those obtained by \citet{solomon1987} and \citet{roman-duval2010}. For example, \citet{solomon1987} mention that their typical GMCs have $\Sigma \sim 170\,M_\odot$\,pc$^{-2}$. This difference might be due to the fact that our method includes the emission down to the sensitivity limit of the data. 
Using a threshold in brightness, like \citet{solomon1987} did (they identified clouds with $T_{\rm B} > 4$\,K), one identifies clouds that are smaller, and their average mass surface density is systematically larger, as all the faint emission around the periphery of the cloud is missed. Possibly for the same reason, the large range in $\Sigma$ reported here contrasts with the general idea that GMCs have a nearly constant value of $\Sigma$ \citep{larson1981,solomon1987}.

In addition, because our technique places almost all of the CO emission into clouds, we identified a large number of faint and smaller clouds that were missed by previous methods. Our analysis confirms the findings of \citet{digel1990} that clouds in the outer Galaxy have a lower $\Sigma$. This is also similar to what is seen in M51, where diffuse and fainter CO emission is detected far away from the main spiral arms \citep{koda2009,pety2013}.

\begin{figure*}
\centering
\includegraphics[draft=false, angle=0]{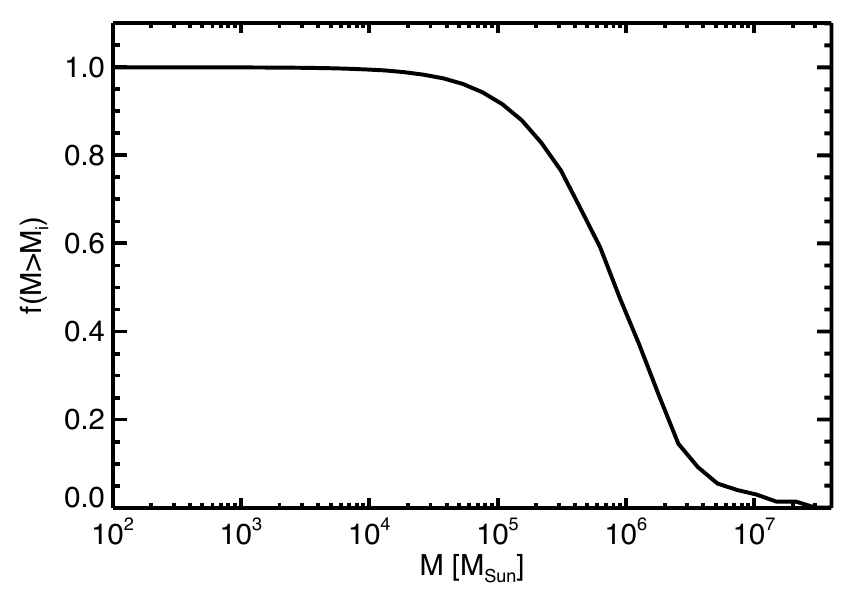}
\includegraphics[draft=false, angle=0]{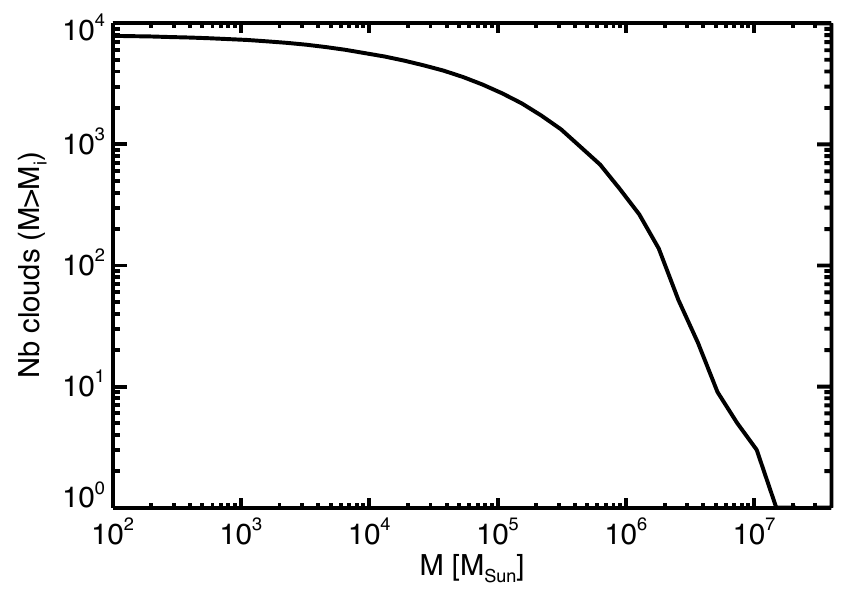}
\caption{\label{fig:cumulative_mass} Cumulative mass distribution functions: fraction of total mass ({\bf left}) and number of clouds ({\bf right}) with $M>M_i$. Half of the molecular gas mass in the Milky Way is contained in clouds with mass greater than $M=8.4\times10^5$\,M$_\odot$. }
\end{figure*}

\subsubsection{Mass}

The total mass in the catalog is $1.6\times 10^9$\,M$_\odot$, corresponding to a total H$_2$ mass of $1.2\times 10^9$\,M$_\odot$ (assuming $\mu=1.36$). This is in good agreement with the review of \citet{heyer2015}, who estimated the total H$_2$ mass to be ($1.0 \pm 0.3) \times 10^9$\,M$_\odot$ based on a compilation of previous studies.

The mass of clouds ranges from a few $M_\odot$ to $2.2 \times 10^7\,M_\odot$. 
The top right panel of Figure~\ref{fig:PDF_mass} shows the distribution $dN/d\log M$ for the entire catalog, as well as for the inner and outer Galaxy. 
The median mass of the sample is $3.8\times 10^4$\,M$_\odot$. This is comparable to the molecular mass of several clouds of the Gould Belt: Taurus \citep{pineda2010}, Perseus \citep{lee2015}, Ophiuchus \citep{loren1989}, and the Coalsack \citep{cambresy1999} all have masses in the range $1-3\times 10^4$\,M$_\odot$.

Most of the clouds are small, but they collectively contain little mass; the $\sim 5000$ clouds with $M\le10^5M_\odot$ contribute less than 10\% of the total molecular gas mass. The cumulative histogram of the cloud mass (Figure~\ref{fig:cumulative_mass}) indicates that half of the mass is in clouds with $M > 8.4 \times 10^5$\,M$_\odot$; there are $\sim460$ such clouds. 
The higher-mass end of the distribution follows a power-law: $dN/d \ln M \sim M^{-2.0 \pm 0.1}$.

There is a significant difference between the cloud mass distributions of the inner and outer Galaxy. About 60\% of the clouds are located in the inner Galaxy, but they make up 85\% of the total molecular mass of the disk. Clouds are indeed significantly more massive in the inner Galaxy: there the median mass is $7.8\times 10^4$\,M$_\odot$ while it is only $1.6\times 10^4$\,M$_\odot$ outside the solar circle. These values are summarized in Table~\ref{tab:avgvalues}.

These statistical properties are similar to what has been estimated with smaller samples and over a much smaller range of scales than we use here
\citep{solomon1987,williams1997,heithausen1998,kramer1998,heyer2001,rosolowsky2005,roman-duval2010}.

\subsubsection{Cloud Physical Radius}

The clouds in the catalog have sizes ranging from less than 1\,pc to about 150\,pc (see Figure~\ref{fig:PDF_mass} bottom left panel). The most probable value of $R$ is $\sim 30$\,pc, typical for GMCs. We note that the maximum size is comparable to the scale height of the molecular disk. 

Interestingly, the distribution of $R$ is very similar in the outer and inner Galaxy. Even though clouds in the inner Galaxy are more massive, have a larger $\Sigma$, and are in a much more crowded region in PPV space than in the outer Galaxy, their size distributions are surprisingly similar.

We also want to point out that, unlike $\Sigma$, the determination of the typical size of a cloud is less sensitive to the brightness threshold used for cloud identification. If $R$ is determined with a brightness-weighted method (see Sect.~\ref{sec:radius}), the low-brightness outskirts of clouds do not contribute significantly to the size. This explains why the values of $R$ reported here are similar to previous studies \citep[e.g.,][]{solomon1987} while there are significant differences in $\Sigma$.

\subsubsection{Cloud Mean Density}

The mean H$_2$ number density of the whole sample is $\langle n_{\rm H2} \rangle = 24.1$\,cm$^{-3}$. The mean values in the inner and outer Galaxy are $33.7$ and $11.0$\,cm$^{-3}$, respectively. Such low values of $n_{\rm H2}$, well below the CO critical excitation density of the order of $10^3$\,cm$^{-3}$, have been reported early on \citep{blitz1980}. They indicate that the filling factor of dense gas in molecular clouds is significantly below unity \citep{perault1985}. 
In other words, the low density values measured here indicate that emitting gas does not fill the CfA survey beam. Indeed, higher-resolution observations tend to find higher values of $n_{\rm H2}$ \citep{roman-duval2010}.

\begin{figure*}
  \centering  
\includegraphics[draft=false, angle=0]{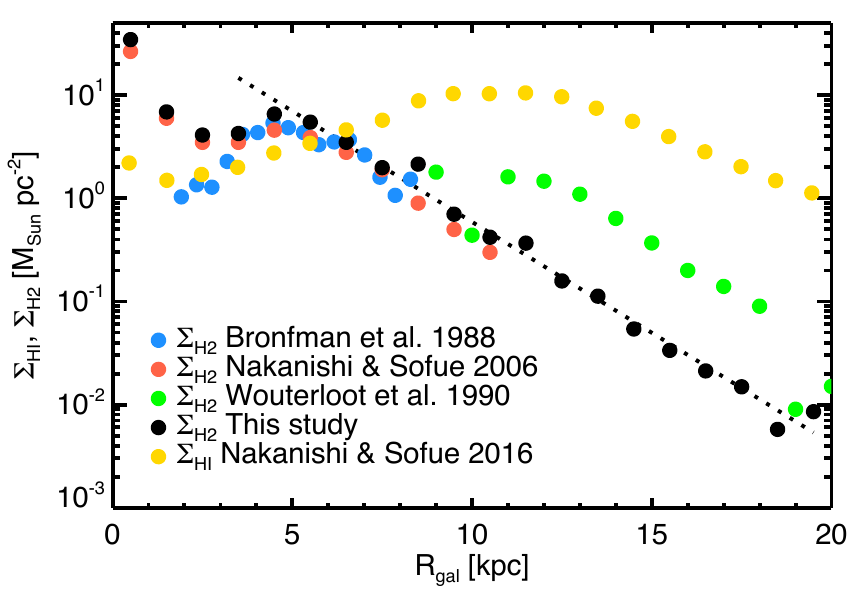}
\includegraphics[draft=false, angle=0]{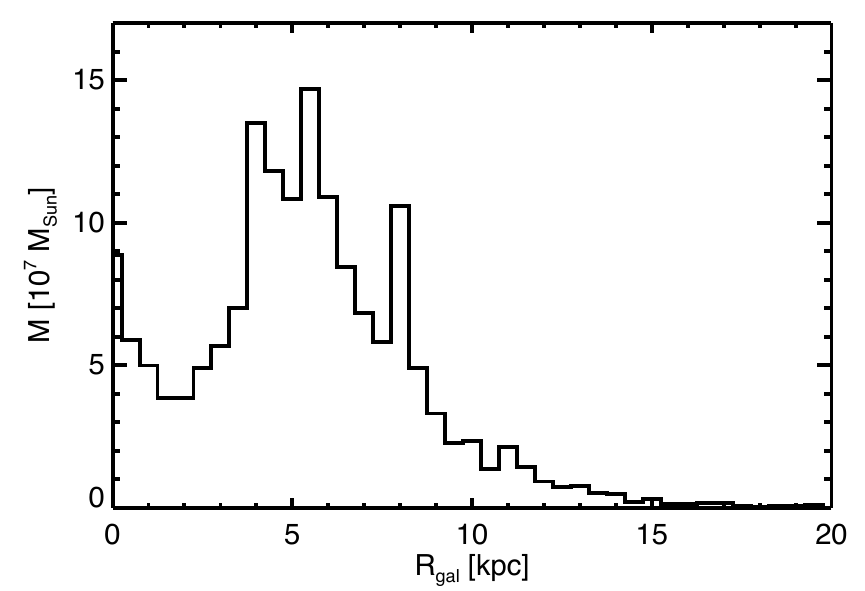}
\includegraphics[draft=false, angle=0]{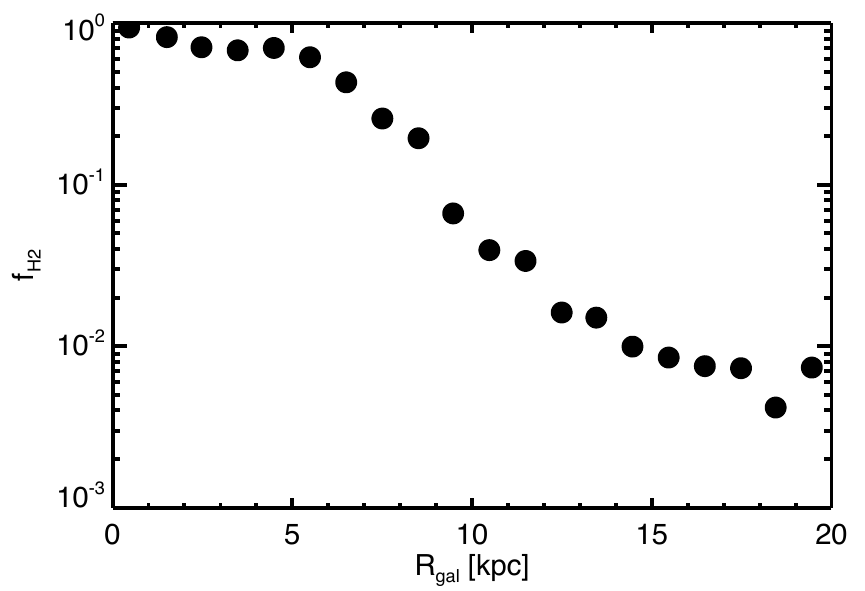}
\includegraphics[draft=false, angle=0]{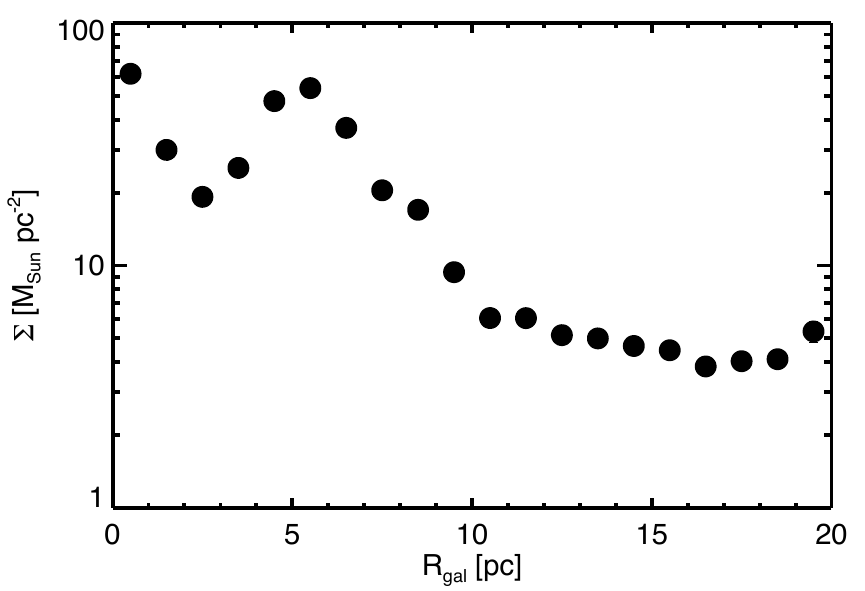}
\caption{\label{fig:Sigma_vs_Rgal} Top left: surface density of H$_2$ mass as a function of galactocentric radius (black filled circles). The blue, red, and green filled circles correspond, respectively to the estimate of \citet{bronfman1988}, \citet{nakanishi2006}, and \citet{wouterloot1990}. The dotted line is the exponential fit over the range $4<R_{\rm gal}<17$\,kpc: $\Sigma = 83\,\exp(-R_{\rm gal}/2.0)$.  The yellow filled circles correponds to the H\,{\sc i} surface density of \citet{nakanishi2016}. {Top right: } total mass of clouds in galactocentric rings (thickness 0.5\,kpc), as a function of $R_{\rm gal}$. {Bottom left:} molecular fraction $f_{\rm H2}=\Sigma_{\rm H2}/(\Sigma_{\rm HI} + \Sigma_{\rm H2})$ built using our estimate of $\Sigma_{\rm H2}$ and $\Sigma_{\rm HI}$ from \citet{nakanishi2016}. {Bottom right: } variation of the median cloud mass surface density $\Sigma$ as a function of galactocentric radius.}
\end{figure*}

\subsection{Large-scale Distribution of Molecular Clouds}

\subsubsection{Surface Density versus Galactic Radius}

It is generally believed that a substantial fraction of the molecular material in the  Milky Way is located in a ring in the range $3\, {\rm kpc} \lesssim R_{\rm gal} \lesssim 7\,{\rm kpc}$. This was established by looking at the radial variation of the mass surface density of H$_2$ gas (i.e., $\Sigma_{\rm H2}$ versus $R_{\rm gal}$) and modeling the CO emission as coming from a disk of varying scale height and midplane density with $R_{\rm gal}$ \citep{bronfman1988,nakanishi2006}.\footnote{Here we make a distinction between the mass surface density of a single cloud, $\Sigma$, and the mass surface density of H$_2$ gas in the Milky Way, $\Sigma_{\rm H2}$.} Because the catalog presented here contains almost all the CO emission, it is possible for the first time to estimate $\Sigma_{\rm H2}$ using a cloud-based approach, i.e., by integrating the mass of all the clouds in an annulus of constant $R_{\rm gal}$. Our results are shown in the top left panel of Figure~\ref{fig:Sigma_vs_Rgal}. They are compatible with most previous studies, especially with the analysis of \citet{nakanishi2006}. Close to the Galactic center, there is a departure from the estimates of \citet{bronfman1988}; this departure is explained by the fact that those authors did not include data from $|l| < 12$ in their analysis. With the same restriction in longitude, our estimate of $\Sigma_{\rm H2}$ (not shown) is compatible with that of \citet{bronfman1988}.

\begin{figure*}
\centering
\includegraphics[draft=false, angle=0, width=18cm]{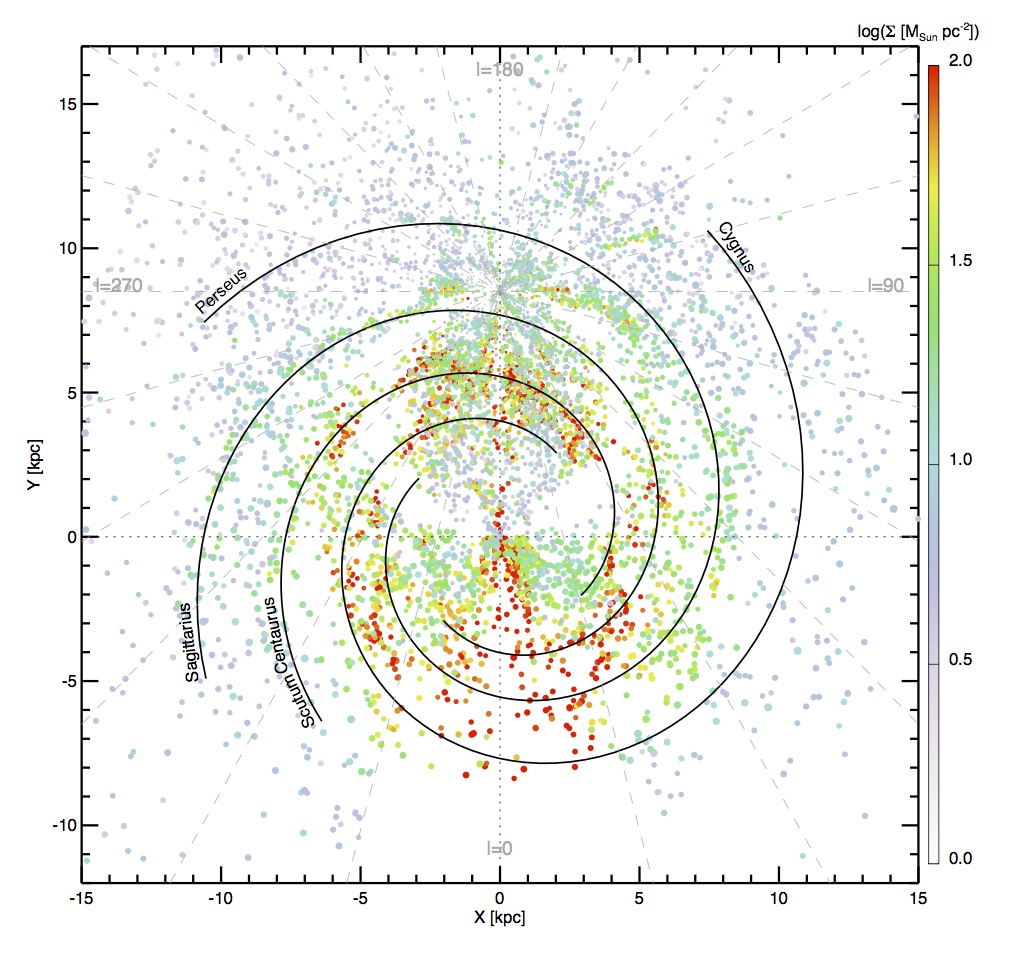}
\caption{\label{fig:xy_surfdens} Face-on view of the H$_2$ surface density.
  The symbol size is proportional to $\log(R)$ while the color indicates $\log(\Sigma)$.
The four spiral arms model of \citet{vallee2008} is overlaid. The dashed lines indicate Galactic longitudes by step of $15^\circ$.}
\end{figure*}

For $R_{\rm gal} > 3$\,kpc, we note that the surface density falls off exponentially with a scale length of 2.0\,kpc. This result is significantly different from the scale length of 3.5\,kpc found by \citet{williams1997}. These authors based their analysis on the outer-Galaxy results of \citet{wouterloot1990} (green filled circles in Figure~\ref{fig:Sigma_vs_Rgal}, top left panel), which are not compatible with our estimates or with those of \citet{nakanishi2006} or \citet{pohl2008} \citep[see also Figure~7 of][]{heyer2015}.

The molecular ring is best seen in the top right panel of Figure~\ref{fig:Sigma_vs_Rgal}, which shows the total mass in rings of thickness 0.5\,kpc as a function of $R_{\rm gal}$. We note that the increase of $\Sigma_{\rm H2}$ and total mass toward the inner Galaxy is produced in part by an increase of the number of clouds, but more significantly by an increase of the average mass surface density of individual clouds (Figure~\ref{fig:Sigma_vs_Rgal}, bottom right panel).

Finally, it is interesting to compare the H$_2$ and H\,{\sc i} surface density as a function of $R_{\rm gal}$. In the top left panel of Figure~\ref{fig:Sigma_vs_Rgal} we also show the $\Sigma_{\rm HI}$ data points of \citet{nakanishi2016}, which are similar to the ones obtained by \citet{koda2016}. In the inner Galaxy the H\,{\sc i} surface density rises from a value of about $1\,M_\odot$\,pc$^{-2}$ to  $10\,M_\odot$\,pc$^{-2}$ at $R_{\rm gal} \sim 10$\,kpc. At larger $R_{\rm gal}$, the H\,{\sc i} surface density falls exponentially with a scale length of $R_{\rm HI}\approx 3.75\,{\rm kpc}$ \citep{kalberla2009}. Both \citet{nakanishi2016} and \citet{koda2016} showed that the fraction of the gas in molecular form, $f_{\rm H2} = \Sigma_{\rm H2}/(\Sigma_{\rm HI} + \Sigma_{\rm H2})$, declines steadily with $R_{\rm gal}$. This is also shown in the bottom left panel of Figure~\ref{fig:Sigma_vs_Rgal} using our estimate of $\Sigma_{\rm H2}$.  In the solar neighborhood, the molecular fraction is about 10\% only.

\subsubsection{Face-on View}

Looking at external galaxies like M51 \citep{koda2009,pety2013}, it is likely that the molecular gas in the Milky Way is not located in a ring, but rather that it follows the spiral arms \citep{dobbs2012a}. Do GMCs follow the spiral structure of the Galaxy? Answering this question is challenging, due to the fact that we are located inside the disk. Velocity crowding, the kinematic distance ambiguity, noncircular motions, and the definition of cloud frontiers all add to the difficulty of the task. Many attempts were made in the past, on portions of the disk \citep{clemens1988,solomon1989}. Recently \citet{rice2016} have shown that the brightest molecular clouds seem to be spatially correlated with the spiral structures of the Galaxy. On the other hand, \citet{koda2016} showed that there is rather modest variation of the fraction of molecular to atomic gas (about 20\%) between the arm and the interarm regions of the Milky Way. These authors noticed that the main change in gas phase occurs at $R_{\rm gal} > 6$\,kpc where the molecular fraction drops below 50\% (see Figure~\ref{fig:Sigma_vs_Rgal}, bottom left panel).

In Figures~\ref{fig:xy_surfdens} and Figures~\ref{fig:xy_mass} to \ref{fig:xy_height} we present face-on views of the Milky Way representing the distribution of $\Sigma$, $M$, $n_{\rm H2}$, as well as the distance $z$ to the midplane. In every diagram, each dot represents a single cloud. The size of the dot is proportional to $\log(R)$. The color scale indicates the value of each parameter. 
Overlaid on these figures is the spiral arm model of \citet{vallee2008} adjusted to the Sun--Galactic center distance $R_{\rm Sun}=8.5$\,kpc. 

\begin{figure*}
\centering
\includegraphics[draft=false, angle=0]{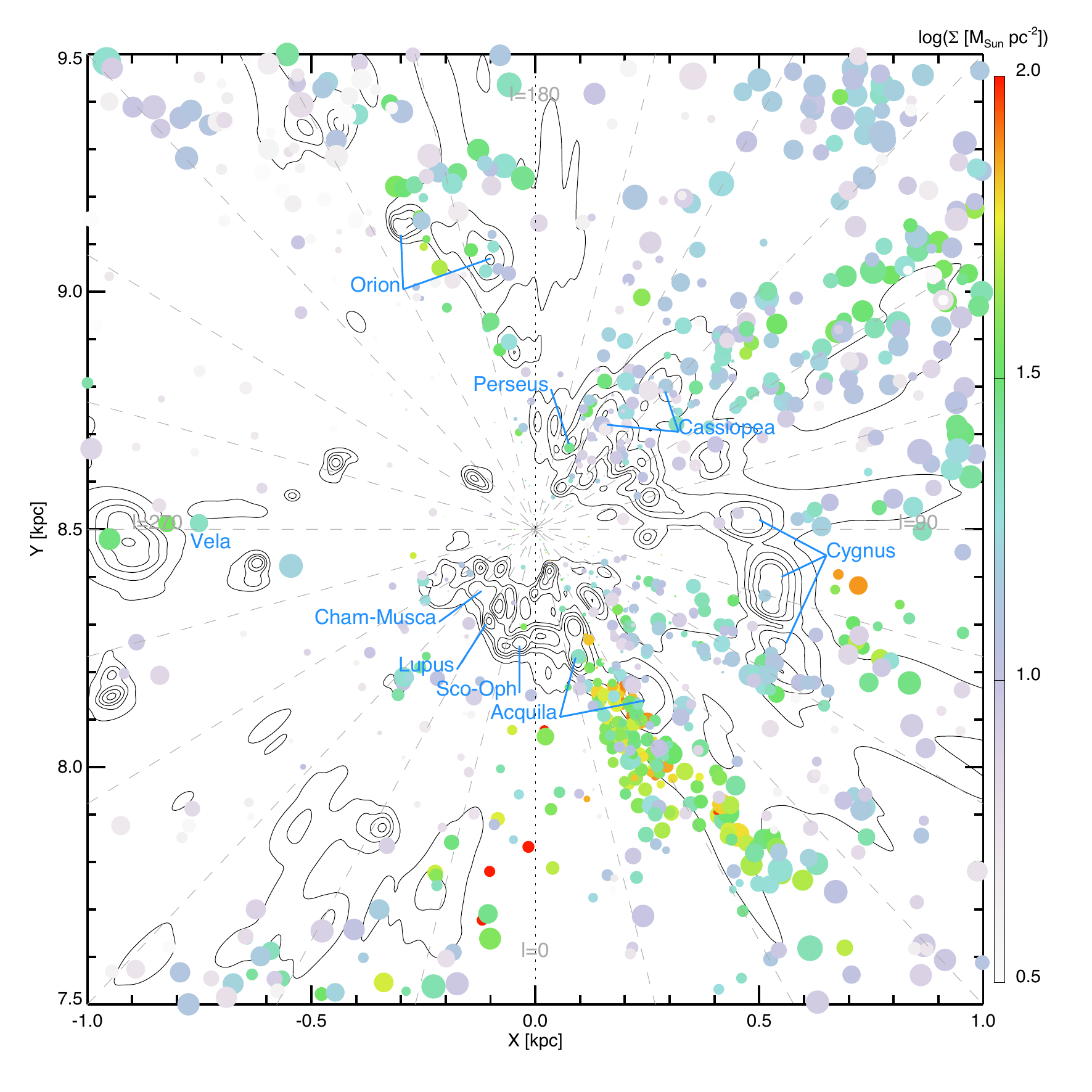}
\caption{\label{fig:xy_localbubble} Face-on view of the H$_2$ surface density in the Solar neighborhood. The symbol size is proportional to $\log(R)$ while the color indicates $\log(\Sigma)$. The contours indicate the distribution of dense gas as deduced from extinction measurements by \citet{lallement2014}. The local bubble can be traced as a paucity of molecular emission (and dust extinction contours) from the location of the Sun (at the center of the plot) to the upper left, towards $l=240^\circ$.}
\end{figure*}

The spatial distribution of $\Sigma$ (Figure~\ref{fig:xy_surfdens}) is probably the most useful one to determine what the Milky Way might look like seen from the outside. Because $\Sigma$ seems to be unrelated to the cloud physical size (see Sect.~\ref{sec:mass-size}), it does not suffer as much as $M$ and $n_{\rm H2}$ do from a distance bias (see Appendix~\ref{sec:distance-bias}). This map shows an increase in cloud surface density close to the four spiral arms, although the association is not clear everywhere. A significant fraction of the clouds are located between the arms. Like \citet{rice2016}, we notice a lack of clouds in the Perseus arm close to the Sun (for $l>90^\circ$). 

Unsurprisingly, the face-on view of $M$ (Figure~\ref{fig:xy_mass}) indicates that more distant clouds are generally bigger and more massive than more nearby clouds.
This is due to the limit imposed by the finite angular resolution and sensitivity of the observations. Naturally, the survey detects larger and  more massive clouds with increasing distance \citep{heyer2001}. The dependence on $D$ of $R$ and $M$ and the detection limit of the catalog are detailed in Figure~\ref{fig:mass_vs_distance} in Appendix~\ref{sec:distance-bias}. 
This effect does not imply that the catalog misses a large fraction of the emission from regions that are farther away but rather that it does not have the ability to resolve faraway complexes into their substructures. 
Molecular clouds have a fractal structure and are not isolated in PPV space. This is clearly seen in CO observations of external galaxies \citep{koda2009,pety2013}. Therefore, in the Milky Way, closer molecular complexes are easily resolved into smaller pieces, while similar complexes that are far away are seen as a single entity. Unfortunately, this is inherent to the fact that we look at the Milky Way from the inside. This has to be kept in mind when looking at distance-dependent statistical properties of clouds.

\begin{figure*}
\centering
\includegraphics[draft=false, angle=0]{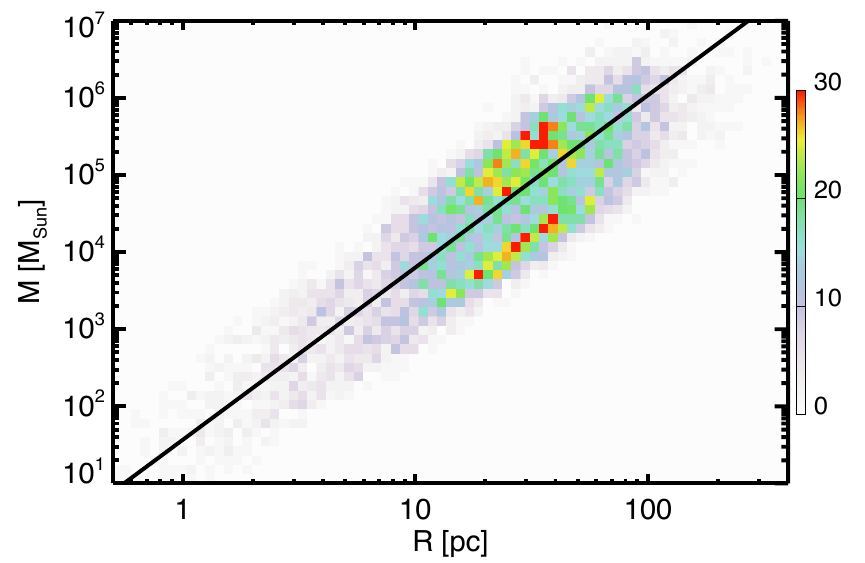}
\includegraphics[draft=false, angle=0]{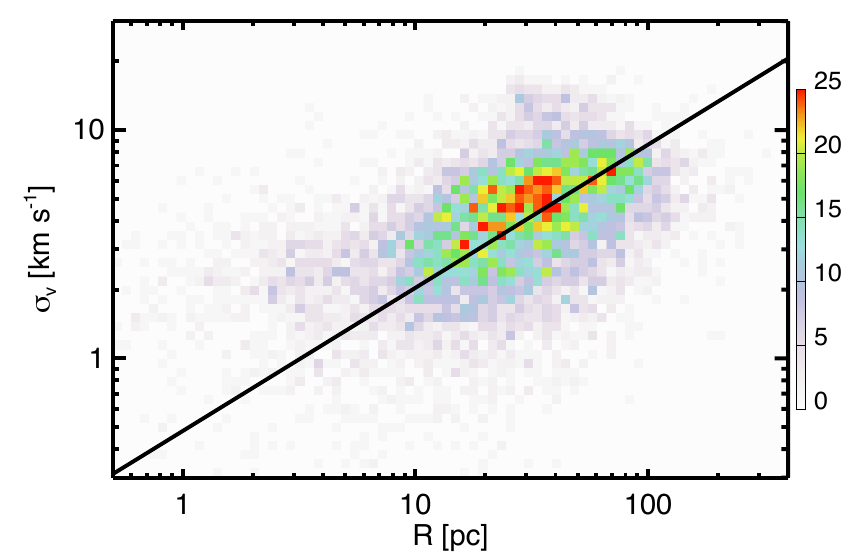}
\includegraphics[draft=false, angle=0]{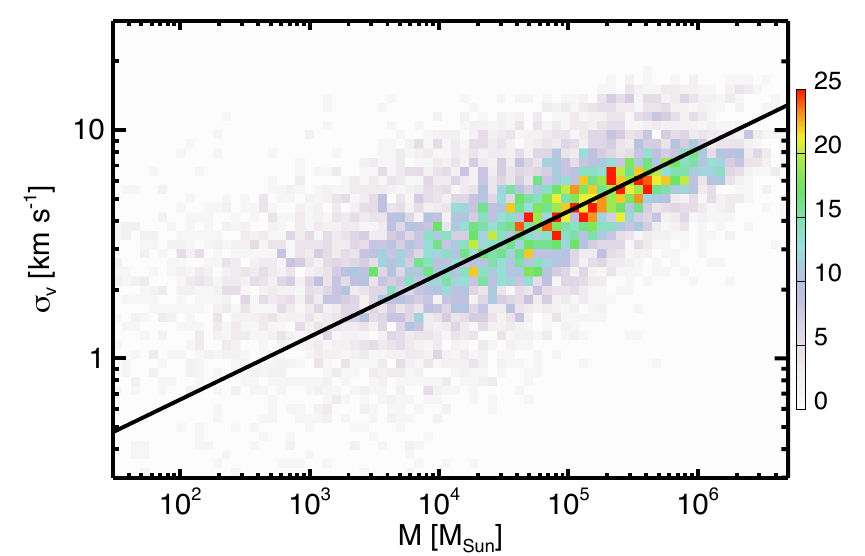}
\includegraphics[draft=false, angle=0]{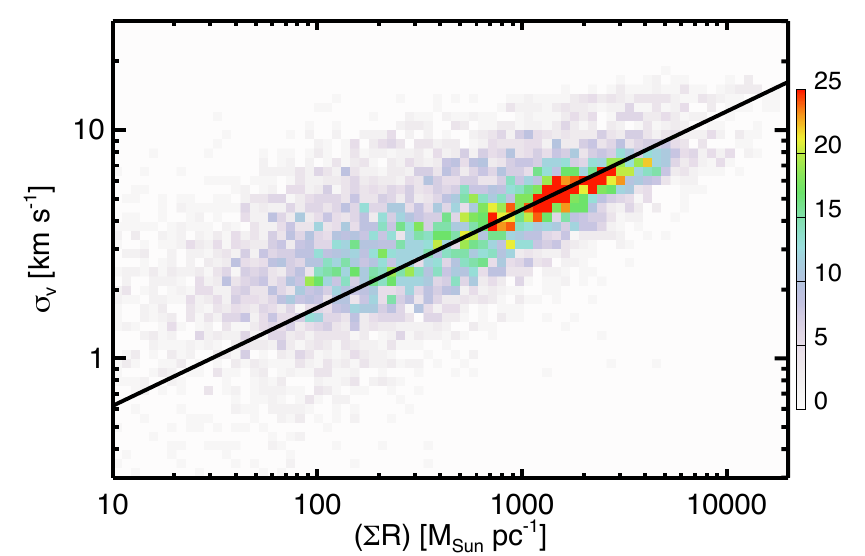}
\caption{\label{fig:m_vs_r} Relations between quantities of the catalog shown as two-dimensional histograms. In all four diagrams, all the clouds in the catalog were used. The solid line corresponds to a linear fit done using the bisector estimator. The uncertainty on the exponent was computed by comparing the bisector result to the results from the $Y$ vs. $X$ and $X$ vs. $Y$ linear fits. The color scale is proportional to the density of points. {Top-left: } $M$ vs. $R$. The fit is $M = 36.7 \, R^{2.2 \pm 0.2}$ and the Pearson correlation coefficient is $0.89$. The two ridgelines traced by the red pixels are produced by inner-Galaxy clouds (the upper ridgeline) and outer-Galaxy clouds (the lower ridgeline); these two populations can also be seen in the top left panel of Figure \ref{fig:PDF_mass}. {Top-right:}  $\sigma_v$ vs. $R$. The fit is $\sigma_v = 0.48\, R^{0.63\pm0.30}$, and the Pearson correlation coefficient is $0.55$. {Bottom-left: } $\sigma_v$ vs. $M$. The fit is $\sigma_v = 0.19\, M^{0.27 \pm 0.10}$, and the Pearson correlation coefficient is $0.68$. {Bottom-right: } $\sigma_v$ vs. ($\Sigma R$). The fit is $\sigma_v = 0.23\, (\Sigma R)^{0.43\pm0.14}$, and the Pearson correlation coefficient is 0.71.}
\end{figure*}

Compared to the mass ($\propto D^2$), the density is less biased by distance ($\propto D^{-1}$). It is true that closer clouds are generally smaller and tend to have a higher density (the distance bias), but the face-on view of $n_{\rm H2}$ (Figure~\ref{fig:xy_density}) also shows a general increase of the average density inside the solar circle. In the outer Galaxy, clouds are about 10 times less dense than in the spiral arms. We notice that the Scutum-Centaurus arm (at $X\approx 0\,{\rm kpc}$, $Y\approx5\,{\rm kpc}$), where a significant fraction of the star-forming activity of the Milky Way takes place, shows a clear increase in gas density relative to other regions.

Finally, Figure~\ref{fig:xy_height} in Appendix~\ref{sec:XYviews} shows the distance to the midplane for each cloud. This diagram provides an interesting view of the Galactic warp seen in molecular emission. In the inner Galaxy, the molecular gas is concentrated in a layer close to $z=0$, with a scale height of the order of 100\,pc. On the other hand, in the outer Galaxy, the molecular layer is clearly flaring and it is warped. Toward $l=90^\circ$ the molecular gas is above the midplane, while it is below toward $l=270^\circ$. This has been identified by many studies since the 1980s \citep[see Figure~6 of][]{heyer2015}, but this figure is the first time that it has been shown over the whole plane with a single coherent sample of clouds.

\subsubsection{The Local Bubble}

Close to the Sun, most of the interstellar matter is seen at higher Galactic latitude. For instance, all the Gould Belt clouds are at intermediate latitudes \citep[$|b| \sim 10-20^\circ$, see][]{perrot2003}. Even though the current study concentrates only on CO emission in a very limited range in latitude ($-5^\circ < b < 5^\circ$), we notice that the global morphology of the solar neighborhood is reproduced. Figure~\ref{fig:xy_localbubble} shows a $1\times 1$\,kpc face-on view of the clouds' surface density, centered on the Sun. Also shown are the isocontours of ${\rm E(B-V)/pc}$ of \citet{lallement2014} obtained by an inversion of stellar color excess measurements obtained in the optical. This method uses parallax or photometric estimates of distance, providing a much more precise determination of the 3D structure of the ISM close to the Sun. The work of \citet{lallement2014} clearly shows the position of nearby clouds with respect to the Sun. Several of these clouds are in the range $-30^\circ < l < 75^\circ$ and at a distance $100-300$\,pc. They are part of the Gould Belt. These correspond to the Ophiuchus, Lupus, and Chamaeleon-Musca clouds, all located at $|b|>5^\circ$. Therefore, they are not present in our catalog.
On the other hand, complexes located at lower latitude seem to show a rather good correspondence with the clouds found here. This is the case for Vela, Cygnus, and Cassiopeia. Even part of the Acquila, Perseus, and Orion complexes are captured. 
One very striking feature of the solar neighborhood is the Local Bubble, a vast region with almost no dense gas that extends toward $l\sim 240^\circ$, corresponding to a diagonal to the upper left in Figure \ref{fig:xy_localbubble}. This is very well captured by the current catalog.

\subsection{Mass--Size Relation}

\label{sec:mass-size}

Figure~\ref{fig:m_vs_r} shows the $M-R$ relation for our sample of molecular clouds. 
The relation found using the bisector estimate is $M = 36.7 \, R^{2.2 \pm 0.2}$. 
The statistical uncertainty on the exponent of the bisector estimator is only 0.01. Here we quote the uncertainty calculated from the difference between the result from the bisector estimate and the results from the $Y$ versus $X$ and $X$ versus $Y$ fits. 

The mass--size relation has two potentially very interesting applications. First, it might reveal properties of interstellar turbulence. The mass--size relation has been investigated in numerical simulations of isothermal flows without gravity \citep{kritsuk2007,federrath2009,audit2010}. Depending on the numerical setup and on the structure identification method used, exponents in the range of $2.0 - 2.5$ were found. It is thought that this relation is linked to the properties of turbulence \citep{kritsuk2007}. A similar range of values is obtained from observations of molecular clouds. \citet{hennebelle2012a} showed that, when compiled, the observational results are compatible with a variation of the exponent with $R$, i.e., non-power-law behavior, a feature that was also noticed in numerical simulations \citep{veltchev2016}. 

The second application is related to the fact that the $M-R$ relation could be used as a way to estimate the distance to clouds. As $M \propto D^2$ and $R \propto D$, an exponent different from 2 in the $M-R$ relation would provide a way to estimate distance. From their analysis of Galactic Ring Survey data, \citet{roman-duval2010} concluded that $M \propto R^{2.36}$, and they used this relation to solve the near/far ambiguity for molecular clouds. On the other hand, other studies \citep[e.g.,][]{larson1981,beaumont2012} found exponents close to 2. 

The exponent of the $M-R$ relation found here is compatible within the error with 2.0. Therefore, we did not use this relation as a distance estimator. The fact that $M\propto R^2$ implies that $\Sigma$ is independent of $R$. Indeed, we do not see any correlation between $\Sigma$ and $R$ (Pearson correlation coefficient of 0.08). This also implies that $n_{\rm H2} \propto R^{-1}$.


\begin{figure}
\centering
\includegraphics[draft=false, angle=0]{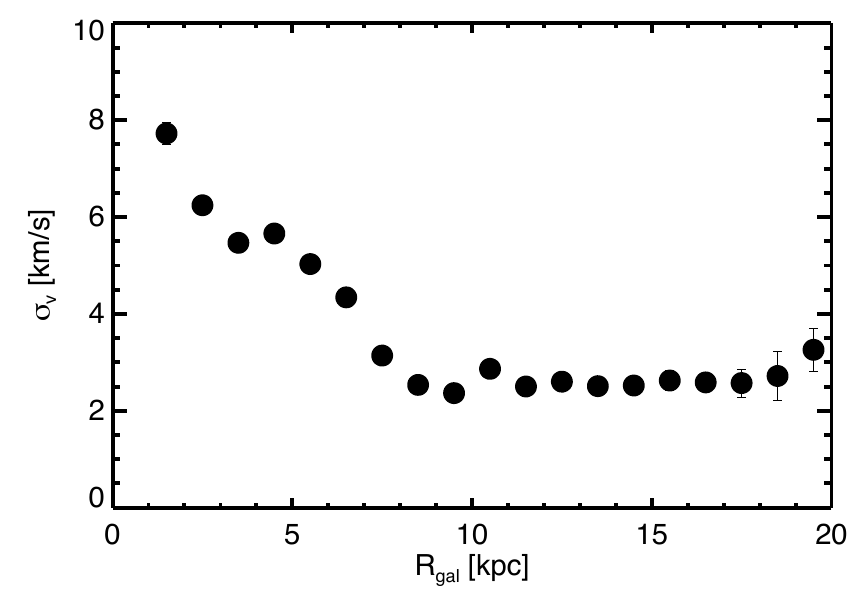}
\includegraphics[draft=false, angle=0]{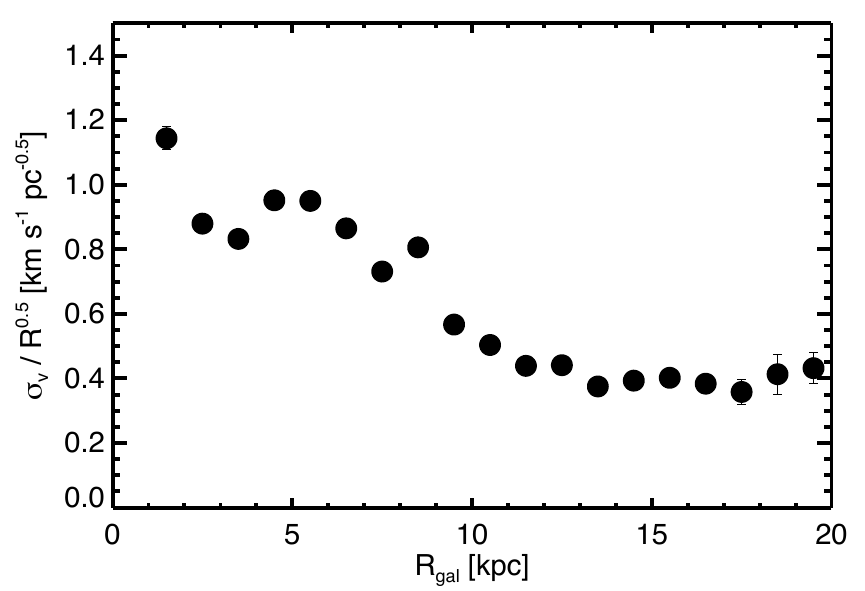}
\includegraphics[draft=false, angle=0]{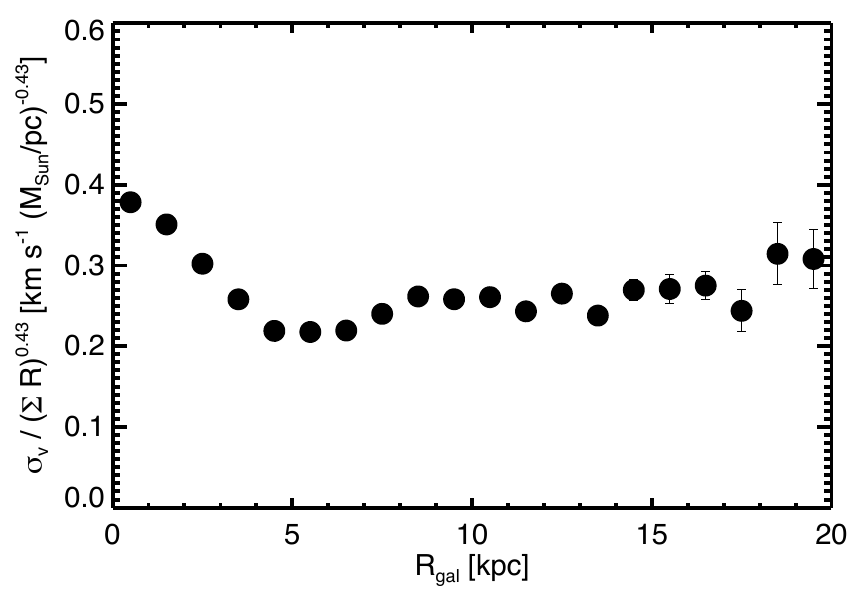}
\caption{\label{fig:sigmav_norm} Variation of $\sigma_v$ (top), $\sigma_v/R^{1/2}$ (middle), and $\sigma_v/(\Sigma R)^{0.43}$ (bottom) as a function of galactocentric radius. In all panels each dot and its associated error bar indicate, respectively, the median and $1\sigma$ dispersion of the values in a bin of $R_{\rm gal}$.}
\end{figure}

\subsection{Velocity--Size Relation}

\label{sec:velocity-size}

The seminal work of \citet{larson1981} provides some foundations on which the theoretical framework of star formation has been built. \citet{larson1981} obtained two main observational relationships: one between the cloud's largest dimension, $L$, and the cloud line width ($\sigma_v = 1.10\,L^{0.38}$), and one between size and volume density ($n_{\rm H2} = 3400\,L^{-1.1}$). Assuming that $M \propto n_{\rm H_2}\,L^3$, these relationships imply that molecular clouds are in virial equilibrium, on average.\footnote{More precisely, it implies that the virial parameter is close to one with a weak dependence on scale: $2\,G\,M/\sigma_v^2 L \propto L^{0.14}$.} These results were seemingly confirmed by the analysis of \citet{solomon1987} on a larger sample of clouds (273 instead of 46). Contrary to \citet{larson1981}, who obtained a $\sigma_v-L$ exponent close to what is expected for incompressible turbulence ($\sigma_v \propto L^{1/3}$), \citet{solomon1987} obtained $\sigma_v \propto L^{1/2}$, which is consistent with expectations for compressible supersonic turbulence \citep{brunt2004} or simply for clouds in virial equilibrium. In the latter case, $\sigma_v$ also depends on surface density \citep{heyer2009}. To this day, the origin of the size--line width relation (gravity or turbulence) is still a matter of debate \citep{ballesteros-paredes2011a}.

The $\sigma_v-R$ relation for our whole sample of clouds is shown in the top right panel in Figure~\ref{fig:m_vs_r}.
The best fit, computed using the bisector estimator, gives
\begin{equation}
\label{eq:sigv_R}
\sigma_v = 0.48\, R^{0.63\pm0.30}.
\end{equation}
Like for the $M-R$ relation, the statistical uncertainty on the exponent is calculated from the difference between the result from the bisector estimate and the results from the $Y$ versus $X$ and $X$ versus $Y$ fits. The Pearson correlation coefficient of this relation is 0.55.

In addition to $\sigma_v-R$, we explored a number of different relations between $\sigma_v$ and other cloud parameters, as well as combinations of parameters. For instance, we found a rather good correlation between $\sigma_v$ and $M$ (Figure~\ref{fig:m_vs_r}, bottom left panel), with a Pearson correlation coefficient of 0.68:
\begin{equation}
\label{eq:sigv_M}
\sigma_v = 0.19\, M^{0.27 \pm 0.10}.
\end{equation}
This result is similar to the relation $\sigma_v = 0.42 M^{0.20}$ found by \citet{larson1981}.

Of all the relations involving $\sigma_v$ that we explored, that between $\sigma_v$ and  $(\Sigma R)$ (Figure~\ref{fig:m_vs_r}, bottom right panel) is the tightest one we found (Pearson coefficient of 0.71). Using a different data set, \citet{heyer2009} reached the same conclusion. Our linear fit gives
\begin{equation}
\label{eq:sigv_SigmaR}
\sigma_v = 0.23\, \left( \Sigma R \right)^{0.43 \pm 0.14}.
\end{equation}
We found that the normalization of this relation ($\sigma_v / (\Sigma R)^{0.43}$) is nearly constant across the Galactic plane, as shown in the bottom panel of Figure~\ref{fig:sigmav_norm}.

\begin{figure}
\centering
\includegraphics[draft=false, angle=0]{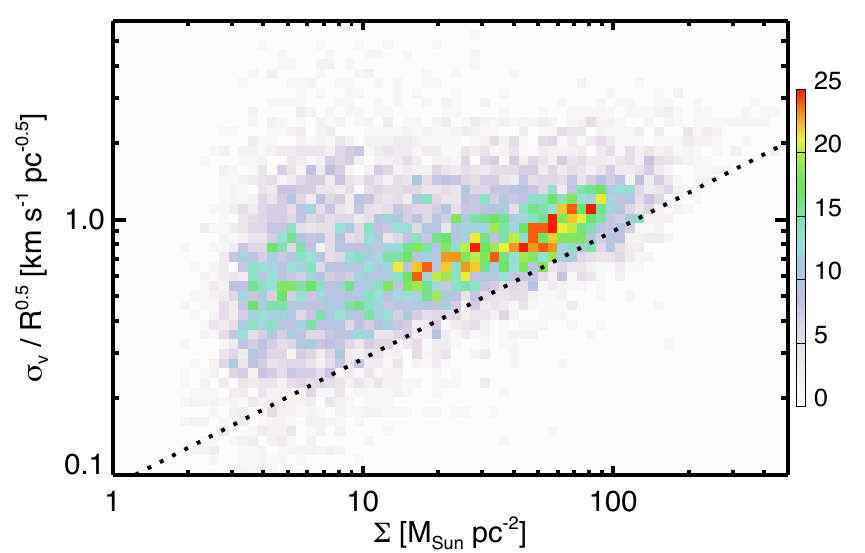}
\caption{\label{fig:sigv_virial} Two-dimensional histogram of $\sigma_v / R^{0.5}$ vs. $\Sigma$. The color scale is proportional to the density of points. The dotted line indicates the locus of gravitational bound structures ($\alpha_{\rm vir}=3$; see Eq.~\ref{eq:alpha}). }
\end{figure}

The almost constant value of this quantity across the Galactic plane contrasts with the variation of $\sigma_v$ with $R_{\rm gal}$ (Figure~\ref{fig:sigmav_norm} top panel). We found that $\sigma_v$, averaged over Galactic annuli, decreases with increasing $R_{\rm gal}$ in the inner Galaxy, while it is nearly independent of $R_{\rm gal}$ outside the solar circle. 

This effect is not due primarily to variations in cloud radius with $R_{\rm gal}$; on average, clouds inside the solar circle are similar in size to those in the outer Galaxy; see the lower left panel of Figure \ref{fig:PDF_mass}, where the orange and blue histograms have similar shapes.
This is demonstrated in the middle panel of Figure \ref{fig:sigmav_norm}. Assuming the scaling velocity with size appropriate for supersonic turbulence ($\sigma_v = \sigma_0 \, R^{1/2}$), the middle panel of Figure~\ref{fig:sigmav_norm} shows the normalization $\sigma_0$ as a function of $R_{\rm gal}$ (see also the face-on view of $\sigma_0$ in Figure~\ref{fig:xy_sigv1pc}). The decrease of $\sigma_0$ with $R_{\rm gal}$ inside the solar circle is less pronounced than $\sigma_v$ but still signifiant. This is compatible with the analysis of \citet{oka2001}, who found that the $\sigma_v-R$ relation in the Galactic center area ($R_{\rm gal} < 3$\,kpc) has a significantly larger normalization than that found at larger Galactic radii. Over the whole sample of clouds, the median value of $\sigma_0$ is $0.75$\,km\,s$^{-1}$, with a standard deviation of $0.43$\,km\,s$^{-1}$. We note that the value in the solar neighborhood ($\sigma_0 \approx 0.8$\,km\,s$^{-1}$) is similar to the one estimated in the more diffuse H\,{\sc i} component \citep{wolfire2003,saury2014}.

Adding $\Sigma$ to $R$ in the parameterization of $\sigma_v$ provides a significant statistical improvement. The normalization of the $\sigma_v$ versus $(\Sigma\,R)$ relation shows significantly less variation with $R_{\rm gal}$ than $\sigma_0$ (see middle and bottom panels in Figure~\ref{fig:sigmav_norm}). Figure~\ref{fig:sigv_virial} shows the two-dimensional histogram of $\sigma_0$ as a function of $\Sigma$. For most of the range in $\Sigma$ sampled here, i.e., for $\Sigma\gtrsim 10\,M_{\rm Sun}\pc^{-2}$, we observe a clear increase of  $\sigma_0$ with $\Sigma$, contrary to what is mentioned by \citet{hennebelle2012a}.

Based on these results, we used the $\sigma_v - (\Sigma R)$ relation, instead of $\sigma_v - R$, to select between near and far distance.

We note that this variation is weaker at low values of $\Sigma\lesssim10\,M_{\rm Sun}\pc^{-2}$, where $\sigma_0$ seems unrelated to $\Sigma$. As seen in Figure~\ref{fig:xy_sigv1pc}, clouds with low values of $\sigma_0$ are generally located at large $R_{\rm gal}$ where there is no near/far distance ambiguity. The behavior seen here is reminiscent of the results of \citet{heyer2001} who found that small clouds in the outer Galaxy have a velocity dispersion independent of size.

\begin{figure}
\centering
\includegraphics[draft=false, angle=0]{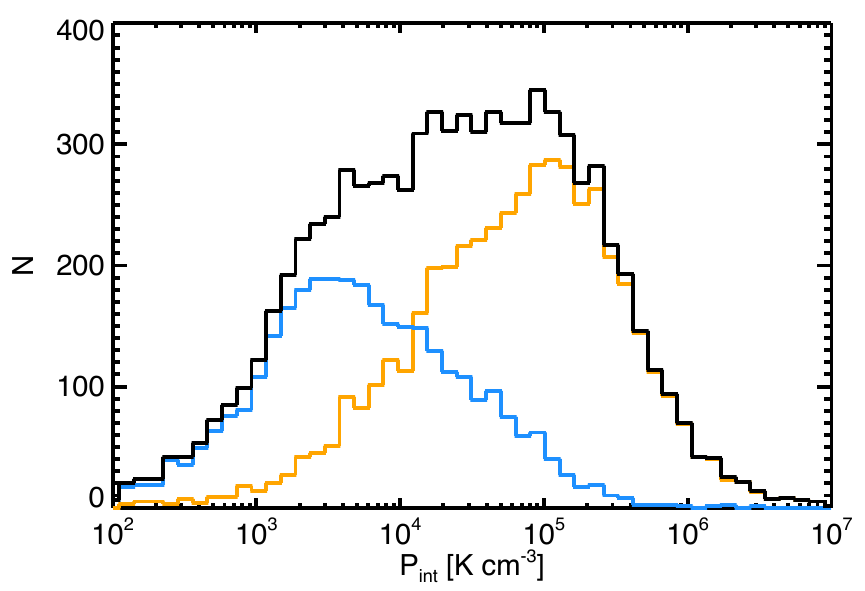}
  \includegraphics[draft=false, angle=0]{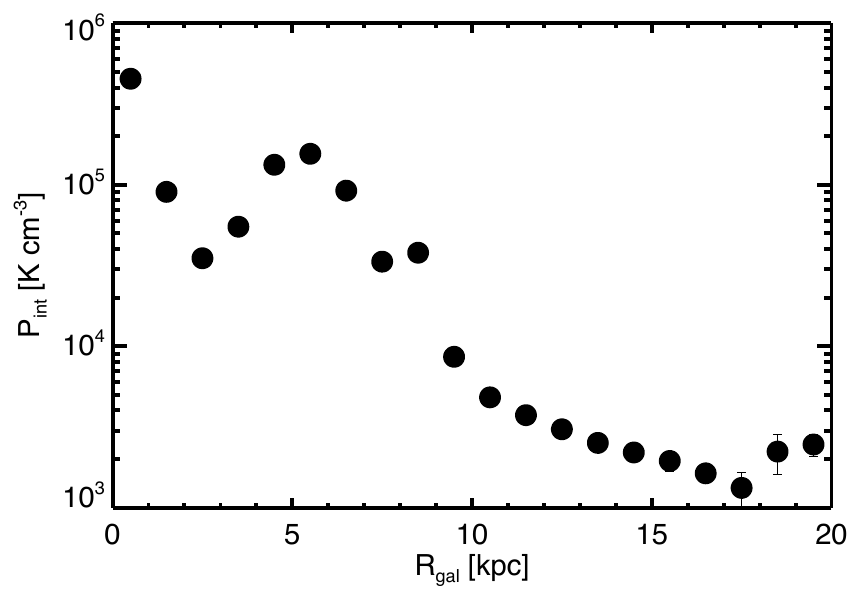}
\caption{\label{fig:PDF_pressure} Internal pressure. {Top:} histogram of $P_{\rm int}$. The solid black line shows the full sample, while the orange and blue lines show, respectively, the clouds located in the inner and outer Galaxy. {Bottom:} median value of $P_{\rm int}$ in  rings of constant $R_{\rm gal}$.}
\end{figure}

\subsection{Advanced Physical Parameters}

In this section we explore more advanced physical quantities to help unravel the physical conditions of molecular clouds and, in particular, the respective role of gravity and turbulence in their evolution. We compute the internal pressure, the virial parameter, the free-fall and dynamical times, and the turbulence energy dissipation and transfer rates.

\subsubsection{Internal Pressure}

For every cloud in the sample, the internal (turbulent) pressure, $P_{\rm int}$ (in units of K\,cm$^{-3}$), is estimated as
\begin{equation}
\label{eq:pressure}
P_{\rm int} = \frac{2\,m_p}{k}\,n_{\rm H2}\,\sigma_v^2,
\end{equation}
where $k$ is the Boltzmann constant. The histogram of $P_{\rm int}$ (Figure~\ref{fig:PDF_pressure}, top panel) illustrates the large range of pressure found: over more than 3 orders of magnitude. The median value is $P_{\rm int} = 3.0\times 10^4$\,K\,cm$^{-3}$, but with a large standard deviation of $5.0\times 10^4$\,K\,cm$^{-3}$. 

From Eq.~\ref{eq:pressure} we note that the total pressure scales like $\Sigma/R \times \sigma_v^2$. As $\sigma_v^2 \propto (\Sigma R)^{0.86}$, the total pressure scales almost like $\Sigma^2$ and is thus very weakly dependent on distance. Therefore, as for $\Sigma$, the clear difference in $P_{\rm int}$ between the inner and outer Galaxy (see orange and blue histograms in Figure~\ref{fig:PDF_pressure}) is physical and not an observational bias. This is also clearly seen in the bottom panel of Figure~\ref{fig:PDF_pressure}, where the variation of $P_{\rm int}$ with $R_{\rm gal}$ appears similar to the variation of $\Sigma$ with $R_{\rm gal}$ (Figure~\ref{fig:Sigma_vs_Rgal}). The similarity between the spatial distribution of $P_{\rm int}$ and $\Sigma$ is also seen in the face-on views (Figures~\ref{fig:xy_surfdens} and \ref{fig:xy_pressure}).

As it has been noticed before \citep[e.g.][]{blitz1993}, the typical internal pressure of molecular clouds is much larger than the typical pressure of the diffuse H\,{\sc i}, $P_{\rm int} = 3.7\times 10^3$\,K\,cm$^{-3}$ in the solar neighborhood according to \citet{jenkins2011}. These authors found excursions in H\,{\sc i} pressure from $10^3$ to $10^4$\,K\,cm$^{-3}$, which is still less than the typical values found here. 

\begin{figure}
\centering
\includegraphics[draft=false, angle=0]{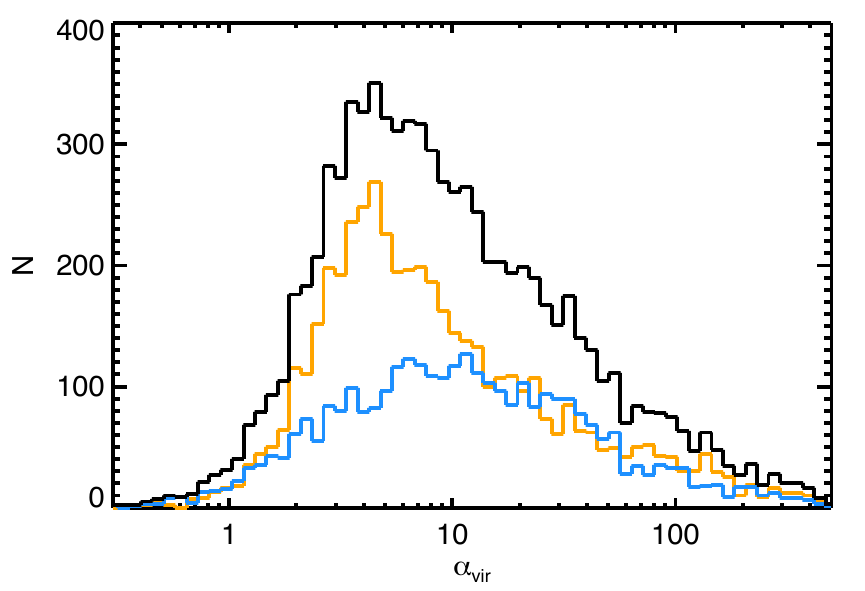}
\includegraphics[draft=false, angle=0]{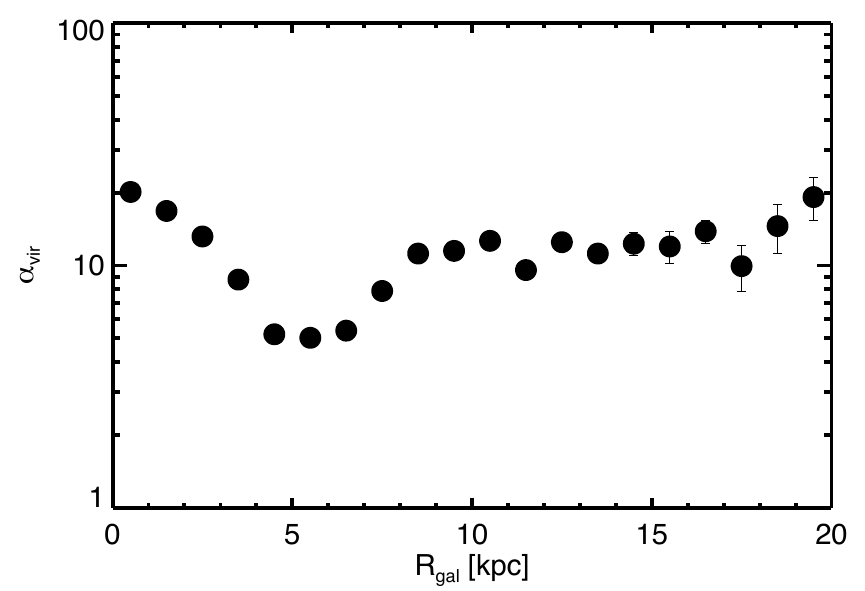}
\caption{\label{fig:PDF_alpha} Virial parameter. {Top:} histogram of $\alpha_{\rm vir}$. The solid black line shows the full sample, while the orange and blue lines show, respectively, the clouds located in the inner- and outer-Galaxy. {Bottom:} median value of $\alpha_{\rm vir}$ in  rings of constant $R_{\rm gal}$.}
\end{figure}

\subsubsection{Virial Parameter}

As mentioned by \citet{heyer2009}, a $\sigma_v - (\Sigma R)$ relationship is expected if gravity is a dominant process in the dynamics. Specifically, one expects $\sigma_v \propto (\Sigma R)^{1/2}$ for clouds in virial equilibirum, i.e., $\alpha_{\rm vir}\approx1-3$ where the virial parameter is defined as 
\begin{equation}
\label{eq:alpha}
\alpha_{\rm vir} = \frac{5\,\sigma_v^2\,R}{G\,M}=a{2{\cal T}\over |W|},
\end{equation}
where ${\cal T}$ is the total kinetic energy of the cloud and $W$ is the gravitational energy associated with the cloud \citep[ignoring tidal effects;][]{bertoldi1992}. We calculate the factor $a$ in Appendix~\ref{sec:virial}, where we show that for fiducial values for the power-law index for the density $\rho\sim r^{-k_\rho}$ ($k_\rho =1$) and for the index in Larson's size--line width relation $\sigma(r)\sim r^p$, with $p=1/2$, that $a\approx 5/3$. In other words, the observed virial parameter $\alpha_{\rm vir}$ will be a factor of $5/3$ larger than the ratio of twice the kinetic energy to the gravitational energy. If the latter is of order 2, meaning equal kinetic and gravitational energies, the observed virial parameter will be $\alpha_{\rm vir}\approx 3$. 

The histogram of $\alpha_{\rm vir}$  for the whole sample of clouds, as well as for the inner- and outer-Galaxy clouds, is shown in Figure~\ref{fig:PDF_alpha} (top panel). The distribution peaks at $\alpha_{\rm vir}$ of around 3-4, where the kinetic energy of the cloud is about equal to the gravitational energy and is strongly positively skewed. 
 Clouds in the outer Galaxy are more likely to have $\alpha_{\rm vir} >> 1$, something already noted by many studies \citep{sodroski1991,heyer2001}. The Galactic profile (Figure~\ref{fig:PDF_alpha}, bottom panel) and the face-on view (Figure~\ref{fig:xy_alpha}) of $\alpha_{\rm vir}$ indicate a slight decrease in the $3 < R_{\rm gal} < 7$\,kpc region. We also note an increase of $\alpha_{\rm vir}$ in the inner part of the Galaxy ($R_{\rm gal} < 3$\,kpc) as well as an unexpected variation of $\alpha_{\rm vir}$ in the outer Galaxy (see Figure~\ref{fig:xy_alpha}). The virial parameter is systematically larger in the third quadrant ($180^\circ < l < 270^\circ$) compared to the second quadrant ($90^\circ < l <180^\circ$). Like $P_{\rm int}$, the virial parameter is very weakly dependent on distance.

\begin{figure}
\centering
\includegraphics[draft=false, angle=0]{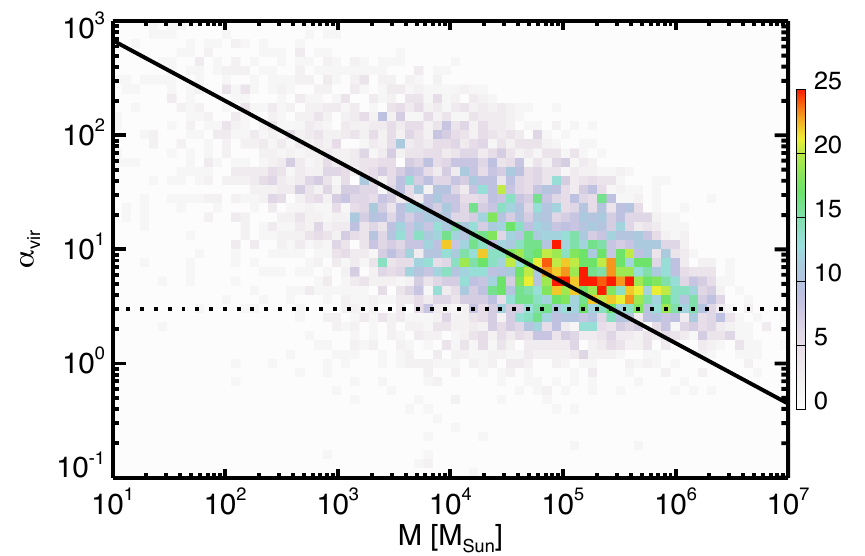}
\caption{\label{fig:alpha_mass} Two-dimensional histogram of $\alpha_{\rm vir}$ vs $M$. The color scale is proportional to the density of points. The dotted line indicate the locus of gravitational bound structures ($\alpha_{\rm vir}=3$, see Eq.~\ref{eq:alpha}). The solid line is the result of the fit: $\alpha_{\rm vir} = 2.3 \times 10^3 \, M^{-0.53\pm0.30}$. The Pearson coefficient of $\alpha_{\rm vir}$ vs $M$ is -0.54.}
\end{figure}

Figure~\ref{fig:alpha_mass} shows the two-dimensional histogram of $\alpha_{\rm vir}$ versus $M$. The correlation between these two quantities is modest (Pearson coefficient of -0.54). Overall, only 15\% of the clouds have $\alpha_{\rm vir}\leq 3$ but they contribute 40\% of the total mass in clouds. 
Assuming that $\alpha_{\rm vir} = 2.3 \times 10^3 \, M^{-0.53\pm0.30}$ (see Figure~\ref{fig:alpha_mass}), one could conclude that clouds with $M > 2.8 \times 10^5$\,M$_\odot$ are gravitionally bound. On the other hand, looking in detail at Figure~\ref{fig:alpha_mass}, there are more unbound clouds with $M > 2.8 \times 10^5$\,M$_\odot$ than bound clouds. Therefore, even though more massive clouds tend to have a lower virial parameter, it seems difficult to conclude that mass is the main criterion that defines the gravitational stability state of a given cloud.

These results are commensurate with previous estimates and with the idea that unbound (high $\alpha_{\rm vir}$) clouds are less efficient at forming stars:  \citet{liszt2010} argue that about 40\% of the Galactic molecular gas is not forming stars, while \citet{goldsmith2008} reached the same conclusion specifically for the Taurus molecular clouds.

\subsubsection{Free-fall Time and Dynamical Time}

\begin{figure}
\centering
\includegraphics[draft=false, angle=0]{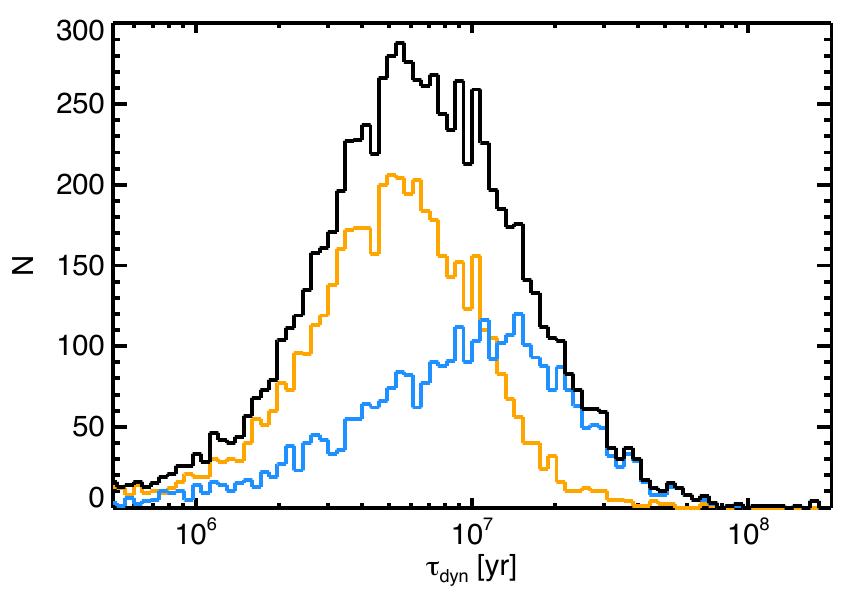}
\includegraphics[draft=false, angle=0]{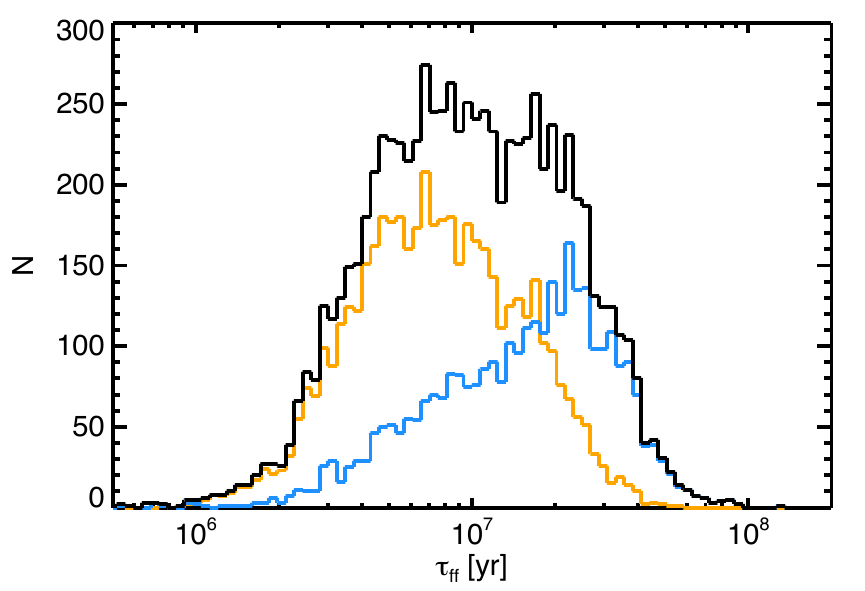}
\caption{\label{fig:PDF_time} Histogram of the crossing time $\tau_{\rm dyn} = \sigma_v/R$ and free-fall time $\tau_{\rm ff} = (3\pi/32\,G\,\rho)^{1/2}$. The solid black line shows the full sample, while the orange and blue lines show, respectively, the clouds located in the inner and outer Galaxy. }
\end{figure}

The comparison of the free-fall time
\begin{equation}
\tau_{\rm ff} = \left( \frac{3\pi}{32 G \rho} \right)^{1/2}
\end{equation}
where $\rho = \mu 2 m_p n_{\rm H2}$,
and the dynamical time (or crossing time)
\begin{equation}
\tau_{\rm dyn} = \frac{R}{\sigma_v}
\end{equation}
provides some insight into the physical processes involved in the efficiency with which molecular gas turns into stars. 

The histograms of $\tau_{\rm ff}$ and $\tau_{\rm dyn}$ are shown in Figure~\ref{fig:PDF_time}. They both have a relatively narrow distribution. The median values are  $\tau_{\rm ff} = 1.0 \times 10^7$\,yr and  $\tau_{\rm dyn} = 6.4 \times 10^6$\,yr. To first order, the free-fall time is about $1.5$ times the value of the dynamical time. Both $\tau_{\rm ff}$ and $\tau_{\rm dyn}$ increase from the inner to the outer Galaxy, but their ratio, like $\alpha_{\rm vir}$, stays about the same. 

The ratio of the total molecular mass of the Milky Way to the typical free-fall time ($1.6 \times 10^9$\,M$_\odot$ /$1.0 \times 10^7$\,yr $=160$\,M$_\odot$\,yr$^{-1}$) provides an estimate of what the star formation rate would be if all molecular clouds were to form stars and if gravity alone were to drive the formation process. A finer estimate of this quantity can be obtained by summing $M / \tau_{\rm ff}$ for all the clouds in the sample. It gives a similar value: $215\,M_\odot$\,yr$^{-1}$. This has to be compared with the actual star formation rate of the Milky Way, which is $\sim 2\,M_\odot$\,yr$^{-1}$, indicating that the star formation efficiency is of the order of 1\% on average.

\subsubsection{Turbulence Energy Dissipation and Transfer Rates}

In order to evaluate what processes drive turbulence in molecular clouds and regulate star formation in galaxies, it is essential to quantify the properties of turbulence. One important parameter is the rate at which energy is injected and dissipated in the turbulent cascade and the rate at which it transfers from large scales to small scales. 

In a three-dimensional turbulent flow, the kinetic energy injected at large scales is transferred to smaller scales through nonlinear interactions, down to the dissipation scale, where it is converted into heat. In an incompressible turbulent flow, \citet{kolmogorov1941} showed that the energy tranfer rate per unit mass, $v^3/l$, is conserved. This means that, for a given turbulent flow with some energy injected at large scales, the energy transfer rate is constant whatever the scale at which it is estimated. 
Later, \citet{lighthill1955} pointed out that, in a compressible turbulent flow, the quantity that is conserved is the mean volume energy transfer rate defined as
\begin{equation}
\label{eq:epsilon}
  2\epsilon = \frac{\rho \, \sigma_v^3}{R}.
  \end{equation}
That led \citet{kritsuk2007} to propose an extension of Kolmogorov's theory \citep{kolmogorov1941} to compressible flow, which applies to molecular clouds. They showed that the quantity $u = \rho^{1/3} v$ has the same statistical properties as the velocity field $v$ in incompressible turbulence, whatever the Mach number of the flow.

It has been seen in several numerical simulations that the turbulent energy decays in one turnover time if it is not maintained \citep[see the review of][]{hennebelle2012a}. 
Following \citet{mac_low1999}, the energy dissipation rate can be estimated as the total kinetic energy divided by the dynamical time:
\begin{equation}
\dot{E}_{\rm dis} = -\frac{1}{2} \, \frac{M\,\sigma_v^3}{R},
\end{equation}
which is simply the kinetic energy transfer rate $\epsilon$ integrated over the whole volume of a cloud.

We computed the $\epsilon$ and $\dot{E}_{\rm dis}$ for each cloud in the catalog. 

\begin{figure}
\centering
\includegraphics[draft=false, angle=0]{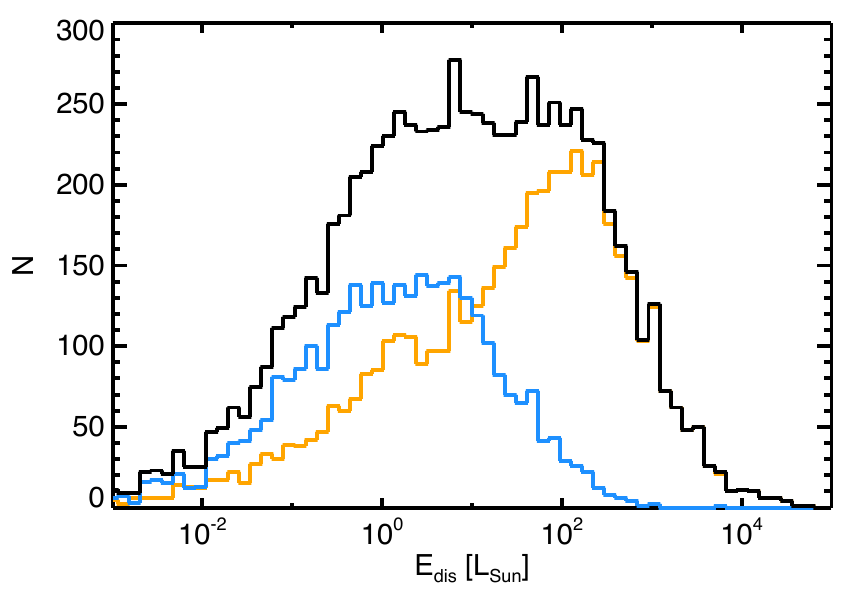}
  \includegraphics[draft=false, angle=0]{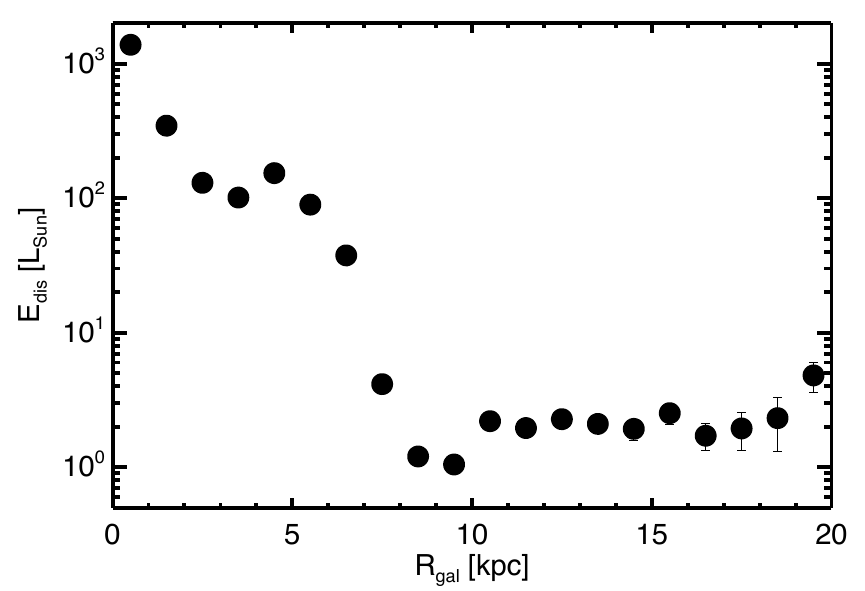}
\caption{\label{fig:PDF_Ediss} Turbulent energy dissipation rate. {Top:} histogram of $\dot{E}_{\rm dis}$. The solid black line shows the full sample, while the orange and blue lines show, respectively, the clouds located in the inner and outer Galaxy. {Bottom:} median value of $\dot{E}_{\rm dis}$ in rings of constant $R_{\rm gal}$.}
\end{figure}

The histogram of $\dot{E}_{\rm dis}$ and its variation with $R_{\rm gal}$ are shown in Figure~\ref{fig:PDF_Ediss}. The face-on view of $\dot{E}_{\rm dis}$ is also shown in Figure~\ref{fig:xy_Edis}. The variation of the median value of $\dot{E}_{\rm dis}$ as a function of $R_{\rm gal}$ is quasi-bimodal, with  $\dot{E}_{\rm dis} \sim 2$\,L$_\odot$ for $R_{\rm gal} > 7$\,kpc and $\dot{E}_{\rm dis} \sim 100$\,L$_\odot$ at smaller $R_{\rm gal}$. 

\begin{figure}
\centering
\includegraphics[draft=false, angle=0]{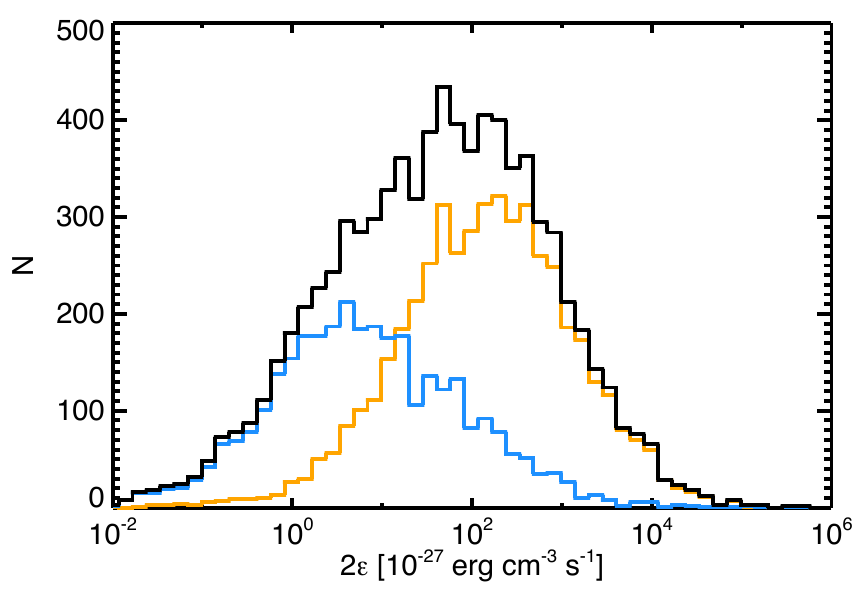}
  \includegraphics[draft=false, angle=0]{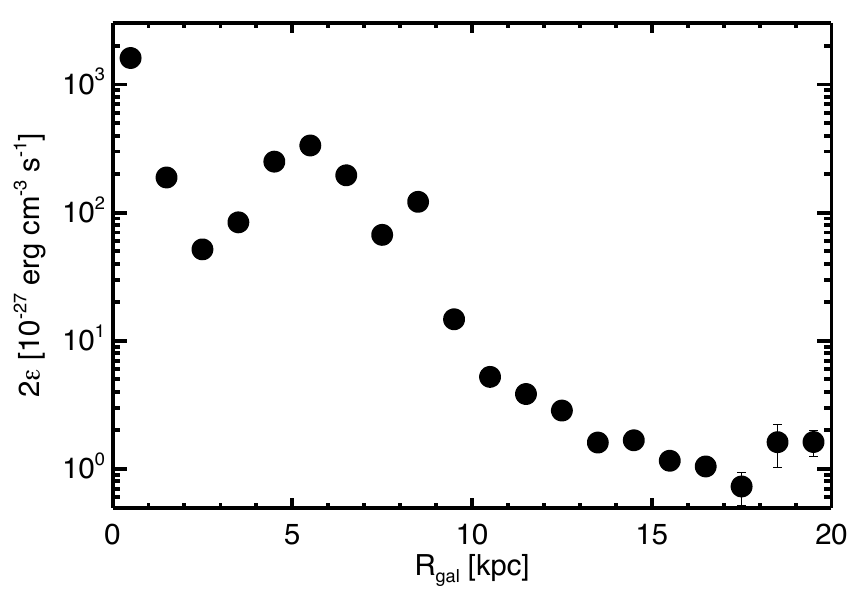}
\caption{\label{fig:PDF_epsilon} Kinetic energy transfer rate. {Top:} Hhistogram of $2\epsilon$. The solid black line shows the full sample, while the orange and blue lines show, respectively, the clouds located in the inner and outer Galaxy. {Bottom:} median value of $2\epsilon$ in rings of constant $R_{\rm gal}$.}
\end{figure}

The histogram of $2\epsilon$ (Figure~\ref{fig:PDF_epsilon}) shows a significant spread in values, of about 4 orders of magnitude, similar to the one of $\dot{E}_{\rm dis}$. There is a clear difference in the value of $\epsilon$ between the inner and outer Galaxy; it drops steadily for $R_{\rm gal} > 5$\,kpc, with an exponential scale length similar to what we found for the mass surface density $\Sigma$ (Figure~\ref{fig:Sigma_vs_Rgal}). For $R_{\rm gal}<9$\,kpc, the value of $2\epsilon \sim 10^{-25}$\,erg\,cm$^{-3}$\,s$^{-1}$ is similar to previous estimates obtained for molecular clouds and H\,{\sc i} clouds in the solar neighborhood \citep{hennebelle2012a}.

We observed no variation of $\epsilon$ with GMC radius $R$, which is not inconsistent with the presence of a turbulent cascade. However, we are unable to evaluate whether $\epsilon$ is constant through scales in individual clouds, as was done in \citet{hennebelle2012a}. As mentioned in \S~\ref{sec:sigv_discussion}, we are only sampling the large-scale properties of compressible turbulence for isolated clouds, not their multiscale nature.

Finally, we note that $\dot{E}_{\rm dis}$ and $\epsilon$ depend significantly on distance with $\dot{E}_{\rm dis} \propto D^{2.3}$ and $\epsilon \propto D^{-0.7}$. One could worry that this would affect the radial distributions shown in Figures~\ref{fig:PDF_Ediss} and \ref{fig:PDF_epsilon}. The fact that these two quantities depend in an opposite manner on $D$ ($\dot{E}_{\rm dis}$ tends to increase with $D$ while $\epsilon$ decreases) provides some guidelines to evaluate the impact of the distance bias on the trends seen here. The 2 orders of magnitude difference between the inner and outer Galaxy, seen for both quantities, cannot be the result of an observational bias. On the other hand, the exact shape of the variations in the range $5 < R_{\rm gal} < 15$\,kpc could be partly due to the different distance bias of $\dot{E}_{\rm dis}$ and $\epsilon$.

\section{Discussion}

\label{sec:discussion}

\subsection{The Cloud Identification Method}

The defining feature of the cloud identification method presented here is that it relies on a Gaussian decomposition of the brightness temperature spectra. This allows us to compress the molecular emission into a smaller number of coefficients, limiting the effect of noise and increasing the contrast of structures in PPV space when the emission of the Gaussian components is concentrated in the centroid velocity channel. All this facilitates the identification of coherent structures. It also allows us to include the emission down to the noise level. 

Compared with similar studies, the current catalog includes a much larger fraction (98\%) of the total observed CO emission. For example, the catalog of clouds covering the full Galactic plane produced by \citet{rice2016} includes 25\% of the total H$_2$ mass. Similarly, the catalogs of \citet{dame1986}, \citet{solomon1987}, and \citet{scoville1987} contain less than 20\% of the total CO emission in a restricted range of Galactic longitude. 

Even though the method is successful at including almost all the CO emission it does not resolve the velocity crowding problem. Structures that are unrelated in 3D but appear connected in PPV space are mistakenly taken as one cloud. This remains a limitation that is difficult to circumvent. One can only hope that the patchy and supersonic nature of the molecular ISM limits this problem \citep{burkhart2013a,pan2015}.

One important validation of the cloud identification and of the distance estimate is the fact that the overall catalog reproduces the H$_2$ surface density as a function of $R_{\rm gal}$ (Figure~\ref{fig:Sigma_vs_Rgal}). This is not a given. The H$_2$ surface density was computed by summing the mass of all clouds within galactocentric rings of increasing $R_{\rm gal}$. The distribution of $\Sigma_{\rm H2}$ versus $R_{\rm gal}$ depends strongly on the distance attributed to each cloud. The comparison with the results of \citet{bronfman1988}, \citet{nakanishi2006}, and \citet{koda2016} is an important validation. These studies do not rely on a cloud identification method. They assumed that the CO emission is coming from rings of constant $R_{\rm gal}$ or Gaussian layers of diffuse molecular gas. In essence they deal with the near/far ambiguity by assuming that there is as much molecular gas on the far side as on the near side of the Galactic center.\footnote{We recall that the difference with the results of \citet{bronfman1988} for $R_{\rm gal} < 3$\,kpc is due to the fact that these authors did not use the CO emission in the region $-12^\circ < l < 12^\circ$.}
Correspondingly, the total H$_2$ mass in the catalog, $1.2\times 10^9$\,M$_\odot$, is in accordance with previous estimates that used different methods \citep{heyer2015}. 

There is another indication that the kinematic distance used here is valid. 
{\citet{lee2016} showed that for 90\% of the clouds for which a star-forming complex could be associated, the near/far resolution based on the $\sigma_v-(\Sigma R)$ relation is in accordance with absorption measurements. }

\subsection{Size--Line Width Relation}

\label{sec:sigv_discussion}

Like many studies before us \citep[see the reviews by][]{hennebelle2012a,heyer2015}, we find a large scatter in the $\sigma_v-R$ plane, which is reflected in the large uncertainty on the slope ($0.3$) and the relatively low Pearson correlation factor (0.55). 
We also found that clouds located in the outer Galaxy seem to have a shallower $\sigma_v-R$ relation, if indeed there is any correlation. Like \citet{heyer2001}, we found that clouds smaller than about 7\,pc seem to have a velocity dispersion unrelated to $R$. 

The interpretation of the $\sigma_v-R$ relation shown in Figure~\ref{fig:m_vs_r} deserves some clarifications about what is really measured here. 

Let us consider a given molecular cloud with a well-developed turbulent cascade fed by some energy input at large scales. The velocity field of such a cloud is described by a power-law power spectrum at scales smaller than the energy input scale. We write this as $\sigma_v = \sigma_{0}\,(R/R_0)^\beta$ where $\beta$ is related to the nature of the turbulence ($\beta=1/3$ for subsonic turbulence, $\beta=1/2$ for supersonic turbulence) and $\sigma_{0}$ to the amplitude of the turbulent motion (or to the Reynolds number). 
This type of statistical signature has been identified in individual clouds \citep[see][]{hennebelle2012a}. We emphasize the fact that the Hennebelle \& Falgarone (2012) type of study, looking at the statistical properties measured on different length scales in single clouds, is different from the $\sigma_v-R$ relations obtained by measuring the same property at the largest scale of a large number of clouds, as in Figure~\ref{fig:m_vs_r} (top right panel). The former corresponds to a Type 3 or Type 4 relation in the nomenclature of \citet{goodman1998}, while the latter corresponds to a Type 2 relation.

For each cloud in our sample, the measured velocity dispersion is that corresponding to the largest scale of the cloud. The interpretation in terms of turbulence of a Type 2 $\sigma_v-R$ relation built this way makes sense only if the Galaxy behaves like a turbulent flow on the scales comparable to or larger than the largest clouds. If that were the case, molecular clouds could be seen as cells of a global turbulent cascade driven on some large scale, e.g., the disk scale height of the Milky Way.

It is certainly possible that local sources of energy injection will modify the normalization of the turbulent cascade in different regions of the Galaxy. If there are such local sources of turbulent energy injection, the Type 2 $\sigma_v-R$ relation built by combining clouds from all over the Galaxy will sample any spatial variations of the energy injection mechanism that exist. In other words, the  $\sigma_v-R$ computed as we do in this paper encompasses information not only on $\beta$, but also on the spatial variations of $\sigma_0$. One might worry that if the amplitude of turbulent energy depends on cloud size (i.e., $\sigma_0=\sigma_{0}(R)$) the measured value of $\beta$ might not be very informative about the properties of turbulence in molecular clouds. 

In this context, one natural explanation of the scatter we find in the $\sigma_v-R$ relation is the fact that our sample of clouds traces variations of $\sigma_0$ across the Galactic plane. Clouds with different levels of turbulent energy input will scatter up and down relative to the general trend indicated by the black line in Figure \ref{fig:m_vs_r} (top right panel). The variations of $\sigma_0$ can also be appreciated in the face-on view shown in Figure~\ref{fig:xy_sigv1pc}.

The fact that $\sigma_0$ decreases systematically with with increasing $R_{\rm gal}$ (middle panel, Figure~\ref{fig:sigmav_norm}) is an indication that the amount of energy injected in the turbulent cascade varies with location in the disk. In other words, turbulence is not universal in the Milky Way. In fact, if the amplitude of the turbulent energy injection in a single cloud is related to feedback or gravitational energy, one would expect larger values of $\sigma_{0}$ toward the Galactic center, where both the binding energy and the star formation rate of the GMCs are larger than those of GMCs outside the solar circle. 

The top right panel of Figure \ref{fig:PDF_mass} shows that GMCs in the inner Galaxy are much more massive than those in the outer Galaxy. This suggests the possibility that the larger velocity dispersion in the inner Galaxy is related to the fact that the GMCs are more massive, at a given GMC radius, than GMCs in the outer Galaxy.

It is known that the star formation rate in the inner Galaxy is higher than in the outer galaxy, which suggests another possible contributor to the larger velocity dispersion in the inner Galaxy.  However, given that so few of the massive clouds host massive star clusters \citep{lee2016}, the self-gravity explanation looks better than the feedback explanation. The fact that the line width correlates best with $M/R$ (Figure~\ref{fig:m_vs_r}, bottom right panel) certainly seems to favor self-gravity as the driver of the increased velocity dispersion at small $R_{\rm gal}$. We discuss this point in the next section.

\subsection{What Drives Turbulence?}

One important question in star formation theory is the nature of the energy driving turbulence in molecular clouds.
Many potential sources of turbulence have been proposed: gravitational instabilities in the disk, either global (e.g., spiral arms, leading to accretion through the disk toward the Galactic center) or local (contraction of or accretion onto GMCs); magnetorotational instability \citep{balbus1998} (again driving gas through the disk toward the center),; or stellar feedback (photons, winds, and supernovae). 

Related to this question, there is still a lively controversy over whether molecular clouds are long-lived entities or not. This is connected to the fact that, in the absence of driving, turbulent motions decay within one dynamical time, even in the presence of a magnetic field. Some argue that stars form as molecular clouds form \citep{hartmann2001}. In this scenario, molecular clouds are dynamical structures, and stars form early in the densest small-scale substructures of the cloud \citep{heitsch2006b}, solving the problem of the origin of turbulent energy; the initial collapse drives turbulence, which decays, but not before the cloud is disrupted. Others are in favor of star formation that would last for $4-10\tau_{\rm dyn}$, where stellar feedback would maintain turbulence \citep{tan2006}. 

In this section we discuss several possibilities for the source of turbulent motions in molecular clouds, in light of the results presented here.

\subsubsection{Mass Accretion}

One might look more globally at the problem of turbulence driving. Throughout the course of its evolution, a galaxy loses energy, by radiation,  and gas, by the star formation process and through galactic scale outflows. Without mass accretion from the circumgalactic medium, the intergalactic medium, or both, the star formation rate and the level of turbulent energy would rapidly die off. 

\citet{klessen2010} suggested that gas accretion is the main process that maintains turbulence at the scale of a galaxy, of a molecular cloud, and of a protoplanetary disk. They argue that what maintains the star formation activity and the level of turbulence in galaxies is the accretion of matter from the intergalactic medium. At the scale of clouds, they proposed that turbulence could be driven by gas accretion due to large-scale motions in the disk.

According to \citet{klessen2010}, the turbulent motions in molecular clouds are inherited from the cloud formation process itself. Dense clouds in galaxies are formed in regions of higher pressure where the gas can cool efficiently \citep{saury2014}. These regions will then become molecular and gravitationally unstable. Converging flows, possibly caused by perturbation of the gravitational potential (e.g., spiral density waves) or supernova explosions, are one likely source of pressure increase. In this scenario the turbulent energy is provided by the continuous accretion onto GMCs induced by the large-scale converging flow. 

Following \citet{klessen2010}, the energy injection by mass accretion is
\begin{equation}
\label{eq:Edot_in}
\dot{E}_{\rm inj} = \frac{1}{2} \, \dot{M}_{\rm acc} \, v_{\rm inf}^2
\end{equation}
where $v_{\rm inf}$ is the infall velocity and 
\begin{equation}
\label{eq:Mdot_in}
\dot{M}_{\rm acc} = 4 \pi \, R^2 \, v_{\rm inf} \, \rho_{\rm ISM}
\end{equation}
is the mass accretion rate of diffuse matter of mass density
$\rho_{\rm ISM}$ onto a cloud of radius $R$.  Assuming a mean ISM volume number
density of $n_{\rm ISM}=1$\,cm$^{-3}$ and a typical molecular cloud
radius of $R=30$\,pc, and scaling to an infall (or converging flow) velocity 
$v_{\rm inf}=10$\,km\,s$^{-1}$ produces an energy injection rate of
$\dot{E}_{\rm inj}\approx32$\,L$_\odot$. We have chosen an inflow velocity corresponding to the sound speed of ionized gas, for reasons that will become apparent.

To see whether this process provides enough energy to maintain turbulence, it has to be compared with the turbulent energy dissipation rate. 
In the outer Galaxy, our results are compatible with a constant value of $\dot{E}_{\rm dis} \approx 2\,L_\odot$ (see Figure~\ref{fig:PDF_Ediss}). 
This is a factor of 15 lower than the input energy by converging flows of $10$\,km\,s$^{-1}$.
The energy conversion efficiency from converging flow to turbulent motion of dense clouds has been estimated to be in the range of $0.01-0.1$ \citep{klessen2010}. We conclude that the turbulent motions of non-star-forming molecular clouds in the outer disk of the Milky Way can be explained by the general process of cloud formation and mass accretion by converging flows of H\,{\sc i} gas.

Of course, there must be a source of free energy to drive the gas inflow onto the cloud. We note that in the outer Galaxy the virial parameter of the clouds is $\sim 10$, so that the potential energy of the GMC is small compared to the kinetic energy of the turbulence, i.e., the gravitational potential energy of the GMC is not a significant source of energy. This is not true in the inner Galaxy, a point we will return to. Since there is very little star formation in the outer Galaxy, we are left with two sources of free energy, accretion onto the disk from the circumgalactic or intergalactic medium, or accretion through the disk, whether driven by the magnetorotational instability or gravity-driven instability, e.g., via spiral arms (which are seen in the outer disk). The gas depletion time of the Milky way is of order $1$ Gyr, much shorter than the age of the Galaxy, suggesting that accretion either onto or through the disk, or both, is ongoing; assuming that the situation is at least near an equilibrium, we scale the accretion to the star formation rate, which is of order $1-2M_\odot\yr^{-1}$. The potential energy released per second by the accretion is roughly
\be
\dot E_{\rm acc}\approx \dot M_{\rm acc} v_c^2\approx2.9\times10^{40}
\left({\dot M_{\rm acc}\over 1M_\odot\yr^{-1}}\right)
\left({v_c\over 220\kms}\right)^2
\erg\s^{-1}.
\ee
This should be compared to the turbulent luminosity of the disk, roughly
\begin{eqnarray}
\dot E_{\rm dis, disk}&&={1\over2}M_g \sigma_v^3/H\approx 10^{39}
\left({M_g\over 2\times10^9M_\odot }\right)\nonumber\\
&&\left({\sigma_v\over 5\kms}\right)^3
\left({H\over 100\pc}\right)^{-1}
\erg\s^{-1}.
\end{eqnarray}
We conclude that accretion energy is sufficient to drive the turbulence in the disk and that accretion of gas onto GMCs in the outer disk is sufficient to drive the turbulence in individual clouds outside the solar radius.

The situation in the inner Galaxy is very different, however. The observed GMC turbulent energy dissipation rate in the inner Galaxy is a factor of 100 larger than that in the outer Galaxy. The mean density of the atomic/ionized ISM is similar in both locations, as are the mean GMC radii, and we expect that the typical infall velocities at large distances from the GMC are similar as well. Thus, it would appear that a new source of turbulent energy is called for.

The virial parameters of clouds inside the solar circle are noticeably smaller than those of clouds in the outer Galaxy; see the bottom panel of Figure \ref{fig:PDF_alpha}. In other words, the potential energy in inner-Galaxy GMCs is comparable to the kinetic energy, in contrast to the GMCs in the outer Galaxy. This suggests that contraction of these GMCs may power the higher levels of turbulence we see.

This does not explain {\em why} GMCs in the inner Galaxy are more tightly bound than those in the outer Galaxy.

More significantly, the fact that the GMCs have low virial parameters and are contracting raises the question, what halts the contraction? It is clearly not the turbulence.

\subsubsection{Stellar Feedback}

Turbulence in molecular clouds that are not forming stars is comparable to that in more actively star-forming clouds. \citet{williams1994} noticed that two molecular clouds, one actively forming stars while the other is not, have very similar size--line width relations and clump mass spectra. This is also found in the LMC by \citet{kawamura2009}. In addition, statistical studies of the velocity field of individual molecular clouds indicate no characteristic scale. Both points support the idea that the turbulent energy injection occurs at scales larger than the size of clouds \citep{ossenkopf2002,brunt2009}, and in particular that feedback does not strongly and directly affect the turbulence at scales below the GMC scale.

To take a specific example, jets from protostars can drive turbulence in molecular gas; \citet{matzner2007} estimates the total impulse produced per stellar mass formed of $v_*\approx40\kms$. Using this value, one finds that $\sim10^{4}M_\odot$ in young (jet-emitting) stellar objects are needed to power the turbulence we find in an inner-Galaxy GMC. In Lee et al. (2016) we show that the only about 50 of the $\sim 500$ GMCs with $M>10^6M_\odot$ have star clusters this massive, assuming a normal initial mass function. In other words, neither the energetics nor (the lack of) a feature in the turbulent spectrum on the scale of jets is compatible with protostellar jets being the source of turbulence in inner Galaxy GMCs.

{This does not mean that other forms of stellar feedback have no role in the generation of turbulence. For example, O star winds, H\,{\sc ii} regions, and radiation pressure produce momentum outputs on much larger scales than do protostellar jets, as do supernovae. Estimates for the first three forms of feedback give $v_*\approx 300\kms$, while supernovae provide $v_*\approx 3000\kms$ \citep[e.g.,][]{ostriker2011}. Thus, the problem with the energetics of jet-driven turbulence might not apply with these other forms of feedback. But while each form of feedback may produce large-scale turbulence, they all act initially on small scales, so that we might expect to see many GMCs with excess turbulence on small scales; such features are not common in any data set we are aware of. Thus, it appears unlikely that stellar feedback is directly responsible for the  high GMC dissipation rates we find in the inner Galaxy.}

However, the normalization of the line width, $\sigma_0 = \sigma_v / R^{1/2}$, increases by a factor of two from the outer Galaxy to the inner Galaxy (Figure~\ref{fig:sigmav_norm}). The variation in the GMC turbulent dissipation rate is far more dramatic; it is a factor of 100 larger in the inner Galaxy, where star formation is active (Figure~\ref{fig:PDF_Ediss}). This implies that the turbulent energy injection rate is also 100 times higher than the injection rate per cloud in the outer Galaxy.

{This increase can be explained by contraction of inner-Galaxy GMCs \citep[e.g.,][]{ballesteros-paredes2011a,ibanez-meja2016}; this is the solution we advocate here. However, contraction of the GMC cannot be the entire solution, since continued contraction would lead to very dense massive GMCs, which do not exist. Something must halt, or rather completely reverse, the contraction.}

{We suggest that some combination of the four feedback processes described above reverses the contraction of GMCs in the inner Galaxy. This reversal must occur over about a GMC dynamical time, since we do not see many GMCs that are clearly in the act of dispersing. This input of turbulent energy on the GMC size scale would necessarily have to equal, or more likely exceed, the amount of turbulent energy dissipated during the earlier collapse.}

{In this scenario, the contraction of GMCs is the proximate or direct driver of the large turbulent dissipation rates we see, but the ultimate source of the turbulent energy is the burning of nuclear fuel in massive stars. Some fraction of the nuclear energy released unbinds the GMC hosting the massive stars, restoring the gravitation potential energy lost in the contraction, and more. This potential energy is then available to power the turbulence in the next generation of GMCs.}

If the increase of turbulent energy injection were due to stellar feedback, we could expect to see an associated increase of the velocity dispersion of the gas, as indeed we do. We note, however, that it is likely that only a fraction of the energy released by stellar feedback goes into random motions. For example, shocks created by supernovae would inject energy more efficiently into compressive than solenoidal modes; the interstellar gas gets more compressed than agitated. Indeed, like the turbulent energy dissipation rate, the internal GMC {\em pressure} in the inner Galaxy is about 2 orders of magnitude larger than in the outer Galaxy.

\section{Conclusion}

In this study we have presented a new catalog of molecular clouds obtained from the segmentation of the $^{12}$CO data of \citet{dame2001}. The cloud identification presented here is based on a hierarchical cluster identification applied to the result of a Gaussian decomposition of each $^{12}$CO spectrum. This new method allowed us to go significantly deeper in brightness. The catalog contains 8107 clouds that include 98\% of the CO emission, compared to 20-25\% for previous studies.
This allowed us to present physical properties of Galactic molecular clouds over the whole Milky Way disk. The main conclusions of the paper are the following.  

\begin{enumerate}
\item The total H$_2$ mass in the catalog is $1.2\times 10^9$\,M$_\odot$, in accordance with previous estimates \citep{heyer2015}.
\item The variation of the H$_2$ surface mass density with $R_{\rm gal}$ is compatible with previous work (but incompatible with \citet{wouterloot1990} and therefore with the scale length of 3.5\,kpc deduced by \citet{williams1997}. The H$_2$ scale length we obtain is 2.0\,kpc.
\item Because the identification method is able to include emission down to the sensitivity limit, a population of low-$\Sigma$ clouds in the outer Galaxy is revealed. We also report a significant cloud-to-cloud variation of $\Sigma$ (from 2 to $300\,M_\odot$\,pc$^{-2}$), in contrast with the general idea that molecular clouds have a constant mass surface density.

\item The median values of the mass and radius are $M=3.8\times 10^4\,M_\odot$ and $R=25$\,pc. These values are small because of the predominance of small clouds in the catalog, but they are actually typical of nearby molecular clouds like Perseus \citep{lee2015} or Taurus \citep{pineda2010}.
  
  Another way to look at typical values is to estimate the cloud mass for which half of the mass in the catalog is in more more (or less) massive clouds: half of the mass of the catalog is in clouds more (or less) massive than $M=8.4\times 10^5$\,M$_\odot$. Clouds with this mass have a typical size of $R=60$\,pc.
  
The inner Galaxy hosts the most massive clouds of the sample: both the average values of $M$ and $\Sigma$ decrease with $R_{\rm gal}$. On the other hand, we note that the median cloud size does not vary significantly from the inner to the outer Galaxy. 
  
\item The typical volume density is $n_{\rm H2} \sim 10$\,cm$^{-3}$, well below the CO critical density. This suggests that molecular clouds are in fact multiphase objects where dense structures occupy a small fraction of the volume.

\item We did not find a tight correlation between line width and size (Pearson correlation coefficient of 0.55). The tightest correlation is found to be $\sigma_v = 0.23\, (\Sigma R)^{0.43\pm 0.14}$, in agreement with \citet{heyer2009}. This is indicative that gravitational energy is playing a role in driving the turbulent motions of molecular clouds. 

\item On average, molecular clouds are on the verge of being in virial equilibrium, with $\alpha_{\rm vir} \sim 3-4$.  There is a very large cloud-to-cloud variation of $\alpha_{\rm vir}$. Only 15\% of the clouds are gravitationnaly bound ($\alpha_{\rm vir} \leq 3$), but they represent 40\% of the molecular mass.
  
  We do not observe a strong variation of $\alpha_{\rm vir}$ with $R_{\rm gal}$. There is a slight decrease of $\alpha_{\rm vir}$ in the molecular ring and an increase toward the Galactic center.

  \item The velocity dispersion normalized at 1\,pc ($\sigma_0 = \sigma_v / (R/1\,{\rm pc})^{1/2}$) varies by only a factor of two (from 0.8 to 0.4\,km\,s$^{-1}$) over $0 < R_{\rm gal} < 20$\,kpc. Over the same range in $R_{\rm gal}$, the turbulent energy dissipation (injection) and the internal pressure vary by almost 2 orders of magnitudes. Both $P_{\rm int}$ and $\dot{E}_{\rm dis}$ increase strongly for $R_{\rm gal} < 6-8$\,kpc. This increase is also seen in the average cloud surface density ($\Sigma$) and in the fraction of molecular gas ($f_{\rm H2}$). 

\end{enumerate}

From the results obtained over the whole Milky Way disk it is possible to draw some conclusions about the nature of molecular clouds. The small number of clouds in virial equilibrium is in favor of the scenario in which molecular clouds are dynamical and short-lived structures. In this scenario the formation of molecular clouds is related to the compression of the diffuse WNM/WIM gas, by supernova shocks, by transonic converging streams of gas, or by gravitational instability, which triggers the formation of cold, dense, and fragmented gas via the thermal instability \citep{field1965,heitsch2006b,saury2014}. 

In the outer Galaxy, the properties of the large number of small molecular clouds can be explained by the accretion of matter in large-scale converging flows \citep{klessen2010}. 
In the inner Galaxy, the turbulent energy injection rate and the internal pressure increase by a factor of almost 100, but with a much smaller increase in the velocity dispersion of the gas. This indicates that most of the kinetic energy goes into compressive modes as opposed to random (solenoidal) motions. Gravitational collapse is a natural source of compressive modes. We cannot rule out a contribution from stellar feedback, as shocks from winds, radiation pressure, and supernovae will also inject more compressive than solenoidal modes into the ISM. 

{Nevertheless, we conclude that it is more likely that gas accretion and hierarchical gravitational collapse are driving the observed nonthermal line widths of molecular clouds. We base this conclusion on the fact that most of the massive clouds we observe have little or no sign of massive young stars, and that those clouds that do contain substantial numbers of O stars do not show larger turbulent velocities or dissipation rates than clouds lacking O stars. }

{Because of the nature of the formation process (the combination of turbulence and thermal instability produces very fragmented small-scale structures), the free-fall time of small-scale structures is much shorter than the free-fall time of the global molecular cloud. In this scenario, stars form along with the cloud. Stellar feedback is then responsible for cloud disruption and the low star formation efficiency and for restoring the energy dissipated in the collapse of the previous generation of molecular clouds, at least in the inner Galaxy.}


\acknowledgements
We thank the anonymous referee for very constructive feedback and comments that helped to improve the paper significanty. 
We thank Rosine Lallement for providing us with the extinction map of the solar neighborhood.
This work was supported by the program Physique et Chimie du Milieu Interstellaire (PCMI) funded by the Conseil National de la Recherche Scientifique (CNRS) and the Centre National d'Etudes Spatiales (CNES) of France.
This research was undertaken, in part, thanks to funding from the Canada Research Chairs program. N.W.M. was supported in part by the Natural Sciences and Engineering Council of Canada.

\begin{appendix}

  \section{Gaussian fitting}

  \label{sec:gaussfit}
  
\begin{figure*}
\centering
\includegraphics[draft=false, angle=0]{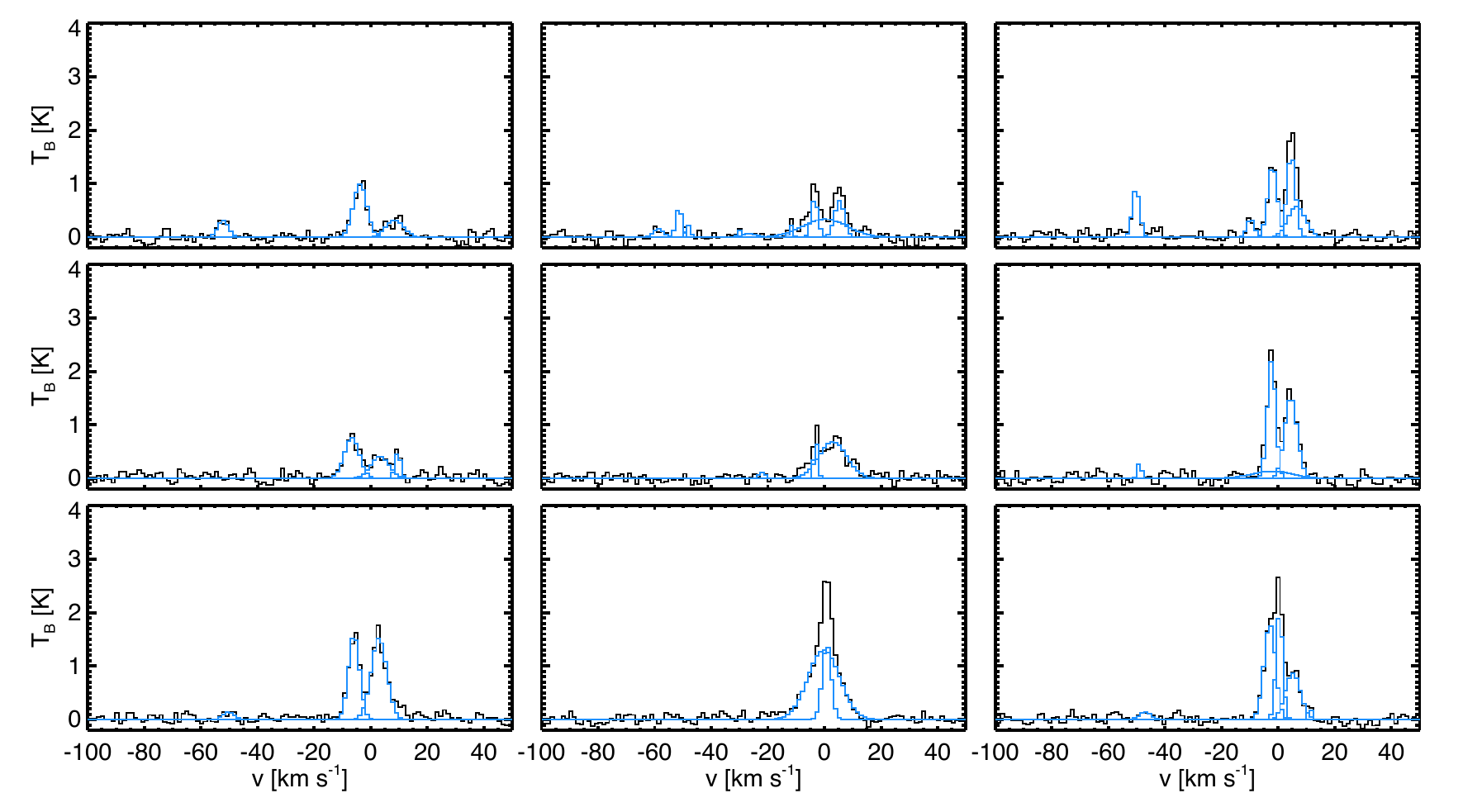}
\includegraphics[draft=false, angle=0]{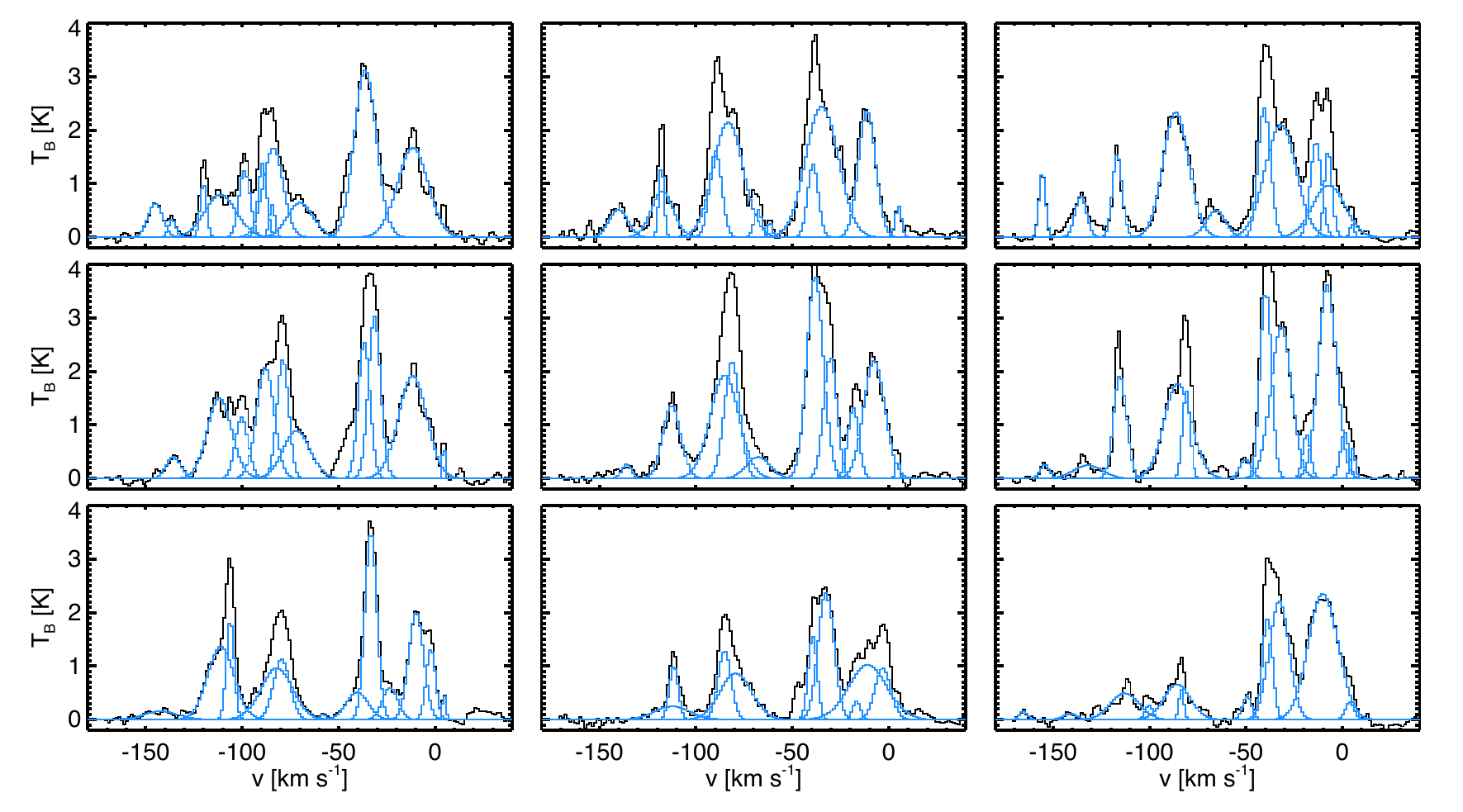}
\caption{\label{fig:example_spectra} Example of the Gaussian decomposition for two different positions on the sky. The top and bottom panels show $^{12}$CO spectra (black) and their Gaussian decomposition (blue) for nine neighboring positions, spaced by 7.5' in $l$ and $b$. {Top:} spectra centered on $l=105^\circ$, $b=0^\circ$. This position is typical of the spectra found in the outer Galaxy. {Bottom:} spectra centered on $l=14^\circ$, $b=0^\circ$. This position shows some of the most complex CO spectra of the data set. }
\end{figure*}

The cloud idenfication presented in this study relies on a Gaussian decomposition of the $^{12}$CO data. 
Given the complexity of the velocity field of molecular clouds and the fact that turbulent motions are likely to be suprathermal, the justification of decomposing $^{12}$CO spectra into a sum of Gaussian components is not obvious. 
We insist on the fact that this decomposition does not mean that there is a direct link between a Gaussian component and a cloud on the line of sight. Clouds can be composed of more than one Gaussian component on a given line of sight producing spectra with an arbitrary shape. Moreover, the fact that we do not associate a single Gaussian with a cloud lifts the problem of the unicity of solution. 

The CO spectra can be very complex (see examples in Figure~\ref{fig:example_spectra}) and it is clear that several solutions can provide similar goodness of fit. 
The unicity of the solution is not critical here because individual Gaussians are not analyzed. Clouds are identified as islands in the integrated emission of the Gaussian components ($W_{\rm CO}$; see Equation~\ref{eq:wco_cube}). What is important is that the fit is a good representation of the data, that variations of the shape of the spectrum are well described by the sum of Gaussians. As mentioned in Section~\ref{sec:clouds}, the goal of the Gaussian decomposition is to project the data on a space with fewer parameters, where noise has less impact and where there is less confusion on the velocity axis. 

\begin{figure}
\centering
\includegraphics[draft=false, angle=0]{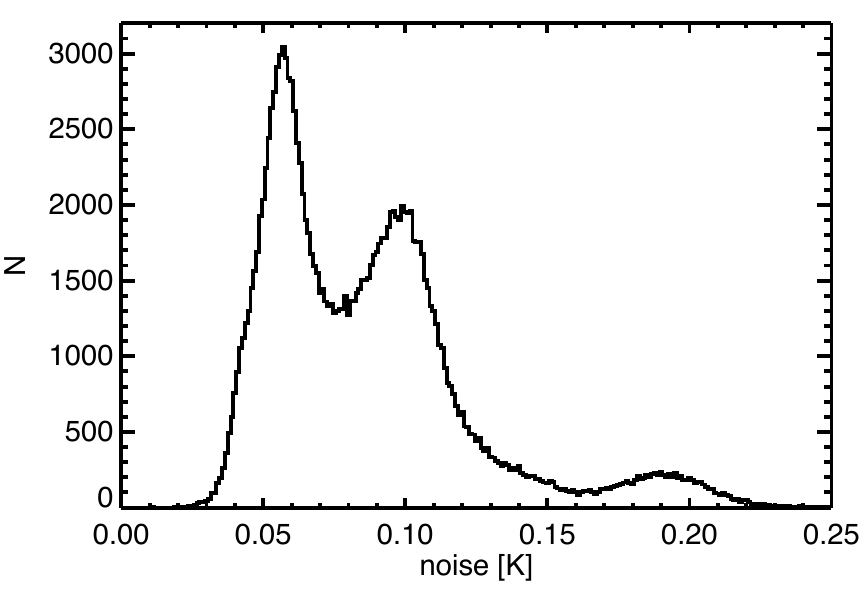}
\caption{\label{fig:PDF_noise} Histogram of the noise value for the whole data set. The noise of a given spectrum was estimated as the width of the Gaussian fit of the histogram of $T_{\rm B}(v)- T_{\rm B}^\prime(v)$.}
\end{figure}

In order to facilitate the identification of clouds in PPV space, we use a method that promotes spatial coherence of the recovered parameters. The Gaussian fit is done using the following procedure. 
  There is a global iteration where all the spectra of the data cube are decomposed into a sum of Gaussian. This is performed several times until the $\chi^2_{\rm red}$ of all pixels is below a given value close to $1.0$. 
The reduced $\chi^2_{\rm red}$ is defined as
\begin{equation}
  \chi^2_{\rm red} = \frac{\sum_v \left[T_{\rm B}(v) - T_{\rm B}^\prime(v)\right]^2}{ \sigma_{\rm noise}^2 N_{\rm free} }
  \end{equation}
where $\sigma_{\rm noise}$ is the noise level of the spectrum and $N_{\rm free}$ is the number of degrees of freedom, defined as
\begin{equation}
N_{\rm free} = N_{\rm sample} - 3N_{\rm comp}
\end{equation}
where $N_{\rm sample}$ is the number of samples in the spectrum and $N_{\rm comp}$ the number of Gaussian components. The noise level is not uniform across the whole data set (Figure~\ref{fig:PDF_noise}) due to the combination of different observation campaigns.

At each iteration, the fit of a given spectrum is done using an initial guess built with the results of neighboring pixels obtained in the previous iteration. This is done in order to promote spatial coherence of the parameters in the perspective of identifying connected structures in PPV space.  
In practice, for the first iteration, the Gaussian component fit of each spectrum is done without an initial guess. Then, for each subsequent iteration, the initial guess used to fit a single spectrum is computed by looking at solutions in the neighboring positions, obtained in the previous iteration.  We look for solutions in a $9 \times 9$\,position grid centered on the spectrum. Only positions that have a reduced $\chi^2_{\rm red}$ below some threshold are used. Components with $\sigma_i$ larger than 10\,channels are also discarded. This is needed to avoid solutions with wide Gaussians on spectra where narrow components seen in neighborhood pixels blend together. For this subset of components, frequent Gaussian components are identified by looking for groups in the $v_i-\sigma_i$ space. Only groups with at least four components are kept. For each group of Gaussian components, we estimate the average values of $[A_i, v_i, \sigma_i]$ as well as their standard deviation. 
These parameters are used as the initial guess and their standard deviations to limit the space each parameter can vary on (each component can always have an amplitude of zero), something made possible with the IDL {\em mpfit} code \citep{markwardt2009}.  
If the result of the fit has a $\chi^2_{\rm red}$ that is not good enough, an extra component with loose constraints is added. Extra components are added until $\chi^2_{\rm red}$ is better than a threshold value or if the number of components is equal to 12. 

The advantage of this procedure is that it looks for a solution with the minimum number of narrow Gaussian components. In addition, the fact that we look for spatially coherent solutions increases the probability that the shape variations described by the Gaussian basis are real and not due to noise.

\section{Parameterization of the cluster identification method}

\label{sec:param-cluster}

Due to the way we resolve the near/far distance ambiguity (using the $\sigma_v-(\Sigma R)$ relation), the position of clouds in the inner Galaxy depends significantly on their angular size. Therefore, the cluster identification method, which sets the angular size of clouds, does have an impact on how clouds are distributed in the Galactic plane, as well as on the total mass and radial distribution of H$_2$ surface density.

The hierarchical method used here is based on a threshold descent that attaches together nearby Gaussian components. At each step of the descent there is also a merging procedure that attaches small clusters to bigger ones when they are obviously located within them in ($l,b,v$) space. We imposed a limit on this process to avoid a runaway association that would lead to the creation of a few unrealistically large clusters. Once a cluster reaches a size of 50 Gaussian components, we assumed that it is a single structure and that it cannot be merged with another big one (with more than 50 Gaussian components). This parameterization of the method is justified by the fact that, when it is introduced, the catalog reproduces important observational quantities deduced from many previous studies that do not suffer from the near/far distance ambiguity. 
In particular, with this parameterization the catalog reproduces better the total molecular mass in the Milky Way and the radial profile of the surface density.

\begin{figure}
\centering
\includegraphics[width=\linewidth, draft=false, angle=0]{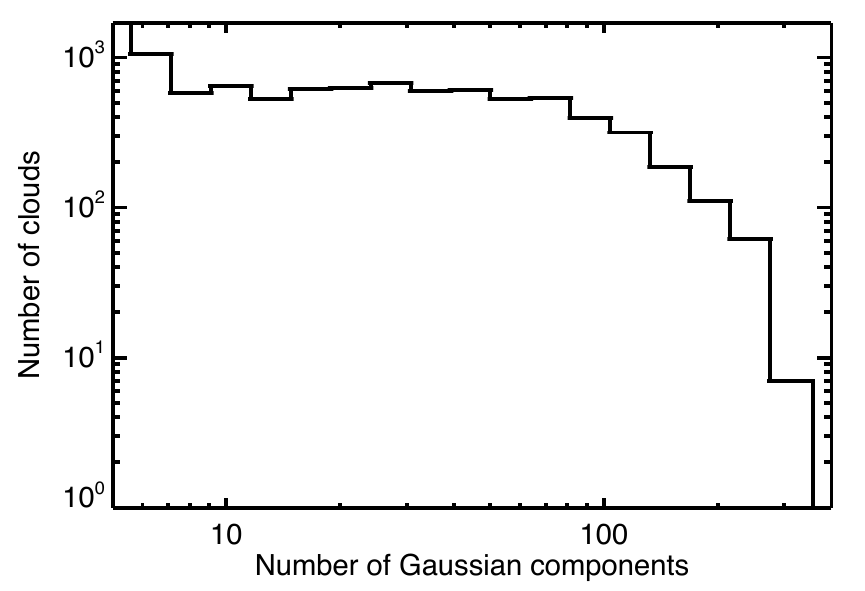}
\caption{\label{fig:nbGauss_in_clouds} Histogram of the number of Gaussian components grouped together to form a cloud. }
\end{figure}

Figure~\ref{fig:nbGauss_in_clouds} shows the histogram of the number of Gaussian components per cloud, with bins equally spaced in log. The distribution ranges from five Gaussian components (the minimum limit to be a cluster) up to more than 300. The parameterization of the merging procedure did not limit the growth of clusters beyond 50 Gaussian components. In fact, there is no specific feature at that value; the distribution is almost flat up to about 80 Gaussian components, where it starts to fall off.

\section{Distance bias}

\label{sec:distance-bias}

\begin{figure}
\centering
\includegraphics[draft=false, angle=0]{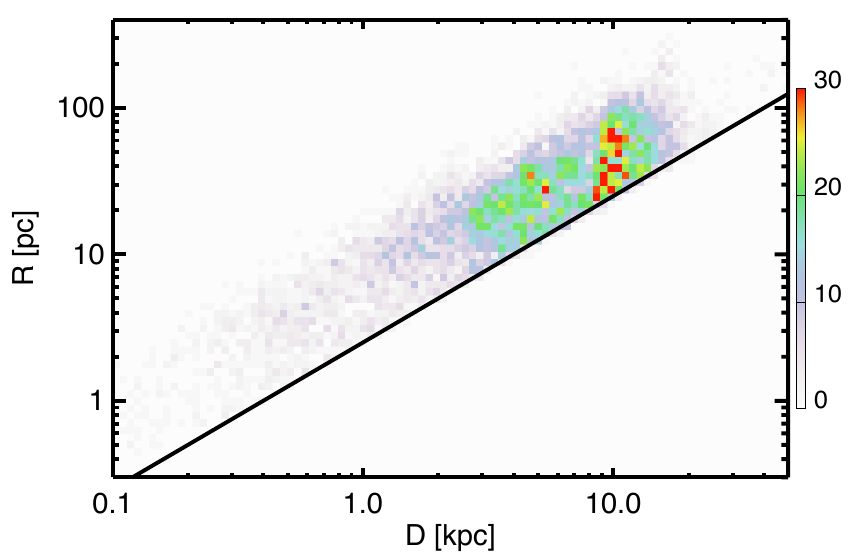}
\includegraphics[draft=false, angle=0]{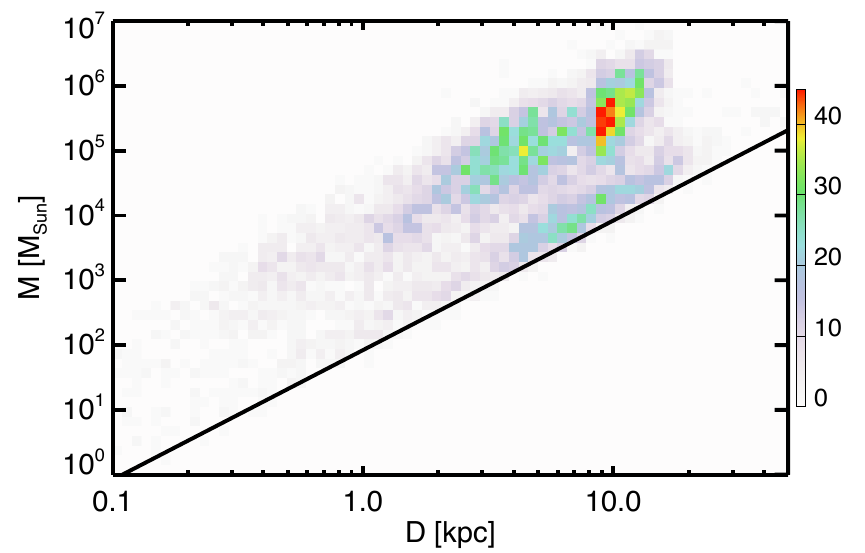}
\caption{\label{fig:mass_vs_distance} Density plot of $R$ vs. $D$ (top) and $M$ vs. $D$ (bottom). The color scale is proportional to the number of clouds. In both plots the black line indicates the minimum radius or mass the catalog is sensitive to at a given distance.}
\end{figure}

Even though the catalog contains almost all the observed CO emission of \citet{dame2001},
the finite angular resolution, finite velocity resolution, and finite sensitivity of the data translate into specific limitations in the parameters of clouds, especially as a function of distance. 
For instance, the minimum radius that can be estimated at a given distance is
\begin{equation}
\label{eq:R_vs_D}
R_{\rm min} = D \, \tan\left(\sqrt{\frac{d\Omega N_{\rm min}}{\pi}}\right)
\end{equation}
where $N_{\rm min}$ is the minimum number of pixels in a cloud. In our case $N_{\rm min} = 5$.

Similarly, the minimum mass is
\begin{equation}
\label{eq:M_vs_D}
M_{\rm min} = D^2 \, \Sigma_{\rm min} \, N_{\rm min} \, d\Omega
\end{equation}
where $\Sigma_{\rm min}$ corresponds to the minimum surface density that can be detected. This value ($3.5\,M_\odot$\,pc$^{-2}$) is linked to the minimum threshold of the cluster identification method: $W_{\rm CO}^{\rm min} = 0.8$\,K\,km\,s$^{-1}$. 

The distribution of $R$ and $M$ of the clouds as a function of their distance is shown in Figure~\ref{fig:mass_vs_distance}. The proportionality with $D$ and $D^2$ for $R$ and $M$, respectively, is clearly visible. In these two plots, the solid lines represent the minimum values given in Equations~\ref{eq:R_vs_D} and \ref{eq:M_vs_D}.

The finite velocity resolution has a lesser impact on the cloud parameters. The observed line width is indeed limited by the velocity resolution of the observations (1.3\,km\s$^{-1}$), but the true $\sigma_v$ of an emission line can be recovered by subtracting quadratically the instrumental broadening to the observed width. With high signal-to-noise ratio, one can estimate values of $\sigma_v$ that are much smaller than the velocity resolution. In other words, the line width of an emission line of arbitrary shape can be deconvolved from the effect of instrumental broadening.

\begin{figure}
\centering
\includegraphics[draft=false, angle=0]{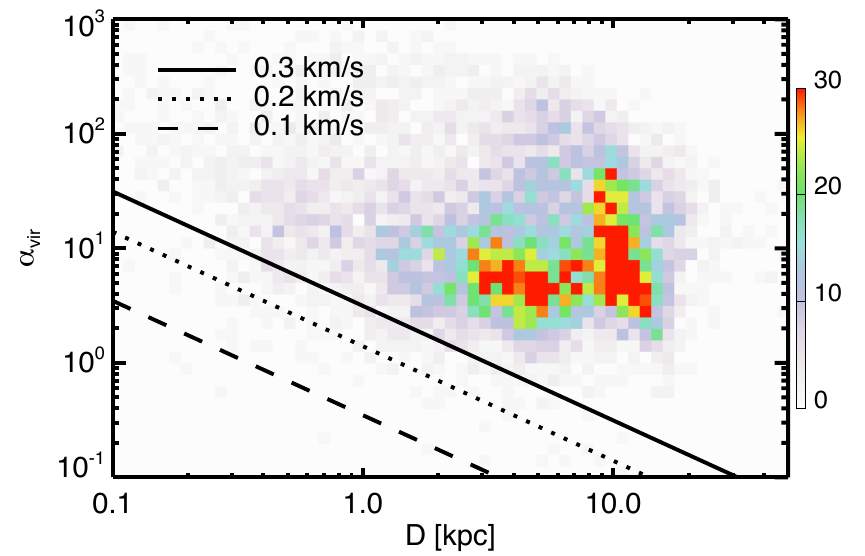}
\caption{\label{fig:alpha_vs_distance} Density plot of $\alpha_{\rm vir}$ vs. $D$. The color scale is proportional to the number of clouds. The three lines indicate $\alpha_{\rm vir,min}$ as a function of $D$ for different values of $\sigma_{v\rm,min}$.}
\end{figure}

Nevertheless, the line width cannot be estimated with infinite precision, and, as mentioned by \citet{heyer2001}, this sets a minimum value for the virial parameter:
\begin{equation}
\alpha_{\rm vir,min} = \frac{5 \sigma_{v\rm,min} R_{\rm min}}{GM_{\rm min}},
\end{equation}
where $R_{\rm min}$ and $M_{\rm min}$ are given by Equations~\ref{eq:R_vs_D} and \ref{eq:M_vs_D}
and $\sigma_{v\rm,min}$ is the minimum line width that can be measured. Through $R_{\rm min}$ and $M_{\rm min}$, $\alpha_{\rm vir,min}$ also depends on distance. This is illustrated in Figure~\ref{fig:alpha_vs_distance}, which shows the 2D histograms of $\alpha_{\rm vir}$ as a function of distance. In this plot, the three lines show $\alpha_{\rm vir,min}$ as a function of $D$ for $\sigma_{v\rm,min}=0.1$, $0.2$, and $0.3$\,km\,s$^{-1}$. There are almost no clouds below the line defined for $\sigma_{v\rm,min}=0.2$\,km\,s$^{-1}$, indicating that it is probably the effective velocity resolution of the catalog.
This plot also indicates that the sensitivity and resolution (spatial and velocity) of the \citet{dame2001} data preclude the identification of gravitationally bound clouds that are nearby ($D < 500$\,pc).

\section{The Virial Parameter}

\label{sec:virial}

\citet{bertoldi1992} introduced the dimensionless virial parameter for a clump or cloud of gas
\be \label{eqn:virial}
\alpha_{\rm vir} \equiv{5\sigma_v^2 R_m\over GM}=a {2{\cal T}\over |W|}.
\ee
The quantity $R_m$ is the average projected cloud size, $M$ is the cloud mass, and $\sigma_v^2$ is the line-of-sight (or 1D) velocity dispersion. The kinetic energy in the cloud is
\be
{\cal T}\equiv {3\over 2}\int _{V_{\rm cl}} \rho \sigma_v^2 dV,
\ee
while the gravitational energy associated with the clump is
\be
W\equiv{1\over 2}\int_{V_{\rm cl}} \rho({\bf x})\Phi({\bf x})d^3x.
\ee
\\
For a triaxial ellipsoidal cloud, they show that
\be
W=-{3\over 5}a_1a_2{GM^2\over R_m}.
\ee
The constant $a_1$ accounts for the effects of nonuniform density, while $a_2$ accounts for the clouds' ellipticity. With the assumption that $\rho(u)\sim u^{-k_\rho}$, they show that
\be
a_1={(1-k_\rho/3)\over (1-2k_\rho/t)},
\ee
while
\be
a_2={R_m\over R}{{\rm arcsinh}(y^2-1)^{1/2}\over (y^2-1)^{1/2}},
\ee
with $y$ being the axis ratio in the case that the cloud is an ellipsoid of revolution.

The line widths in our larger clouds are dominated by nonthermal motions; \citet{larson1981} showed that in individual clouds, $\sigma_v(r)\sim r^p$; a quick survey in our clouds suggests a similar scaling. Estimates for the power-law index are in the range $1/3<p<1/2$, as we noted above. The scaling of line width with size implies that the kinetic energy in the cloud will be modified from the usually assumed case $p=0$. We find
\be
{\cal T}=5 a_3 {\sigma_v^2(R)R\over GM},
\ee
with
\be
a_3={(3-k_\rho)\over (3-k_\rho+2p)}.
\ee

For the fiducial value $k_\rho=1$, \citet{bertoldi1992} find $a_1=10/9$ and $a_2\approx1$. For the same value of $k_\rho$ and $p=1/2$, we find $a_3=2/3$.

From the definition (\ref{eqn:virial}),
\be
\alpha_{\rm vir}={5\sigma_v^2 R_m\over GM}={a_1a_2\over a_3}{2{\cal T}\over |W|}
\approx {5\over 3}{2{\cal T}\over |W|}.
\ee
It follows that the average value of the observed virial parameter is a factor of $5/3$ larger than the ratio $2{\cal T}/|W|$. When the kinetic and gravitational energies are equal, $\alpha_{\rm vir}=10/3\approx 3.3$.

\section{Large-scale structure of the physical parameters}

\label{sec:XYviews}

\begin{figure*}
\centering
\includegraphics[draft=false, angle=0, width=11cm]{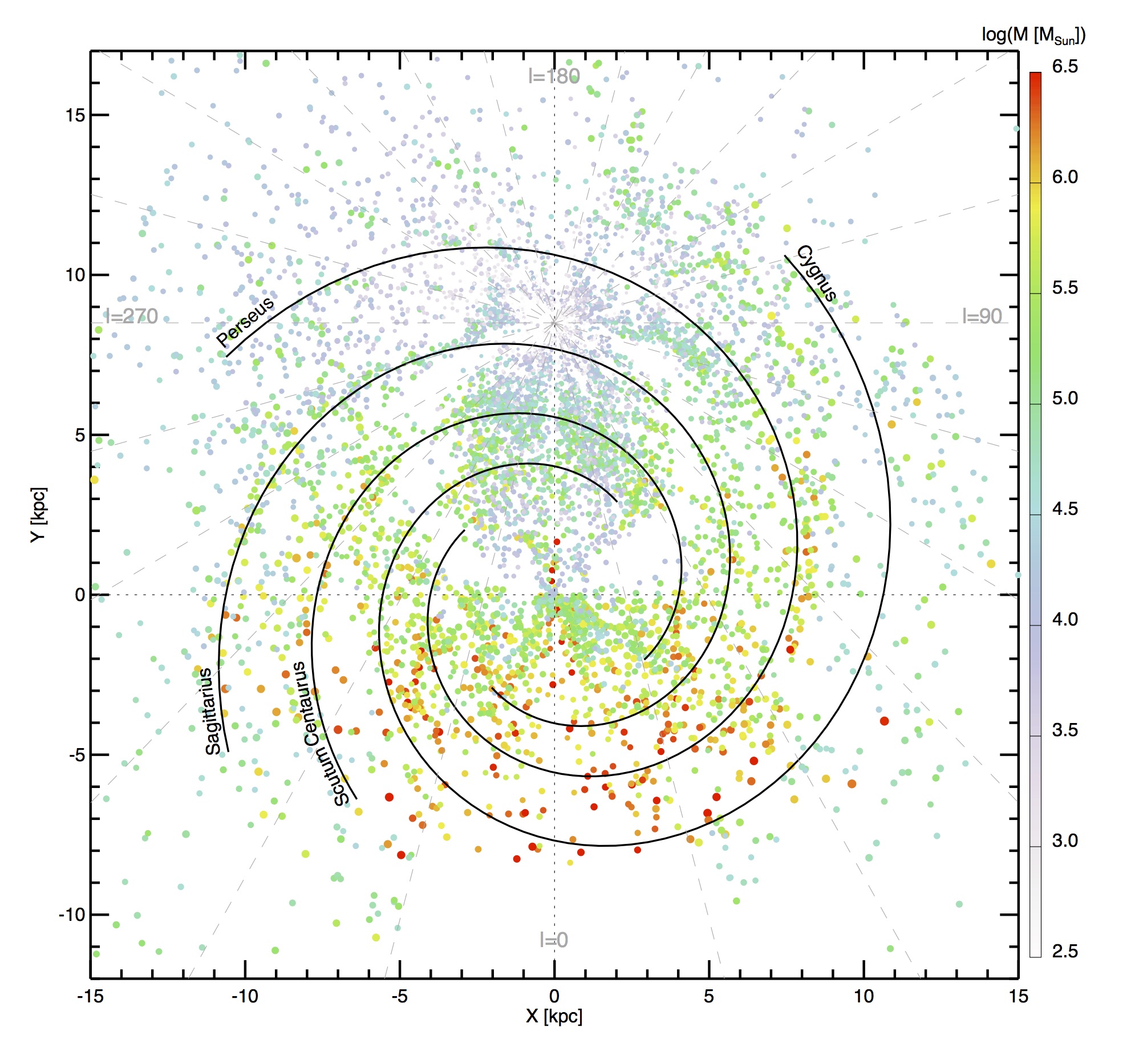}
\caption{\label{fig:xy_mass} Face-on view of the H$_2$ mass.
  The symbol size is proportional to $\log(R)$, while the color indicates $\log(M)$.}
\end{figure*}

\begin{figure*}
\centering
\includegraphics[draft=false, angle=0, width=11cm]{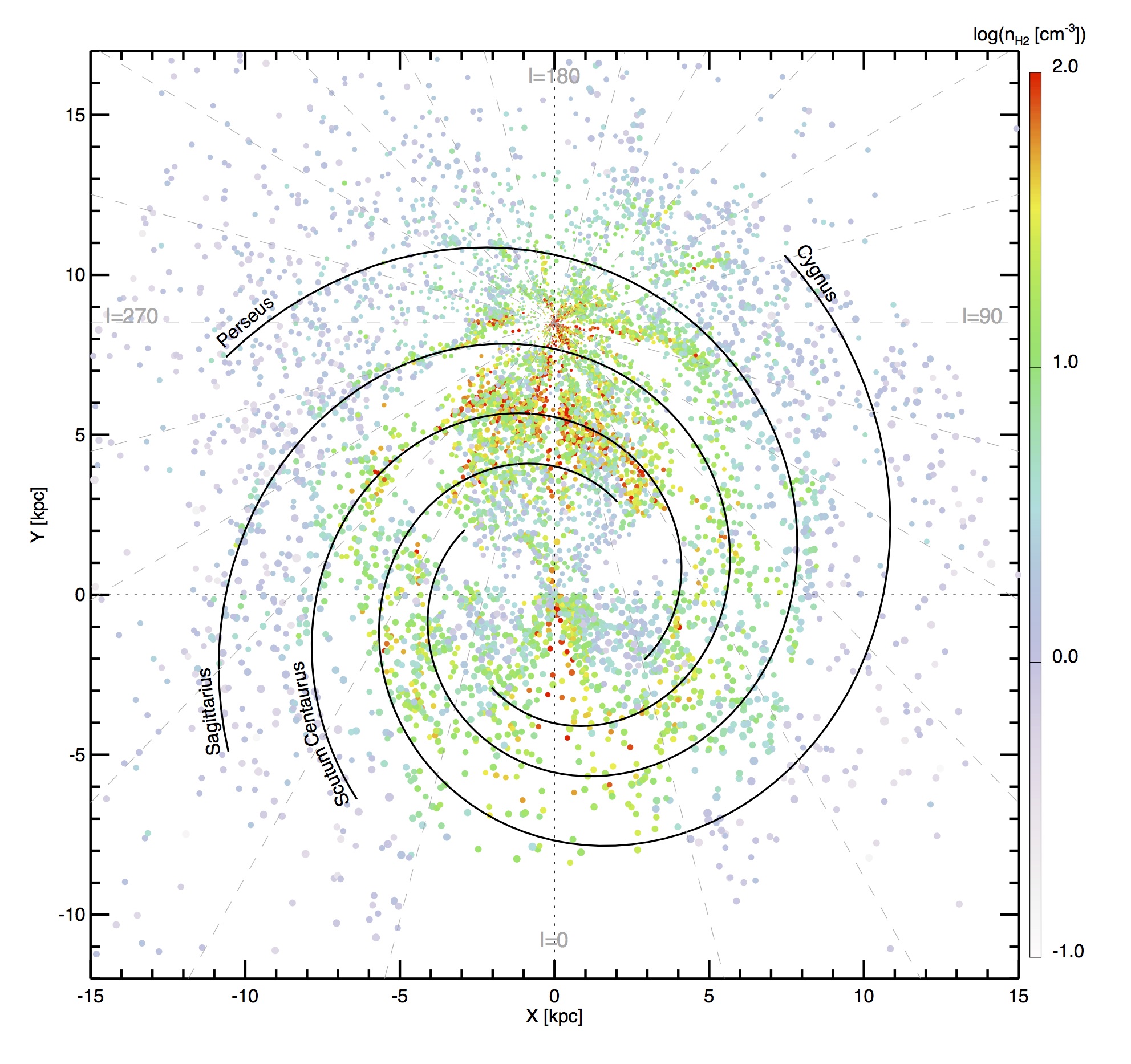}
\caption{\label{fig:xy_density} Face-on view of the H$_2$ density.
  The symbol size is proportional to $\log(R)$, while the color indicates $\log(n_{\rm H2})$.}
\end{figure*}

\begin{figure*}
\centering
\includegraphics[draft=false, angle=0, width=11cm]{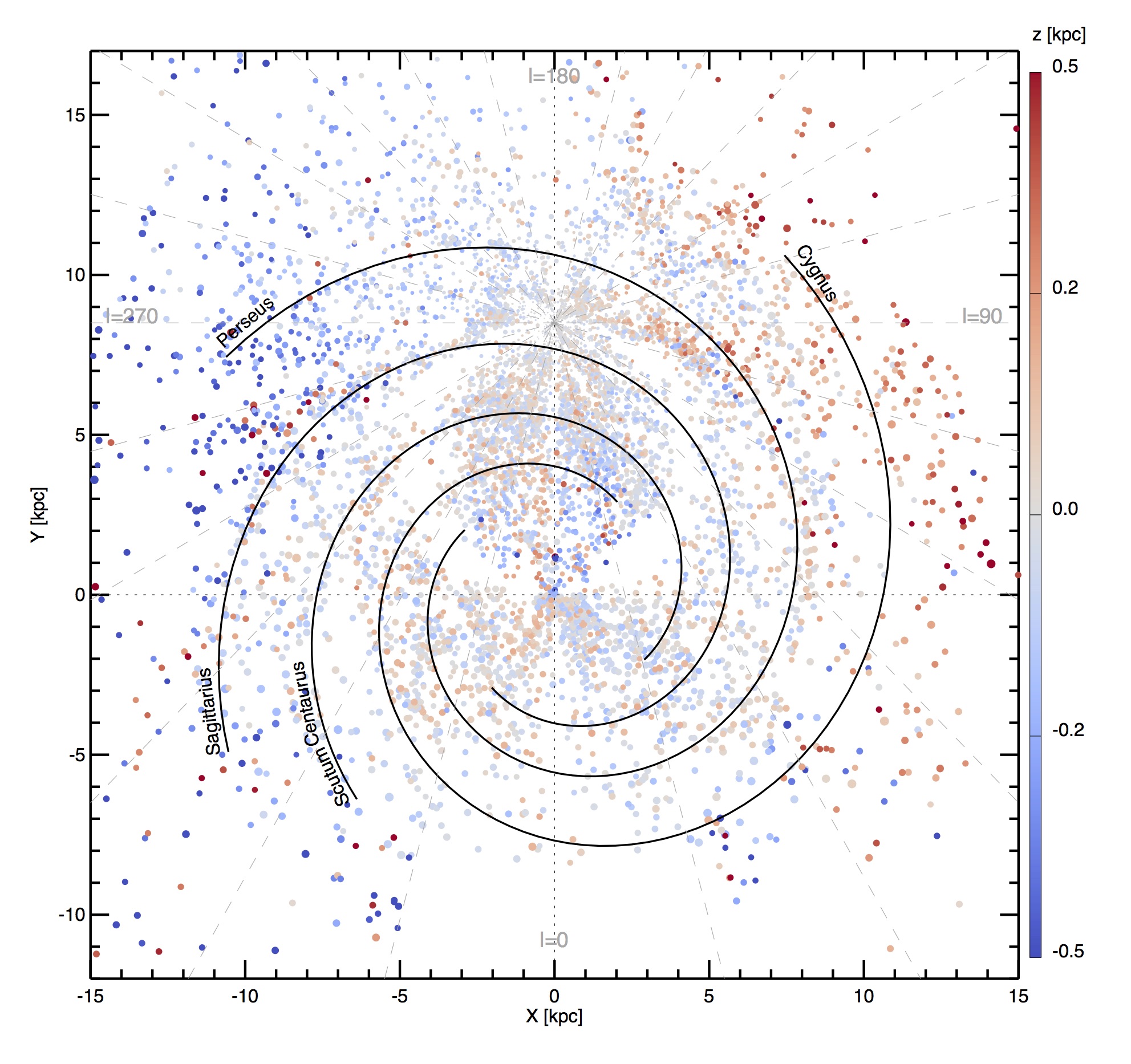}
\caption{\label{fig:xy_height} Face-on view of the $z$ position.
    The symbol size is proportional to $\log(R)$, while the color indicates $z$.}
\end{figure*}

\begin{figure*}
\centering
\includegraphics[draft=false, angle=0, width=11cm]{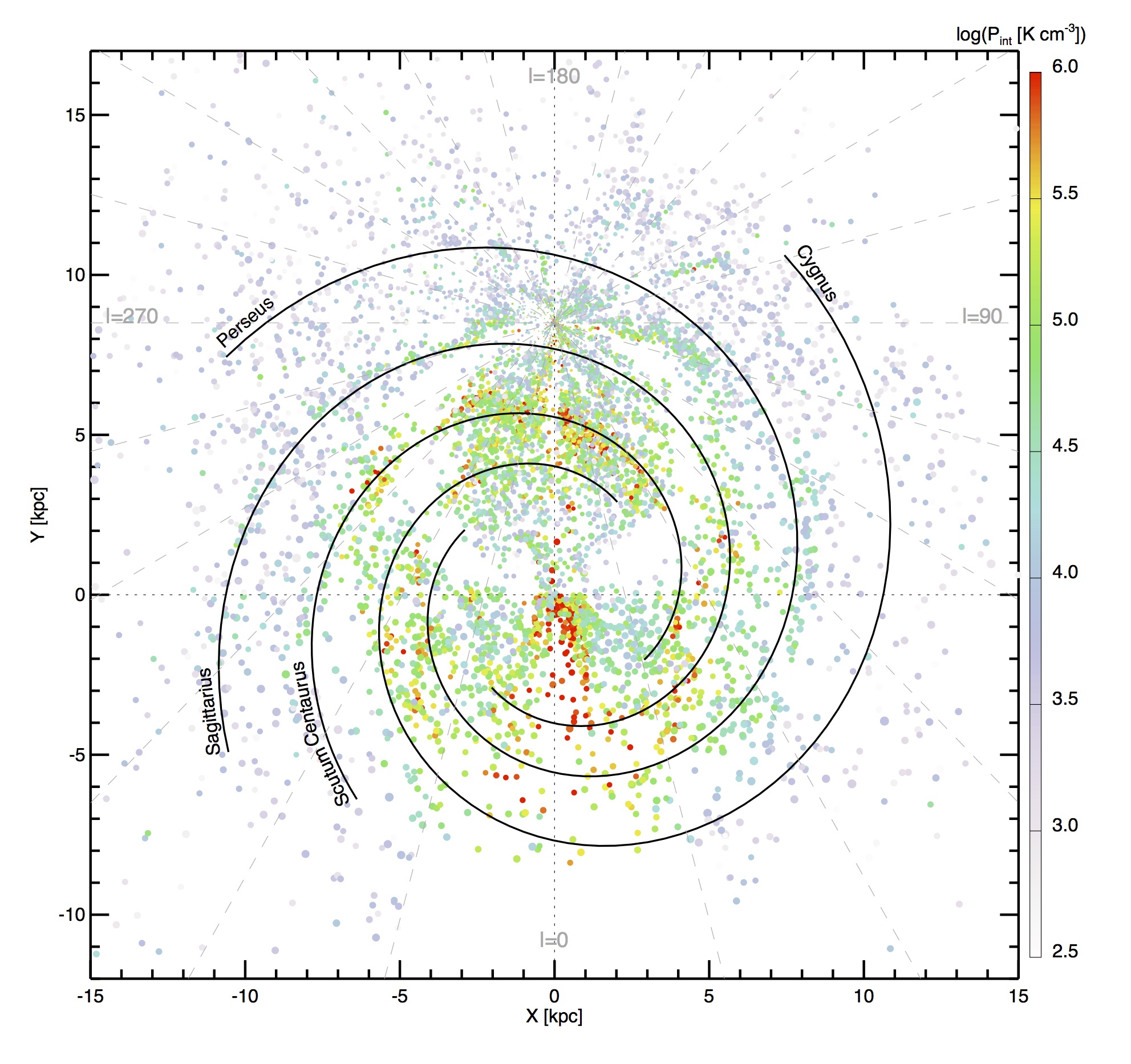}
\caption{\label{fig:xy_pressure} Face-on view of the pressure.
    The symbol size is proportional to $\log(R)$, while the color 
    indicates $P_{\rm int}$.}
\end{figure*}

\begin{figure*}
\centering
\includegraphics[draft=false, angle=0, width=11cm]{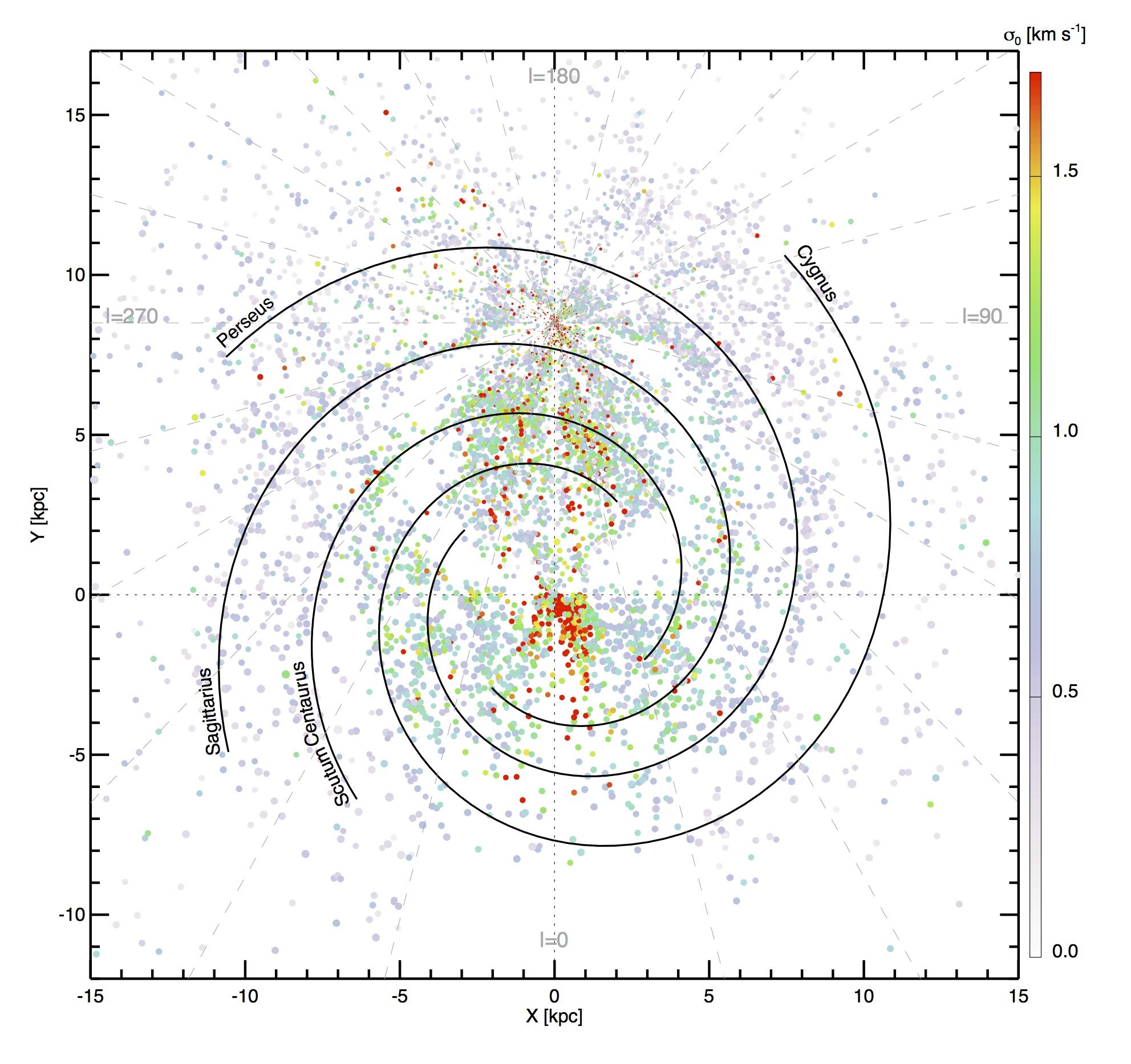}
\caption{\label{fig:xy_sigv1pc} Face-on view of the velocity dispersion normalized to 1\,pc: $\sigma_0 = \sigma_v / R^{1/2}$. The symbol size is proportional to $\log(R)$ while the color indicates $\sigma_0$.}
\end{figure*}

\begin{figure*}
\centering
\includegraphics[draft=false, angle=0, width=11cm]{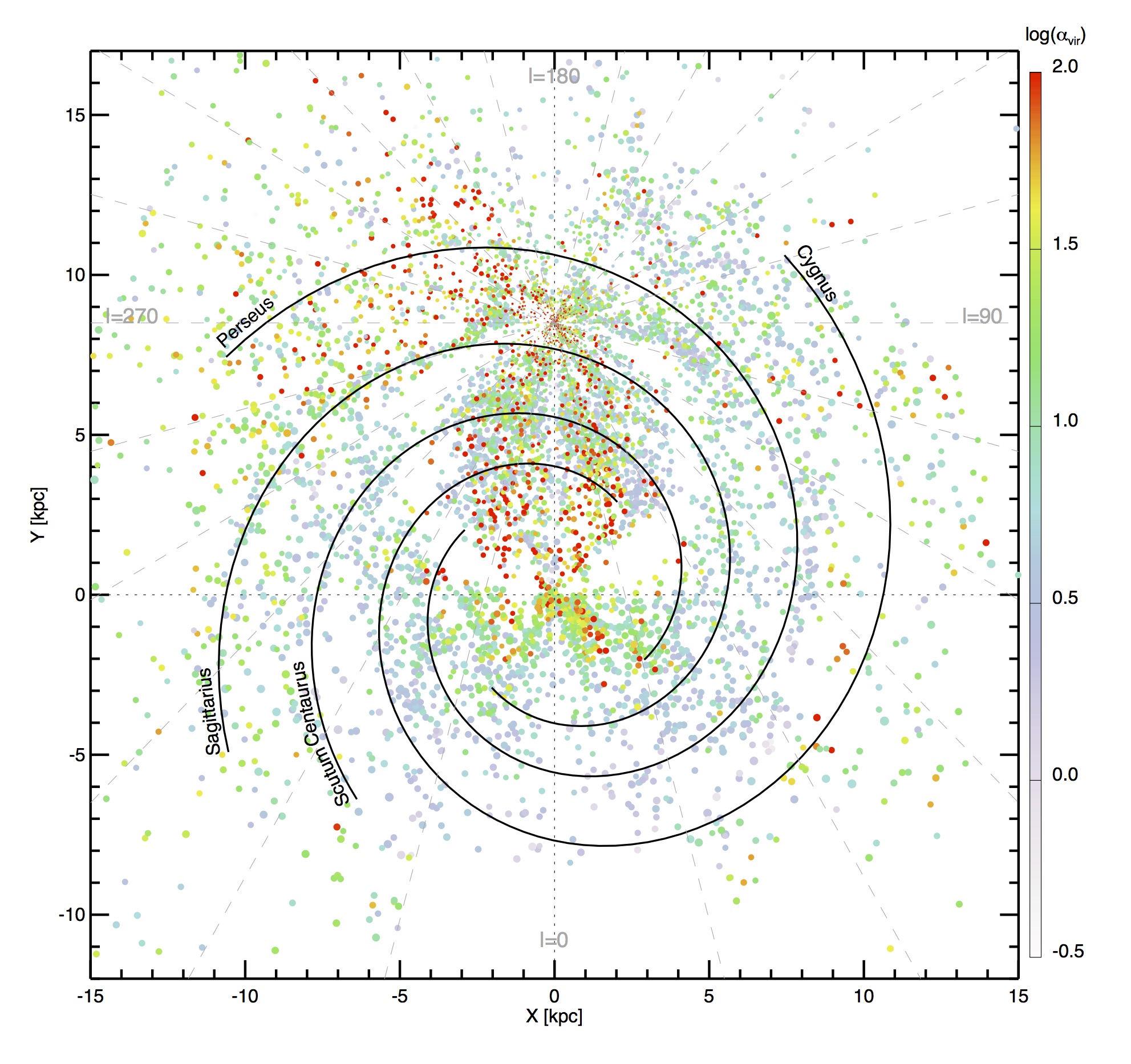}
\caption{\label{fig:xy_alpha} Face-on view of the virial parameter.
    The symbol size is proportional to $\log(R)$, while the color indicates $\log(\alpha_{\rm vir})$.}
\end{figure*}

\begin{figure*}
\centering
\includegraphics[draft=false, angle=0, width=11cm]{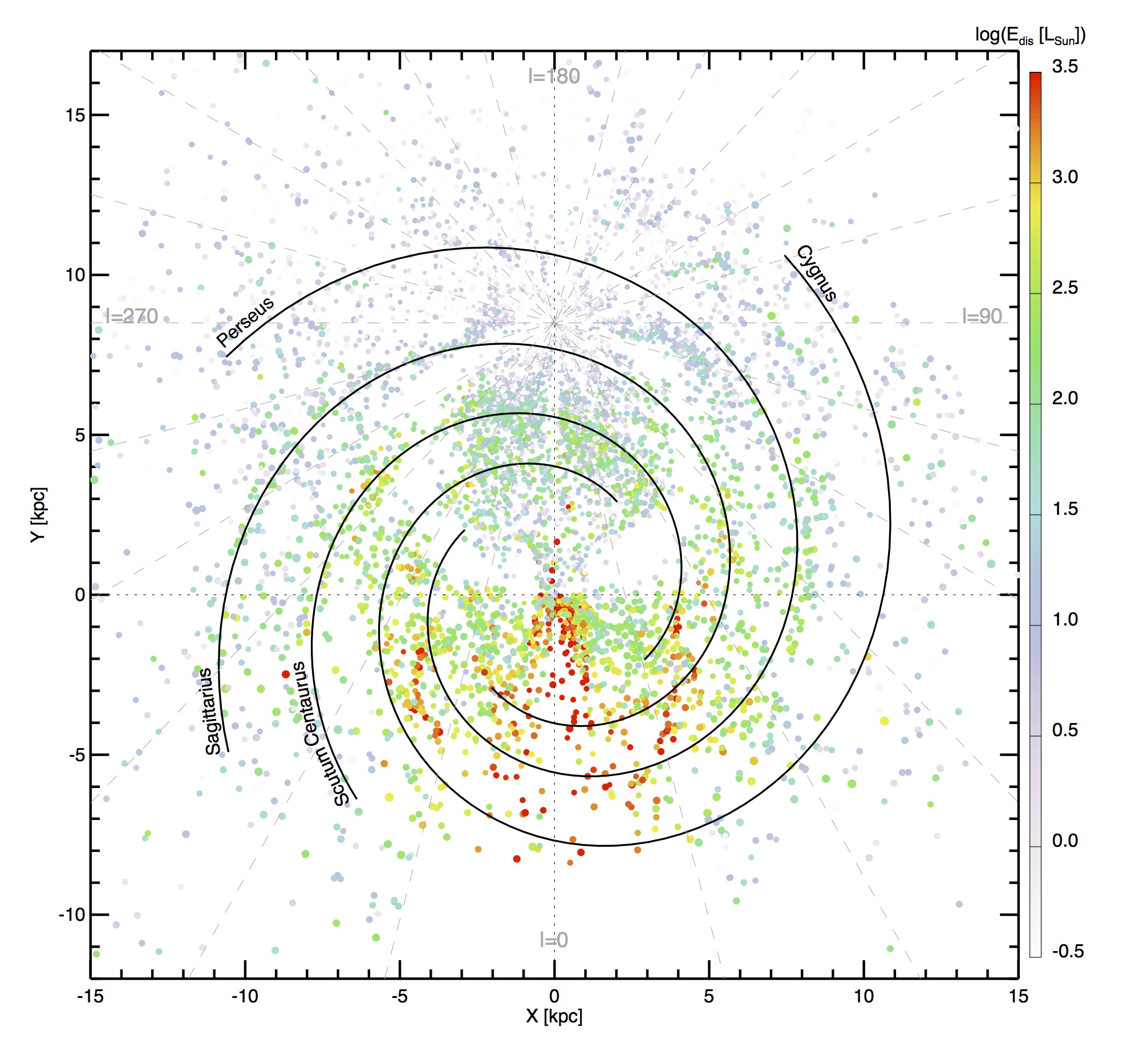}
\caption{\label{fig:xy_Edis} Face-on view of the turbulent energy dissipation rate.
    The symbol size is proportional to $\log(R)$, while the color indicates $\log(\dot{E}_{\rm dis})$.}
\end{figure*}

This appendix presents face-on views of the following parameters: $M$ (Figure~\ref{fig:xy_mass}), $n_{\rm H2}$ (Figure~\ref{fig:xy_density}), $z$ (Figure~\ref{fig:xy_height}), $P_{\rm int}$ (Figure~\ref{fig:xy_pressure}), $\sigma_0$ (Figure~\ref{fig:xy_sigv1pc}), $\alpha_{\rm vir}$ (Figure~\ref{fig:xy_alpha}), and $\dot{E}_{\rm dis}$ (Figure~\ref{fig:xy_Edis}).

  \end{appendix}


\clearpage

\bibliographystyle{aasjournal}
\bibliography{draft_v9}

\end{document}